\magnification=\magstephalf
%%%\Title{Multiple Singular Emission in Gauge Theories}
%%%\input header
%%% revised Dec 5-19 2003 to match final Phys Rev version + removal of comments
%%% about FDH scheme
\newbox\SlashedBox 
\def\slashed#1{\setbox\SlashedBox=\hbox{#1}
\hbox to 0pt{\hbox to 1\wd\SlashedBox{\hfil/\hfil}\hss}{#1}}
\def\hboxtosizeof#1#2{\setbox\SlashedBox=\hbox{#1}
\hbox to 1\wd\SlashedBox{#2}}

% The following is necessary so that we can get a partial slash
% inside a math display... sigh.
\def\mathslashed#1{\setbox\SlashedBox=\hbox{$#1$}
\hbox to 0pt{\hbox to 1\wd\SlashedBox{\hfil/\hfil}\hss}#1}

\def\ifsmall{\iffalse}  % default is unreduced.
\def\titlepagefont{}  % default is ordinary font.

% the ps: landscape must be the first special command in order
% to get the first page in landscape mode -- so we go through some
% contortions to define TeXgraphics in the default case.
\def\DefineTeXgraphics{%
\special{ps::[global] /TeXgraphics { } def}}  % No need to do anything

\def\today{\ifcase\month\or January\or February\or March\or April\or May
\or June\or July\or August\or September\or October\or November\or
December\fi\space\number\day, \number\year}
\def\eatPrefix19{}
\def\Year{\expandafter\eatPrefix\the\year}
\newcount\hours \newcount\minutes
\def\monthname{\ifcase\month\or
January\or February\or March\or April\or May\or June\or July\or
August\or September\or October\or November\or December\fi}
\def\shortmonthname{\ifcase\month\or
Jan\or Feb\or Mar\or Apr\or May\or Jun\or Jul\or
Aug\or Sep\or Oct\or Nov\or Dec\fi}

\def\TimeStamp{\hours\the\time\divide\hours by60%
\minutes -\the\time\divide\minutes by60\multiply\minutes by60%
\advance\minutes by\the\time%
${\rm \shortmonthname}\cdot\if\day<10{}0\fi\the\day\cdot\the\year%
\qquad\the\hours:\if\minutes<10{}0\fi\the\minutes$}

%\DefineTeXgraphics}

%\DefineTeXgraphics}

%\DefineTeXgraphics}

\def\Title#1{%
\vskip 1in{\titlefont\centerline{#1}}\vskip .5in}
%\DefineTeXgraphics}
 
\def\Date#1{\leftline{#1}\tenrm\supereject%
\global\hsize=\hsbody\global\hoffset=\hbodyoffset%
\footline={\hss\tenrm\folio\hss}}% restores pagenumbers

\newif\ifdraftmode
\newif\ifleftlabels  % Labels in left margins as well, for European-size paper

% Stolen from harvmac.tex 04/08/92
%       use \nolabels to get rid of eqn, ref, and fig labels in draft mode
\def\nolabels{\def\wrlabeL##1{}\def\eqlabeL##1{}\def\reflabeL##1{}}
\def\writelabels{\def\wrlabeL##1{\leavevmode\vadjust{\rlap{\smash%
{\line{{\escapechar=` \hfill\rlap{\sevenrm\hskip.03in\string##1}}}}}}}%
\def\eqlabeL##1{{\escapechar-1\rlap{\sevenrm\hskip.05in\string##1}}}%
\def\reflabeL##1{\noexpand\rlap{\noexpand\sevenrm[\string##1]}}}
\def\writeleftlabels{\def\wrlabeL##1{\leavevmode\vadjust{\rlap{\smash%
{\line{{\escapechar=` \hfill\rlap{\sevenrm\hskip.03in\string##1}}}}}}}%
\def\eqlabeL##1{{\escapechar-1%
\rlap{\sixrm\hskip.05in\string##1}%
\llap{\sevenrm\string##1\hskip.03in\hbox to \hsize{}}}}%
\def\reflabeL##1{\noexpand\rlap{\noexpand\sevenrm[\string##1]}}}
\nolabels

\input hyperbasics.tex

\newdimen\fullhsize
\newdimen\hstitle
\hstitle=\hsize % default
\newdimen\hsbody
\hsbody=\hsize % default
\newdimen\hbodyoffset
\hbodyoffset=\hoffset % default
\newbox\leftpage
\def\abstract#1{#1}
\def\rotated{\special{ps: landscape}
\magnification=1000  % This line must come before we change vsize,
                     % since \magnification sets it to a fixed value.
\baselineskip=14pt
\global\hstitle=9truein\global\hsbody=4.75truein
\global\vsize=7truein\global\voffset=-.31truein
\global\hoffset=-0.54in\global\hbodyoffset=-.54truein
\global\fullhsize=10truein
\def\DefineTeXgraphics{%
\special{ps::[global] 
/TeXgraphics {currentpoint translate 0.7 0.7 scale
              -80 0.72 mul -1000 0.72 mul translate} def}}
 % 0.7 is slightly less than the ratio of horizontal sizes: 4.75 to 6.5
\let\lr=L
\def\ifsmall{\iftrue}
\def\titlepagefont{\twelvepoint}
\trueseventeenpoint
\def\almostshipout##1{\if L\lr \count1=1
      \global\setbox\leftpage=##1 \global\let\lr=R
   \else \count1=2
      \shipout\vbox{\hbox to\fullhsize{\box\leftpage\hfil##1}}
      \global\let\lr=L\fi}

\output={\ifnum\count0=1 %%% This is the HUTP version
 \shipout\vbox{\hbox to \fullhsize{\hfill\pagebody\hfill}}\advancepageno
 \else
 \almostshipout{\leftline{\vbox{\pagebody\makefootline}}}\advancepageno 
 \fi}

\def\abstract##1{{\leftskip=1.5in\rightskip=1.5in ##1\par}} }

% Messages on lines by themselves
\def\linemessage#1{\immediate\write16{#1}}

% tagged sec numbers
\global\newcount\secno \global\secno=0
\global\newcount\appno \global\appno=0
\global\newcount\meqno \global\meqno=1
\global\newcount\subsecno \global\subsecno=0
% and figure numbers
\global\newcount\figno \global\figno=0

\newif\ifAnyCounterChanged
% If we are comparing numbers, there's no special problem.
% But if we are comparing roman numerals, we must be careful, because
% stuff read in from the aux file would be made up of ordinary
% characters (category code = 11), whereas \romannumeral generates
% characters with category code = 12..., so the stuff from the
% current run won't appear equal to the previous definition, as far
% as \warnIfChanged is concerned.
% To get around this, we have a macro \makeNormal, which converts
% letters `ivxlcdmIVXLCDM' to normal letters, no matter what their category
% code.  The macro has the convoluted form it does, with aftergroup's & all,
% to avoid blowing up TeX...
% The macro is used below in makeNormalizedRomappno, by which means we
% define the appendix counters to be strings containing vanilla versions
% of the letters... Sigh
\let\terminator=\relax
% The string to be normalized must not contain { and } tokens...
\def\normalize#1{\ifx#1\terminator\let\next=\relax\else%
\if#1i\aftergroup i\else\if#1v\aftergroup v\else\if#1x\aftergroup x%
\else\if#1l\aftergroup l\else\if#1c\aftergroup c\else%
\if#1m\aftergroup m\else%
\if#1I\aftergroup I\else\if#1V\aftergroup V\else\if#1X\aftergroup X%
\else\if#1L\aftergroup L\else\if#1C\aftergroup C\else%
\if#1M\aftergroup M\else\aftergroup#1\fi\fi\fi\fi\fi\fi\fi\fi\fi\fi\fi\fi%
\let\next=\normalize\fi%
\next}
% makes #1 a normalized version of #2...
\def\makeNormal#1#2{\def\doNormalDef{\edef#1}\begingroup%
\aftergroup\doNormalDef\aftergroup{\normalize#2\terminator\aftergroup}%
\endgroup}
% makes a normalized version of its argument:

\def\warnIfChanged#1#2{%
\ifundef#1% skip it
\else\begingroup%
\edef\oldDefinitionOfCounter{#1}\edef\newDefinitionOfCounter{#2}%
%\message{old: \oldDefinitionOfCounter}%
%\message{new: \newDefinitionOfCounter}%
\ifx\oldDefinitionOfCounter\newDefinitionOfCounter%
\else%
\linemessage{Warning: definition of \noexpand#1 has changed.}%
\global\AnyCounterChangedtrue\fi\endgroup\fi}

\def\Section#1{\global\advance\secno by1\relax\global\meqno=1%
\global\subsecno=0%
\bigbreak\bigskip% (combination \goodbreak\bigskip\bigskip)
\centerline{\twelvepoint \bf %
\the\secno. #1}%
\par\nobreak\medskip\nobreak}
\def\tagsection#1{%
\warnIfChanged#1{\the\secno}%
\xdef#1{\the\secno}%
\ifWritingAuxFile\immediate\write\auxfile{\noexpand\xdef\noexpand#1{#1}}\fi%
}
\def\section{\Section}
\def\Subsection#1{\global\advance\subsecno by1\relax\medskip %
\leftline{\bf\the\secno.\the\subsecno\ #1}%
\par\nobreak\smallskip\nobreak}
\def\tagsubsection#1{%
\warnIfChanged#1{\the\secno.\the\subsecno}%
\xdef#1{\the\secno.\the\subsecno}%
\ifWritingAuxFile\immediate\write\auxfile{\noexpand\xdef\noexpand#1{#1}}\fi%
}

\def\subsection{\Subsection}

\def\romappno{\uppercase\expandafter{\romannumeral\appno}}
\def\makeNormalizedRomappno{%
\expandafter\makeNormal\expandafter\normalizedromappno%
\expandafter{\romannumeral\appno}%
\edef\normalizedromappno{\uppercase{\normalizedromappno}}}
\def\Appendix#1{\global\advance\appno by1\relax\global\meqno=1\global\secno=0%
\global\subsecno=0%
\bigbreak\bigskip% (combination \goodbreak\bigskip\bigskip)
\centerline{\twelvepoint \bf Appendix %
\romappno. #1}%
\par\nobreak\medskip\nobreak}
\def\tagappendix#1{\makeNormalizedRomappno%
\warnIfChanged#1{\normalizedromappno}%
\xdef#1{\normalizedromappno}%
\ifWritingAuxFile\immediate\write\auxfile{\noexpand\xdef\noexpand#1{#1}}\fi%
}
\def\appendix{\Appendix}
\def\Subappendix#1{\global\advance\subsecno by1\relax\medskip %
\leftline{\bf\romappno.\the\subsecno\ #1}%
\par\nobreak\smallskip\nobreak}
\def\tagsubappendix#1{\makeNormalizedRomappno%
\warnIfChanged#1{\normalizedromappno.\the\subsecno}%
\xdef#1{\normalizedromappno.\the\subsecno}%
\ifWritingAuxFile\immediate\write\auxfile{\noexpand\xdef\noexpand#1{#1}}\fi%
}

\def\eqn#1{\makeNormalizedRomappno%
\ifnum\secno>0%
  \warnIfChanged#1{\the\secno.\the\meqno}%
  \eqno(\the\secno.\the\meqno)\xdef#1{\the\secno.\the\meqno}%
     \global\advance\meqno by1
\else\ifnum\appno>0%
  \warnIfChanged#1{\normalizedromappno.\the\meqno}%
  \eqno({\rm\romappno}.\the\meqno)%
      \xdef#1{\normalizedromappno.\the\meqno}%
     \global\advance\meqno by1
\else%
  \warnIfChanged#1{\the\meqno}%
  \eqno(\the\meqno)\xdef#1{\the\meqno}%
     \global\advance\meqno by1
\fi\fi%
\eqlabeL#1%
\ifWritingAuxFile\immediate\write\auxfile{\noexpand\xdef\noexpand#1{#1}}\fi%
}
\def\defeqn#1{\makeNormalizedRomappno%
\ifnum\secno>0%
  \warnIfChanged#1{\the\secno.\the\meqno}%
  \xdef#1{\the\secno.\the\meqno}%
     \global\advance\meqno by1
\else\ifnum\appno>0%
  \warnIfChanged#1{\normalizedromappno.\the\meqno}%
  \xdef#1{\normalizedromappno.\the\meqno}%
     \global\advance\meqno by1
\else%
  \warnIfChanged#1{\the\meqno}%
  \xdef#1{\the\meqno}%
     \global\advance\meqno by1
\fi\fi%
\eqlabeL#1%
\ifWritingAuxFile\immediate\write\auxfile{\noexpand\xdef\noexpand#1{#1}}\fi%
}
\def\anoneqn{\makeNormalizedRomappno%
\ifnum\secno>0
  \eqno(\the\secno.\the\meqno)%
     \global\advance\meqno by1
\else\ifnum\appno>0
  \eqno({\rm\normalizedromappno}.\the\meqno)%
     \global\advance\meqno by1
\else
  \eqno(\the\meqno)%
     \global\advance\meqno by1
\fi\fi%
}
\def\mfig#1#2{\ifx#20%unnumbered figure
\else\global\advance\figno by1%
\relax#1\the\figno%
\warnIfChanged#2{\the\figno}%
\xdef#2{\the\figno}%
\reflabeL#2%
\ifWritingAuxFile\immediate\write\auxfile{\noexpand\xdef\noexpand#2{#2}}\fi\fi%
}

\catcode`@=11 % borrow the private macros of PLAIN (with care)

% \LoadFigure is used to put a figure into the text.  Its first argument
% is the symbolic name for the figure (if it isn't defined, a new number
% will be assigned);  the second argument is a caption;
% the third argument size information in the form 
% \epsfxsize=3.0in\epsfysize=3.5in (this argument may be blank and
% may contain any valid preparatory argument used by the epsf package);
% the fourth and last argument is the name of the file which contains the 
% figure.
% The macro is basically just a front-end for \epsfbox; its purpose is 
% to allow figures to be switched from placement in the running text
% to placement on a separate page at the end of the text.  This choice
% is made using the flag \FiguresInText{true,false}; in the latter case,
% figures are placed at the end, size information is ignored (figures
% will be full-size), and the captions are listed separately on a page
% when the \listfigs command is invoked, followed by the figures, each
% on a separate page.  
%  The epsf package must be loaded by the user.
%  To change the size of captions in the text, redefine \captionsize.
\newif\ifFiguresInText\FiguresInTexttrue
\newif\if@FigureFileCreated
\newwrite\capfile
\newwrite\figfile

%default
\newif\ifcaption
\captiontrue
\def\captionsize{\tenrm}
\def\PlaceTextFigure#1#2#3#4{%
\vskip 0.5truein%
#3\hfil\epsfbox{#4}\hfil\break%
\ifcaption\vskip 5pt\hfil\vbox{\captionsize Figure #1. #2}\hfil\fi%
\vskip10pt}
\def\PlaceEndFigure#1#2{%
\epsfxsize=\hsize\epsfbox{#2}\vfill\centerline{Figure #1.}\eject}

\def\LoadFigure#1#2#3#4{%
\ifundef#1{\phantom{\mfig{}#1}}\else%  Write out definition only if it's new.
% 10/08/97, presumably this was because something broke, but for forward
% references we have to write out old ones too...
%\warnIfChanged#1{\the\secno}  always warns -- why?
\ifx#10% unnumbered figure
\else%\warnIfChanged#1{\the\figno}%
\ifWritingAuxFile\immediate\write\auxfile{\noexpand\xdef\noexpand#1{#1}}\fi\fi
\fi%
\ifFiguresInText% Figure is immediate
\PlaceTextFigure{#1}{#2}{#3}{#4}%
\else% Figure is at the end
\if@FigureFileCreated\else%
\immediate\openout\capfile=\jobname.caps%
\immediate\openout\figfile=\jobname.figs%
@FigureFileCreatedtrue\fi%
\immediate\write\capfile{\noexpand\item{Figure \noexpand#1.\ }{#2}\vskip10pt}%
\immediate\write\figfile{\noexpand\PlaceEndFigure\noexpand#1{\noexpand#4}}%
\fi}

\def\listfigs{\ifFiguresInText\else%
\vfill\eject\immediate\closeout\capfile%\parindent=20pt
\immediate\closeout\figfile%
\centerline{{\bf Figures}}\bigskip\frenchspacing%
\catcode`@=11 % borrow the private macros of PLAIN (with care)
\def\captionsize{\tenrm}
\input \jobname.caps\vfill\eject\nonfrenchspacing%
\catcode`\@=\active
\catcode`@=12  % No longer.
\input\jobname.figs\fi}

%\font\titlefont=cmr10 at 16pt
\font\ninerm=cmr9
\font\eightrm=cmr8
\font\sixrm=cmr6

\def\loadtrueseventeenpoint{
 \font\seventeenrm=cmr10 at 17.28truept
 \font\seventeeni=cmmi10 at 17.28truept
 \font\seventeenbf=cmbx10 at 17.28truept
 \font\seventeenit=cmti10 at 17.28truept
 \font\seventeensl=cmsl10 at 17.28truept
 \font\seventeensy=cmsy10 at 17.28truept
}
\def\loadfourteenpoint{
\font\fourteenrm=cmr10 at 14.4pt
\font\fourteeni=cmmi10 at 14.4pt
\font\fourteenit=cmti10 at 14.4pt
\font\fourteensl=cmsl10 at 14.4pt
\font\fourteensy=cmsy10 at 14.4pt
\font\fourteenbf=cmbx10 at 14.4pt
}
\def\loadtruetwelvepoint{
\font\twelverm=cmr10 at 12truept
\font\twelvei=cmmi10 at 12truept
\font\twelveit=cmti10 at 12truept
\font\twelvesl=cmsl10 at 12truept
\font\twelvesy=cmsy10 at 12truept
\font\twelvebf=cmbx10 at 12truept
\font\twelvesc=cmcsc10 at 12truept
}

\font\ninei=cmmi9
\font\eighti=cmmi8
\font\sixi=cmmi6
\skewchar\ninei='177 \skewchar\eighti='177 \skewchar\sixi='177

\font\ninesy=cmsy9
\font\eightsy=cmsy8
\font\sixsy=cmsy6
\skewchar\ninesy='60 \skewchar\eightsy='60 \skewchar\sixsy='60

\font\ninebf=cmbx9
\font\eightbf=cmbx8
\font\sixbf=cmbx6

\font\ninett=cmtt9
\font\eighttt=cmtt8

\hyphenchar\tentt=-1 % inhibit hyphenation in typewriter type
\hyphenchar\ninett=-1
\hyphenchar\eighttt=-1         

\font\ninesl=cmsl9
\font\eightsl=cmsl8

\font\nineit=cmti9
\font\eightit=cmti8
\font\sevenit=cmti7

% Added 07/09/97, initially, 10-point...
\scriptfont\itfam=\sevenit

 % unslanted text italic
                      
\newskip\ttglue
\def\tenpoint{\def\rm{\fam0\tenrm}%
  \textfont0=\tenrm \scriptfont0=\sevenrm \scriptscriptfont0=\fiverm
  \textfont1=\teni \scriptfont1=\seveni \scriptscriptfont1=\fivei
  \textfont2=\tensy \scriptfont2=\sevensy \scriptscriptfont2=\fivesy
  \textfont3=\tenex \scriptfont3=\tenex \scriptscriptfont3=\tenex
  \def\it{\fam\itfam\tenit}%
      \textfont\itfam=\tenit\scriptfont\itfam=\sevenit
  \def\sl{\fam\slfam\tensl}\textfont\slfam=\tensl
  \def\bf{\fam\bffam\tenbf}\textfont\bffam=\tenbf \scriptfont\bffam=\sevenbf
  \scriptscriptfont\bffam=\fivebf
  \normalbaselineskip=12pt
  \let\sc=\eightrm
  \let\big=\tenbig
  \setbox\strutbox=\hbox{\vrule height8.5pt depth3.5pt width\z@}%
  \normalbaselines\rm}

\def\twelvepoint{\def\rm{\fam0\twelverm}%
  \textfont0=\twelverm \scriptfont0=\ninerm \scriptscriptfont0=\sevenrm
  \textfont1=\twelvei \scriptfont1=\ninei \scriptscriptfont1=\seveni
  \textfont2=\twelvesy \scriptfont2=\ninesy \scriptscriptfont2=\sevensy
  \textfont3=\tenex \scriptfont3=\tenex \scriptscriptfont3=\tenex
  \def\it{\fam\itfam\twelveit}\textfont\itfam=\twelveit
  \def\sl{\fam\slfam\twelvesl}\textfont\slfam=\twelvesl
  \def\bf{\fam\bffam\twelvebf}\textfont\bffam=\twelvebf%
  \scriptfont\bffam=\ninebf
  \scriptscriptfont\bffam=\sevenbf
  \normalbaselineskip=12pt
  \let\sc=\eightrm
  \let\big=\tenbig
  \setbox\strutbox=\hbox{\vrule height8.5pt depth3.5pt width\z@}%
  \normalbaselines\rm}

\def\fourteenpoint{\def\rm{\fam0\fourteenrm}%
  \textfont0=\fourteenrm \scriptfont0=\tenrm \scriptscriptfont0=\sevenrm
  \textfont1=\fourteeni \scriptfont1=\teni \scriptscriptfont1=\seveni
  \textfont2=\fourteensy \scriptfont2=\tensy \scriptscriptfont2=\sevensy
  \textfont3=\tenex \scriptfont3=\tenex \scriptscriptfont3=\tenex
  \def\it{\fam\itfam\fourteenit}\textfont\itfam=\fourteenit
  \def\sl{\fam\slfam\fourteensl}\textfont\slfam=\fourteensl
  \def\bf{\fam\bffam\fourteenbf}\textfont\bffam=\fourteenbf%
  \scriptfont\bffam=\tenbf
  \scriptscriptfont\bffam=\sevenbf
  \normalbaselineskip=17pt
  \let\sc=\elevenrm
  \let\big=\tenbig                                          
  \setbox\strutbox=\hbox{\vrule height8.5pt depth3.5pt width\z@}%
  \normalbaselines\rm}

\def\seventeenpoint{\def\rm{\fam0\seventeenrm}%
  \textfont0=\seventeenrm \scriptfont0=\fourteenrm \scriptscriptfont0=\tenrm
  \textfont1=\seventeeni \scriptfont1=\fourteeni \scriptscriptfont1=\teni
  \textfont2=\seventeensy \scriptfont2=\fourteensy \scriptscriptfont2=\tensy
  \textfont3=\tenex \scriptfont3=\tenex \scriptscriptfont3=\tenex
  \def\it{\fam\itfam\seventeenit}\textfont\itfam=\seventeenit
  \def\sl{\fam\slfam\seventeensl}\textfont\slfam=\seventeensl
  \def\bf{\fam\bffam\seventeenbf}\textfont\bffam=\seventeenbf%
  \scriptfont\bffam=\fourteenbf
  \scriptscriptfont\bffam=\twelvebf
  \normalbaselineskip=21pt
  \let\sc=\fourteenrm
  \let\big=\tenbig                                          
  \setbox\strutbox=\hbox{\vrule height 12pt depth 6pt width\z@}%
  \normalbaselines\rm}

\def\ninepoint{\def\rm{\fam0\ninerm}%
  \textfont0=\ninerm \scriptfont0=\sixrm \scriptscriptfont0=\fiverm
  \textfont1=\ninei \scriptfont1=\sixi \scriptscriptfont1=\fivei
  \textfont2=\ninesy \scriptfont2=\sixsy \scriptscriptfont2=\fivesy
  \textfont3=\tenex \scriptfont3=\tenex \scriptscriptfont3=\tenex
  \def\it{\fam\itfam\nineit}\textfont\itfam=\nineit
  \def\sl{\fam\slfam\ninesl}\textfont\slfam=\ninesl
  \def\bf{\fam\bffam\ninebf}\textfont\bffam=\ninebf \scriptfont\bffam=\sixbf
  \scriptscriptfont\bffam=\fivebf
  \normalbaselineskip=11pt
  \let\sc=\sevenrm
  \let\big=\ninebig
  \setbox\strutbox=\hbox{\vrule height8pt depth3pt width\z@}%
  \normalbaselines\rm}

\def\eightpoint{\def\rm{\fam0\eightrm}%
  \textfont0=\eightrm \scriptfont0=\sixrm \scriptscriptfont0=\fiverm%
  \textfont1=\eighti \scriptfont1=\sixi \scriptscriptfont1=\fivei%
  \textfont2=\eightsy \scriptfont2=\sixsy \scriptscriptfont2=\fivesy%
  \textfont3=\tenex \scriptfont3=\tenex \scriptscriptfont3=\tenex%
  \def\it{\fam\itfam\eightit}\textfont\itfam=\eightit%
  \def\sl{\fam\slfam\eightsl}\textfont\slfam=\eightsl%
  \def\bf{\fam\bffam\eightbf}\textfont\bffam=\eightbf \scriptfont\bffam=\sixbf%
  \scriptscriptfont\bffam=\fivebf%
  \normalbaselineskip=9pt%
  \let\sc=\sixrm%
  \let\big=\eightbig%
  \setbox\strutbox=\hbox{\vrule height7pt depth2pt width\z@}%
  \normalbaselines\rm}
  \let\sc=\eightrm

 % use after $ in ninepoint sections
\def\tenbig#1{{\hbox{$\left#1\vbox to8.5pt{}\right.\n@space$}}}
\def\ninebig#1{{\hbox{$\textfont0=\tenrm\textfont2=\tensy
  \left#1\vbox to7.25pt{}\right.\n@space$}}}
\def\eightbig#1{{\hbox{$\textfont0=\ninerm\textfont2=\ninesy
  \left#1\vbox to6.5pt{}\right.\n@space$}}}

% Page layout
%\newinsert\footins
\def\footnote#1{\edef\@sf{\spacefactor\the\spacefactor}#1\@sf
      \insert\footins\bgroup\eightpoint
      \interlinepenalty100 \let\par=\endgraf
        \leftskip=\z@skip \rightskip=\z@skip
        \splittopskip=10pt plus 1pt minus 1pt \floatingpenalty=20000
        \smallskip\item{#1}\bgroup\strut\aftergroup\@foot\let\next}
\skip\footins=12pt plus 2pt minus 4pt % space added when footnote is present
%\count\footins=1000 % footnote magnification factor (1 to 1)
\dimen\footins=30pc % maximum footnotes per page

\newinsert\margin
\dimen\margin=\maxdimen
%\count\margin=0 \skip\margin=0pt % marginal inserts take up no space
\def\titlefont{\seventeenpoint}
\loadtruetwelvepoint % At FNAL...
\loadtrueseventeenpoint

% \use\cs
% puts in the expansion of `\cs' if it's defined, the literal "\cs" otherwise.
\def\eatOne#1{}
\def\ifundef#1{\expandafter\ifx%
\csname\expandafter\eatOne\string#1\endcsname\relax}
\def\notTrue{\iffalse}\def\isTrue{\iftrue}
\def\ifdef#1{{\ifundef#1%
\aftergroup\notTrue\else\aftergroup\isTrue\fi}}
\def\use#1{\ifundef#1\linemessage{Warning: \string#1 is undefined.}%
{\tt \string#1}\else#1\fi}

%     \ref\label{text}
% generates a number, assigns it to \label, generates an entry.
% To list the refs on a separate page,  \listrefs
% \nref does the same without generating any text at the reference
% point
% June 26 1994: \preref postpones the generation of an entry, along with
% the text, until the first use of the reference

% 09/14/95: Added html...
%\def\hyperref#1#2{\special{html:<a href=\quote#1\quote>}{#2}\special{html:</a>}}
% 09/25/95: Now using Tanmoy Battacharya's macros...

%
\catcode`"=11
\let\quote="
\catcode`"=12
\chardef\foo="22
\global\newcount\refno \global\refno=1
\newwrite\rfile
\newlinechar=`\^^J
\def\@ref#1#2{\the\refno\n@ref#1{#2}}
% Added 09/14/95{\the\refno\n@ref#1{#2}}
\def\h@ref#1#2#3{\href{#3}{\the\refno}\n@ref#1{#2}}
\def\n@ref#1#2{\xdef#1{\the\refno}%
\ifnum\refno=1\immediate\openout\rfile=\jobname.refs\fi%
\immediate\write\rfile{\noexpand\item{[\noexpand#1]\ }#2.}%
\global\advance\refno by1}
\def\nref{\n@ref} % Hide to allow redefinitions of \ref,\nref to \preref
\def\ref{\@ref}   % without breaking the latter...
\def\hrref{\h@ref}
% To start a long reference...
\def\lref#1#2{\the\refno\xdef#1{\the\refno}%
\ifnum\refno=1\immediate\openout\rfile=\jobname.refs\fi%
\immediate\write\rfile{\noexpand\item{[\noexpand#1]\ }#2\semi}%
\global\advance\refno by1}
% To continue a long reference...
\def\cref#1{\immediate\write\rfile{#1\semi}}
% To end a long reference...

\def\preref#1#2{\gdef#1{\@ref#1{#2}}}

\def\semi{;\hfil\noexpand\break}

\def\listrefs{\vfill\eject\immediate\closeout\rfile%\parindent=20pt
\centerline{{\bf References}}\bigskip\frenchspacing%
\input \jobname.refs\vfill\eject\nonfrenchspacing}

\def\inputAuxIfPresent#1{\immediate\openin1=#1
\ifeof1\message{No file \auxfileName; I'll create one.
}\else\closein1\relax\input\auxfileName\fi%
}
% For references, some journal names

%and archives...

%\def\hepphref#1{\hyperref{http://xxx.lanl.gov/abs/hep-ph/#1}{archive}%
%{hep-ph/#1}{hep-ph/#1}}

%\def\hepphref#1{\href{http://xxx.lanl.gov/abs/hep-ph/#1}{hep-ph/#1}}

% An .aux file --- for forward references...
\newif\ifWritingAuxFile
\newwrite\auxfile
\def\SetUpAuxFile{%
\xdef\auxfileName{\jobname.aux}%
% Read it in if it exists
\inputAuxIfPresent{\auxfileName}%
% Now write a new one.
\WritingAuxFiletrue%
\immediate\openout\auxfile=\auxfileName}

% Some generally useful notation
\def\L{\left(}\def\R{\right)}
\def\LP{\left.}\def\RP{\right.}
\def\LB{\left[}\def\RB{\right]}

\def\RV{\right|}

% Warn about changed counters...
\def\bye{\par\vfill\supereject%
\ifAnyCounterChanged\linemessage{
Some counters have changed.  Re-run tex to fix them up.}\fi%
\end}

\catcode`\@=\active
\catcode`@=12  % No longer.
\catcode`\"=\active

%%%end header
%%%\input gaugedefs
\def\L{\left(}
\def\R{\right)}

\def\Tr{\mathop{\rm Tr}\nolimits}

\def\LP{\left.}\def\RP{\right.}

\def\pol{\varepsilon}

\def\dl^#1_#2{\delta^{#1}{}_{#2}}

\catcode`@=11  % Make @ letter.
\def\meqalign#1{\,\vcenter{\openup1\jot\m@th
   \ialign{\strut\hfil$\displaystyle{##}$ && $\displaystyle{{}##}$\hfil
             \crcr#1\crcr}}\,}
\catcode`@=12  % No longer.

% counters

% General parameters
\baselineskip 15pt
\overfullrule 0.5pt
%%%end gaugedefs
%%%\input spinordef

\def\Tr{\mathop{\rm Tr}\nolimits}

\def\pol{\varepsilon}

\def\L{\left(}\def\R{\right)}
\def\LP{\left.}\def\RP{\right.}
\def\spa#1.#2{\left\langle#1\,#2\right\rangle}
\def\spb#1.#2{\left[#1\,#2\right]}
\def\lor#1.#2{\left(#1\,#2\right)}
\def\sand#1.#2.#3{%
\left\langle\smash{#1}{\vphantom1}^{-}\right|{#2}%
\left|\smash{#3}{\vphantom1}^{-}\right\rangle}
\def\sandp#1.#2.#3{%
\left\langle\smash{#1}{\vphantom1}^{-}\right|{#2}%
\left|\smash{#3}{\vphantom1}^{+}\right\rangle}
\def\sandpp#1.#2.#3{%
\left\langle\smash{#1}{\vphantom1}^{+}\right|{#2}%
\left|\smash{#3}{\vphantom1}^{+}\right\rangle}
\def\sandpm#1.#2.#3{%
\left\langle\smash{#1}{\vphantom1}^{+}\right|{#2}%
\left|\smash{#3}{\vphantom1}^{-}\right\rangle}
\def\sandmp#1.#2.#3{%
   \left\langle\smash{#1}{\vphantom1}^{-}\right|{#2}%
    \left|\smash{#3}{\vphantom1}^{+}\right\rangle}
\catcode`@=11  % Make @ letter.
\def\meqalign#1{\,\vcenter{\openup1\jot\m@th
   \ialign{\strut\hfil$\displaystyle{##}$ && $\displaystyle{{}##}$\hfil
             \crcr#1\crcr}}\,}
\catcode`@=12  % No longer.

%%%end spinordef
%%%\input epsf
%%%end epsf

%%%% Paper Body start
\SetUpAuxFile
\hfuzz 20pt
\overfullrule 0pt

%%%% Some Definitions

%=================================================================
% necessary fonts
%-----------------------------------------------------------------
%   \font\fivebf  =cmbx10  scaled 500 % five point bold
%   \font\sevenbf =cmbx10  scaled 700 % seven point bold
%   \font\tenbf   =cmbx10             % ten point bold
%   \font\fivemb  =cmmib10 scaled 500 % five point math bold
       % five point math bold
%   \font\sevenmb =cmmib10 scaled 700 % seven point math bold
       % seven point math bold
               % ten point math bold
%=================================================================
% Math in Bold
% example $\boldmath{..}$   { math characters will be bold }
%-----------------------------------------------------------------

%=================================================================
\def\e{\epsilon}
\def\tree{{\rm tree\vphantom{p}}}
\def\dash{\hbox{-\kern-.02em}}

\def\Split{\mathop{\rm C}\nolimits}
\def\Soft{\mathop{\rm Soft}\nolimits}

\def\Ant{\mathop{\rm Ant}\nolimits}
\def\Ctree{\Split^\tree}

\def\phpol{{\rm ph.\ pol.}}
\def\llongrightarrow{%
\relbar\mskip-0.5mu\joinrel\mskip-0.5mu\relbar\mskip-0.5mu\joinrel\longrightarrow}
\def\inlimit^#1{\buildrel#1\over\llongrightarrow}
\def\frac#1#2{{#1\over #2}}

%%% A bibliography
\preref\DixonTASI{L.\ Dixon, in 
{\it QCD \& Beyond: Proceedings of TASI '95}, 
ed. D.\ E.\ Soper (World Scientific, 1996) [hep-ph/9601359]}
%\DelDucaHA
\preref\DDFM{
V.\ Del Duca, A.\ Frizzo and F.\ Maltoni,
%``Factorization of tree QCD amplitudes in the high-energy limit and in  the collinear limit,''
Nucl.\ Phys.\ B568:211 (2000) [hep-ph/9909464]%
%%CITATION = HEP-PH 9909464;%%
}
\preref\GloverCampbell{
J.\ M.\ Campbell and E.\ W.\ N.\ Glover,
%``Double unresolved approximations to multiparton scattering amplitudes,''
Nucl.\ Phys.\ B527:264 (1998) [hep-ph/9710255]%
%%CITATION = HEP-PH 9710255;%%
}
\preref\Color{%
F.\ A.\ Berends and W.\ T.\ Giele,
Nucl.\ Phys.\ B294:700 (1987)\semi
D.\ A.\ Kosower, B.-H.\ Lee and V.\ P.\ Nair, Phys.\ Lett.\ 201B:85 (1988)\semi
M.\ Mangano, S.\ Parke and Z.\ Xu, Nucl.\ Phys.\ B298:653 (1988)\semi
Z.\ Bern and D.\ A.\ Kosower, Nucl.\ Phys.\ B362:389 (1991)}

\preref\Calkul{%
F.\ A.\ Berends, R.\ Kleiss, P.\ De Causmaecker, R.\ Gastmans, and T.\ T.\ Wu,
        Phys.\ Lett.\ 103B:124 (1981)\semi
P.\ De Causmaeker, R.\ Gastmans,  W.\ Troost, and  T.\ T.\ Wu,
Nucl.\ Phys.\ B206:53 (1982)}
\preref\Spinor{Z.\ Xu, D.-H.\ Zhang, L.\ Chang, Tsinghua University
                  preprint TUTP--84/3 (1984), unpublished\semi
Z.\ Xu, D.-H.\ Zhang, and L.\ Chang, Nucl.\ Phys.\ B291:392 (1987)}
\preref\OtherSpinor{R.\ Kleiss and W.\ J.\ Stirling, 
   Nucl.\ Phys.\ B262:235 (1985)\semi
   J.\ F.\ Gunion and Z.\ Kunszt, Phys.\ Lett.\ 161B:333 (1985)} 
\preref\Recurrence{F.\ A.\ Berends and W.\ T.\ Giele, 
Nucl.\ Phys.\ B306:759 (1988)}
\preref\HelicityRecurrence{
D.\ A.\ Kosower,
%``Light Cone Recurrence Relations For QCD Amplitudes,''
Nucl.\ Phys.\ B335:23 (1990)%
%%CITATION = NUPHA,B335,23;%%
}
\preref\BernChalmers{
Z.\ Bern and G.\ Chalmers,
%``Factorization in one loop gauge theory,''
Nucl.\ Phys.\ B447:465 (1995) [hep-ph/9503236]%
%%CITATION = HEP-PH 9503236;%%
}
\preref\MP{M.\ Mangano and S.\ J.\ Parke, Phys.\ Rep.\ 200:301 (1991)}
\preref\StringBased{Z.\ Bern and D.\ A.\ Kosower, Phys.\ Rev.\ Lett.\ 66:1669 (1991)\semi
Z.\ Bern and D.\ A.\ Kosower, Nucl.\ Phys.\ B379:451 (1992)\semi
Z.\ Bern and D.\ C.\ Dunbar, Nucl.\ Phys.\ B379:562 (1992)}
\preref\SingleAntenna{
D.\ A.\ Kosower,
%``Antenna factorization of gauge-theory amplitudes,''
Phys.\ Rev.\ D57:5410 (1998) [hep-ph/9710213]%
%%CITATION = HEP-PH 9710213;%%
}
\preref\Unitarity{Z.\ Bern, L.\ Dixon, D.\ C.\ Dunbar, and D.\ A.\ Kosower, 
Nucl.\ Phys.\ B425:217 (1994) [hep-ph/9403226]\semi
Z.\ Bern, L.\ Dixon, D.\ C.\ Dunbar, and D.\ A.\ Kosower, Nucl.\ Phys.\ B435:59 (1995) 
[hep-ph/9409265]\semi
Z.\ Bern, L.\ Dixon, and D.\ A.\ Kosower,
Ann.\ Rev.\ Nucl.\ Part.\ Sci.\ 46:109 (1996) [hep-ph/9602280]}

% Two-loop helicity amplitudes
\preref\Twoloop{
Z.\ Bern, A.\ De Freitas and L.\ Dixon,
%``Two-loop helicity amplitudes for gluon gluon scattering in QCD and  supersymmetric Yang-Mills theory,''
JHEP 0203:018 (2002) [hep-ph/0201161]%
%%CITATION = HEP-PH 0201161;%%
}
\preref\TwoloopRegulator{
Z.\ Bern, A.\ De Freitas, L.\ Dixon and H.\ L.\ Wong,
%``Supersymmetric regularization, two-loop QCD amplitudes and coupling  shifts,''
Phys.\ Rev.\ D66:085002 (2002) [hep-ph/0202271]%
%%CITATION = HEP-PH 0202271;%%
}

\preref\TwoloopAllPlus{
Z.\ Bern, L.\ J.\ Dixon and D.\ A.\ Kosower,
%``A two-loop four-gluon helicity amplitude in QCD,''
JHEP 0001:027 (2000) [hep-ph/0001001]%
%%CITATION = HEP-PH 0001001;%%
}

\preref\CataniSeymour{
S.\ Catani and M.\ H.\ Seymour,
%``The Dipole Formalism for the Calculation of QCD Jet Cross Sections at Next-to-Leading Order,''
Phys.\ Lett.\ B378:287 (1996) [hep-ph/9602277]\semi
%%CITATION = HEP-PH 9602277;%%
S.\ Catani and M.\ H.\ Seymour,
%``A general algorithm for calculating jet cross sections in NLO QCD,''
Nucl.\ Phys.\ B485:291 (1997); erratum-ibid.\ B510:503 (1997) [hep-ph/9605323]%
%%CITATION = HEP-PH 9605323;%%
}
\preref\AP{G.\ Altarelli and G.\ Parisi, Nucl.\ Phys.\ B126:298 (1977)}
\preref\BerendsGieleSoft{
F.\ A.\ Berends and W.\ T.\ Giele,
%``Multiple Soft Gluon Radiation In Parton Processes,''
Nucl.\ Phys.\ B313:595 (1989)%
%%CITATION = NUPHA,B313,595;%%
}
\preref\CataniGrazzini{
S.\ Catani and M.\ Grazzini,
%``Collinear factorization and splitting functions for  next-to-next-to-leading order {QCD} calculations,''
Phys.\ Lett.\ B446:143 (1999) [hep-ph/9810389]\semi
%%CITATION = HEP-PH 9810389;%%
S.\ Catani and M.\ Grazzini,
%``Infrared factorization of tree level QCD amplitudes at the  next-to-next-to-leading order and beyond,''
Nucl.\ Phys.\ B570:287 (2000) [hep-ph/9908523]%
%%CITATION = HEP-PH 9908523;%%
}

% Two-loop integrals
\preref\Smirnov{
V.\ A.\ Smirnov,
%``Analytical result for dimensionally regularized massless on-shell  double box,''
Phys.\ Lett.\ B460:397 (1999) [hep-ph/9905323]%
%%CITATION = HEP-PH 9905323;%%
}
\preref\Tausk{
J.\ B.\ Tausk,
%``Non-planar massless two-loop Feynman diagrams with four on-shell legs,''
Phys.\ Lett.\ B469:225 (1999) [hep-ph/9909506]%
%%CITATION = HEP-PH 9909506;%%
}

% Tensor reductions
\preref\SmirnovVeretin{
V.\ A.\ Smirnov and O.\ L.\ Veretin,
%``Analytical results for dimensionally regularized massless on-shell  double boxes with arbitrary indices and numerators,''
Nucl.\ Phys.\ B566:469 (2000) [hep-ph/9907385]%
%%CITATION = HEP-PH 9907385;%%
}

\preref\NonPlanarReduction{
C.\ Anastasiou, T.\ Gehrmann, C.\ Oleari, E.\ Remiddi and J.\ B.\ Tausk,
%``The tensor reduction and master integrals of the two-loop massless  crossed box with light-like legs,''
Nucl.\ Phys.\ B580:577 (2000) [hep-ph/0003261]%
%%CITATION = HEP-PH 0003261;%%
}
\preref\OtherTwoLoopIntegrals{
C.\ Anastasiou, E.\ W.\ N.\ Glover, and C.\ Oleari, 
Nucl.\ Phys.\ B575:416 (2000) [Erratum---
ibid.\ B585:763 (2000)] [hep-ph/9912251]\semi
T.\ Gehrmann and E.\ Remiddi, 
Nucl.\ Phys.\ Proc.\ Suppl.\ 89:251 (2000) [hep-ph/0005232]\semi
C.\ Anastasiou, J.\ B.\ Tausk, and M.\ E.\ Tejeda-Yeomans, 
Nucl.\ Phys.\ Proc.\ Suppl.\ 89:262 (2000) [hep-ph/0005328]%
}

\preref\NewerIntegralTechniques{
K.\ G.\ Chetyrkin, A.\ L.\ Kataev, and F.\ V.\ Tkachov, 
Nucl.\ Phys.\ B174:345 (1980)\semi
K.\ G.\ Chetyrkin and F.\ V.\ Tkachov, 
Nucl.\ Phys.\ B192:159  (1981)\semi
T.\ Gehrmann and E.\ Remiddi, 
Nucl.\ Phys.\ B580:485 (2000) [hep-ph/9912329]\semi
S.\ Laporta,
%``Calculation of master integrals by difference equations,''
Phys.\ Lett.\ B504:188 (2001) [hep-ph/0102032]\semi
%%CITATION = HEP-PH 0102032;%%
S.\ Laporta,
%``High-precision calculation of multi-loop Feynman integrals by  difference equations,''
Int.\ J.\ Mod.\ Phys.\ A15:5087 (2000) [hep-ph/0102033]\semi
%%CITATION = HEP-PH 0102033;%%
S.\ Moch, P.\ Uwer, and S.\ Weinzierl,
%``Nested sums, expansion of transcendental functions and multi-scale  multi-loop integrals,''
J.\ Math.\ Phys.\  43:3363 (2002) [hep-ph/0110083]%
%%CITATION = HEP-PH 0110083;%%
}

\preref\TwoLoopFourPoint{
C.\ Anastasiou, E.\ W.\ N.\ Glover, C.\ Oleari, and M.\ E.\ Tejeda-Yeomans, 
Nucl.\ Phys.\ B601:318 (2001) [hep-ph/0010212]\semi
C.\ Anastasiou, E.\ W.\ N.\ Glover, C.\ Oleari, and M.\ E.\ Tejeda-Yeomans,
%``Two-loop QCD corrections to q anti-q $\to$ q anti-q,''
Nucl.\ Phys.\ B601:341 (2001) [hep-ph/0011094]\semi
%%CITATION = HEP-PH 0011094;%%
C.\ Anastasiou, E.\ W.\ N.\ Glover, C.\ Oleari, and M.\ E.\ Tejeda-Yeomans,
%``Two-loop QCD corrections to massless quark gluon scattering,''
Nucl.\ Phys.\ B605:486 (2001) [hep-ph/0101304]\semi
%%CITATION = HEP-PH 0101304;%%
E.\ W.\ N.\ Glover, C.\ Oleari, and M.\ E.\ Tejeda-Yeomans,
%``Two-loop QCD corrections to gluon gluon scattering,''
Nucl.\ Phys.\ B605:467 (2001) [hep-ph/0102201]\semi
%%CITATION = HEP-PH 0102201;%%
L.\ W.\ Garland, T.\ Gehrmann, E.\ W.\ N.\ Glover, A.\ Koukoutsakis, and E.\ Remiddi,
%``The two-loop QCD matrix element for e+ e- $\to$ 3jets,''
Nucl.\ Phys.\ B627:107 (2002) [hep-ph/0112081]\semi
%%CITATION = HEP-PH 0112081;%%
L.\ W.\ Garland, T.\ Gehrmann, E.\ W.\ N.\ Glover, A.\ Koukoutsakis, and E.\ Remiddi,
%``Two-loop QCD helicity amplitudes for e+ e- $\to$ 3jets,''
Nucl.\ Phys.\ B642:227 (2002) [hep-ph/0206067]\semi
%%CITATION = HEP-PH 0206067;%%
S.\ Moch, P.\ Uwer, and S.\ Weinzierl,
%``Two-loop amplitudes with nested sums: Fermionic contributions to  e+ e- $\to$ q anti-q g,''
hep-ph/0207043\semi
%%CITATION = HEP-PH 0207043;%%
Z.\ Bern, A.\ De Freitas, and L.\ Dixon,
%``Two-loop corrections to $gg \to \gamma \gamma$,''
hep-ph/0211344%
%%CITATION = HEP-PH 0211344;%%
}

\preref\CataniConjecture{
S.\ Catani,
%``The singular behaviour of {QCD} amplitudes at two-loop order,''
Phys.\ Lett.\ B427:161 (1998) [hep-ph/9802439]%
%%CITATION = HEP-PH 9802439;%%
}

\preref\GieleGlover{W.\ T.\ Giele and E.\ W.\ N.\ Glover, 
Phys.\ Rev.\ D46:1980 (1992)}

% Phase-space mapping
\preref\NumericalPhaseSpace{
S.\ Weinzierl and D.\ A.\ Kosower,
%``{QCD} corrections to four-jet production and three-jet structure in e+ e-  annihilation,''
Phys.\ Rev.\ D60:054028 (1999) [hep-ph/9901277]%
%%CITATION = HEP-PH 9901277;%%
}

% FDH (String paper)
\preref\FDH{
Z.\ Bern and D.\ A.\ Kosower,
%``The Computation of loop amplitudes in gauge theories,''
Nucl.\ Phys.\ B379:451 (1992)%
%%CITATION = NUPHA,B379,451;%%
}

\preref\CollinsBook{J.C.\ Collins, {\it Renormalization}
(Cambridge University Press, 1984)}

\preref\SusyIdentities{%
M.\ T.\ Grisaru, H.\ N.\ Pendleton and P.\ van Nieuwenhuizen,
Phys.\ Rev.\ D15:996 (1977)\semi
M.\ T.\ Grisaru and H.\ N.\ Pendleton, Nucl.\ Phys.\ B124:81 (1977)\semi
S.\ J.\ Parke and T.\ Taylor, Phys.\ Lett.\ 157B:81 (1985)\semi
Z.\ Kunszt, Nucl.\ Phys.\ B271:333 (1986)}

%%%

\loadfourteenpoint

%%%%%%%%%%%%%%%%%%%%%%%%%%%%%%%%%%%%%%%%%%%%%%%%%%%%%%%%%
\noindent\nopagenumbers
[hep-ph/0212097] \hfill{Saclay/SPhT--T02/174}

\leftlabelstrue
\vskip -0.7 in
\Title{Multiple Singular Emission in Gauge Theories}
% PACS: 11.15.-q, 12.38.Bx
\vskip 10pt

\baselineskip17truept
\centerline{David A. Kosower}
\baselineskip12truept
\centerline{\it Service de Physique Th\'eorique${}^{\natural}$}
\centerline{\it Centre d'Etudes de Saclay}
\centerline{\it F-91191 Gif-sur-Yvette cedex, France}
\centerline{\tt kosower@spht.saclay.cea.fr}

\vskip 0.2in\baselineskip13truept

\vskip 0.5truein
\centerline{\bf Abstract}
{\narrower 
I derive a class of functions unifying all singular limits for the emission
of a given number of soft or collinear gluons in tree-level gauge-theory amplitudes.  
Each function is a
generalization of the single-emission antenna function of ref.~[\use\SingleAntenna].
The helicity-summed squares of these functions are thus also generalizations
to multiple singular emission of the Catani--Seymour dipole factorization function.

}
\vskip 0.3truein

%\centerline{\it Submitted to Physical Review D}

\vfill
\vskip 0.1in
\noindent\hrule width 3.6in\hfil\break
\noindent
${}^{\natural}$Laboratory of the
{\it Direction des Sciences de la Mati\`ere\/}
of the {\it Commissariat \`a l'Energie Atomique\/} of France.\hfil\break

\Date{}

\line{}

\baselineskip17pt
%

%%%%%%%%%%%%%%%%%%%%%%%%%%%%%%%%%%%%%%%%%%%%%%%%%%%%%%%%%%%%%%%%
\def\tp{\!+\!}\def\tm{\!-\!}
\def\tc{\!\cdot\!}
\section{Introduction}
\vskip 10pt

The computation of higher-order corrections in perturbative QCD is important to the
program of high-energy collider experiments, particularly at the Tevatron and LHC.
Such computations involve a variety of technical complications, including the need
to handle what would be a large number of diagrams in a conventional Feynman
diagram approach.  In the past decade, a number of new approaches have been
developed to cope with this complexity, including the color decomposition~[\use\Color],
ideas based on string theory~[\use\StringBased], and the 
unitarity-based method~[\use\Unitarity].
The latter technique has been applied to numerous calculations, most recently
the two-loop calculation of all helicity amplitudes for gluon--gluon 
scattering~[\use\TwoloopAllPlus,\use\Twoloop].

The subject of two-loop calculations has seen tremendous progress in the last three
years.   Smirnov [\use\Smirnov] gave a closed-form expression for the planar double box,
and Tausk~[\use\Tausk] one for the non-planar integral.  
Smirnov and Veretin~[\use\SmirnovVeretin] and Anastasiou et al.~[\use\NonPlanarReduction]
provided algorithms for reducing tensor integrals.  
(More general reduction and evaluation
techniques for integrals have followed as well~[\use\NewerIntegralTechniques].)
These computations, along with
the other integrals required for two-loop 
amplitudes~[\use\NonPlanarReduction,\use\OtherTwoLoopIntegrals]
has in turn led to a long series of computations of four-point 
amplitudes~[\use\TwoLoopFourPoint,\use\Twoloop].  
These amplitudes and matrix elements are one of the building blocks of 
next-to-next-leading order (NNLO) computations in perturbative QCD, in particular of
the cornerstone processes $e^+ e^- \rightarrow 3{\rm\ jets}$ and 
$p{\overline p} \rightarrow 2{\rm\ jets}$.

In order to construct numerical programs from these and other amplitudes, we must
confront another class of technical complications, that of handling 
infrared
divergences.  In the framework of dimensional regularization,
gauge-theory loop amplitudes have poles of infrared origin in the regulator $\e$,
up to two powers per loop.  These poles are canceled in physical differential
cross sections by divergences arising from integrations over infrared-singular
regions of the phase space for real emission of additional partons
from corresponding lower-loop amplitudes.

At next-to-leading order, we need to consider two types of singularities,
soft and collinear.  The former arises when a gluon four-momentum
vanishes, $k_s\rightarrow 0$; the latter when the momenta of
two massless particles become proportional, $k_a\rightarrow z (k_a+k_b)$,
$k_b\rightarrow (1-z) (k_a+k_b)$.  It is helpful to combine these
two into a single function describing both limits, as proposed by
Catani and Seymour~[\use\CataniSeymour] for the square of the matrix element.  I wrote
down an {\it antenna\/} function or amplitude~[\use\SingleAntenna]
 providing a similar unification 
at the amplitude level, within the framework of a color decomposition.  

At next-to-next-to-leading order, in addition to single-emission singularities
in one-loop amplitudes, we must now handle several types of double-emission
singularities in tree level amplitudes,
with correspondingly complicated internal boundaries in phase space: 
double-soft~[\use\BerendsGieleSoft],
soft-collinear~[\use\GloverCampbell,\use\CataniGrazzini], 
and triple-collinear~[\use\GloverCampbell,\use\CataniGrazzini,\use\DDFM]. 
The aim of the present paper is to
extend the notions of ref.~[\use\SingleAntenna] to double- and 
multiple-singular emission,
and to provide a single function describing the factorization of a color-ordered
amplitudes in all the different singular limits of a color-sequential
set of momenta.  The integrals of such functions over phase space will provide
the appropriate generalization of the integrated Catani--Seymour functions to 
NNLO computations.  It is worth noting that Catani has 
predicted~[\use\CataniConjecture]
the IR poles to be expected in the pure two-loop virtual corrections, which
must be canceled by the sum
of the double-emission amplitudes and the single-emission one-loop amplitudes.

  The properties of non-Abelian 
gauge-theory amplitudes in singular limits are easiest to understand
in the context of a color decomposition~[\use\Color].  In the present
paper, I will concentrate on tree-level all-gluon amplitudes, though the formalism
readily extends to amplitudes with quarks and (colored) scalars as well. For 
tree-level all-gluon amplitudes in an $SU(N)$ gauge theory 
the color decomposition has the form,
$$
{\cal A}_n^\tree(\{k_i,\lambda_i,a_i\}) = 
\sum_{\sigma \in S_n/Z_n} \Tr(T^{a_{\sigma(1)}}\cdots T^{a_{\sigma(n)}})\,
A_n^\tree(\sigma(1^{\lambda_1},\ldots,n^{\lambda_n}))\,,
\eqn\ColorDecomposition$$
where $S_n/Z_n$ is the group of non-cyclic permutations
on $n$ symbols, and $j^{\lambda_j}$ denotes the $j$-th momentum
and helicity.  As is by now standard,
I use the normalization $\Tr(T^a T^b) = \delta^{ab}$.
One can write analogous formul\ae\ for amplitudes
with quark-antiquark pairs or uncolored external lines.
The color-ordered or partial amplitude $A_n$ is gauge invariant, and has
simple factorization properties in both the soft and collinear limits,
$$
\eqalign{
A_{n}^\tree(\ldots,a^{\lambda_a},b^{\lambda_b},\ldots)
 &\inlimit^{a \parallel b}
\sum_{\lambda=\pm}  
  \Split^{\rm tree}_{-\lambda}(a^{\lambda_a},b^{\lambda_b};z)\,
      A_{n-1}^\tree(\ldots,(a+b)^\lambda,\ldots)\,,\cr
A_{n}^\tree(\ldots,a,s^{\lambda_s},b,\ldots)
 &\inlimit^{k_s\rightarrow 0}
  \Soft^\tree(a,s^{\lambda_s},b)\,
      A_{n-1}^\tree(\ldots,a,b,\ldots)\,.\cr
}\eqn\Factorization
$$

The collinear splitting amplitude $\Split^{\rm tree}$, 
squared and summed over helicities,
gives the usual unpolarized Altarelli--Parisi splitting function~[\use\AP].
It depends on the collinear momentum fraction $z$ (here made 
explicit) in addition to invariants built out of the collinear momenta.
While the complete amplitude also factorizes in the collinear limit,
the same is not true of the soft limit; the eikonal factors $\Soft^\tree$
get tangled up with the color structure.  It is for this reason
that the color decomposition is useful.  

I will review the unification of these functions into the single-emission
antenna amplitude 
in section~\use\SingleEmissionSection.
  The derivations and formalism of the present paper are based
on use of recurrence relations for gauge-theory amplitudes, which I review
in the next section.  In section~\use\MultipleCollinearSection, 
I consider a subtlety that arises in the
derivation of factorization functions for multiply-collinear emission;
the collinear splitting amplitudes for $1\rightarrow2$ and $1\rightarrow 3$
configurations are given in the appendix.
The use of a unified factorization requires {\it reconstruction\/} functions
describing the factorized hard legs; I review the $3\rightarrow2$ reconstruction
functions in section~\use\SingleReconstructionSection,
and describe the generalization to $n\rightarrow 2$ 
in section~\use\GeneralReconstructionSection.  
I construct the antenna amplitude for double singular emission
in section~\use\DoubleEmissionSection, and that for 
multiple singular emission
in section~\use\MultipleEmissionSection.  I give  
explicit expressions for specific helicities of the
single- and double-emission functions 
in section~\use\DoubleEmissionHelicitySection.  
For use in integrations over
singular phase space, it is convenient to have the squares of the 
antenna amplitudes in dimensional regularization.
I provide such expressions for the squares of the single- and double-emission
antenna amplitudes
in section~\use\DimensionalRegularizationSection.

\section{Recurrence Relations}
\tagsection\RecurrenceSection
\vskip 10pt

I will base the derivations in this paper on the recurrence relations
formalism of Berends and Giele~[\use\Recurrence], starting with the 
form given by Dixon~[\use\DixonTASI] with a slightly different notation.  
(For a helicity-based form of recurrence relations, see 
ref.~[\use\HelicityRecurrence].)  

The recurrence relations define an $n$-point color-ordered
gluon current, $J^\mu(1,\ldots,n\tm1;y)$,
with the leg indexed by $\mu$ off shell, and legs $1,\ldots, n\tm1$ on shell.  
In Dixon's notation, 
$J^\mu(1,\ldots,n\tm1;0)$ here is $-J^\mu(1,\ldots,n\tm1)$.  All momenta are taken
to be outgoing, and I have introduced
an additional argument $y$ representing a momentum excess,
$k_y = -(k_1+\cdots+k_n+k_x)$, for reasons which will become clear in
later sections.  For now, the reader may imagine that this additional
argument is always zero.  I will use labels interchangeably with momenta
carrying those labels as arguments.  For later use, it will also be convenient
to define a contracted form making the last momentum an explicit
argument, $J(1,\ldots,n;P;y) \equiv 
\pol_P\cdot J(1,\ldots,n;y)$.

\def\phpol{{\rm ph.\ pol.}}
\def\back{\hskip -15mm}
\def\spacer{ \hphantom{-\sum_{j_1=1}^{n-1}\sum_{j_2=j_1+1}^{n-1} 
        V_4^{\mu'\nu'\rho'\lambda'} d } }
Define $K_{i,j} = k_i + \cdots + k_j$; the recurrence relations then
have the form,
$$\eqalign{
J_\mu(1,\ldots,n;k_y) &= 
\cr -{i d_{\mu\mu'}(K_{1,n}) \over K_{1,n}^2}
&\LB \sum_{j=1}^{n-1} 
V_3^{\mu'\nu'\rho'}(K_{1,j}+\zeta_1 k_y,K_{j\tp1,n}+\zeta_2 k_y,-K_{1,n}+\zeta_3 k_y)
d_{\nu\nu'}(K_{1,j}) d_{\rho\rho'}(K_{j\tp1,n}) 
\RP\cr&\back\spacer\times\LP
J^\nu(1,\ldots,j;\alpha_1 k_y)
J^\rho(j\tp1,\ldots,n;\alpha_2 k_y)
\RP\cr&\back\hphantom{\LB\vphantom{\sum_{j=1}^1}\RP}\LP
-\sum_{j_1=1}^{n-1}\sum_{j_2=j_1+1}^{n-1}
V_4^{\mu'\nu'\rho'\lambda'}
d_{\nu\nu'}(K_{1,j_1}) d_{\rho\rho'}(K_{j_1\tp1,j_2}) d_{\rho\rho'}(K_{j_2\tp1,n})
\RP\cr&\back\spacer
  \times\LP\vphantom{\sum_{j=1}^1}
J^\nu(1,\ldots,j_1;\beta_1 k_y)
J^\rho(j_1\tp1,\ldots,j_2;\beta_2 k_y)
J^\lambda(j_2\tp1,\ldots,n;\beta_3 k_y)
\RB
\cr
}\eqn\RecurrenceRelations$$
where the three-point vertex is
$$
V_3^{\mu\nu\rho}(P_1,P_2,P_3) = {i\over\sqrt2}
\LB g^{\nu\rho} (P_1-P_2)^\mu +2 g^{\rho\mu} P_2^\nu - 2 g^{\mu\nu} P_1^\rho\RB\,,
\eqn\ThreePointVertex$$
the four-point vertex is
$$
V_4^{\mu\nu\rho\lambda} = {i\over2}\LB 2 g^{\mu\rho} g^{\nu\lambda}
-g^{\mu\nu} g^{\rho\lambda} - g^{\mu\lambda} g^{\nu\rho}\RB\,,
\eqn\FourPointVertex$$
and $d_{\mu\nu}$ is the gluon helicity projector; in the
background-field form used here, $d_{\mu\nu} = -g_{\mu\nu}$.
(Note that the current as defined here has the opposite sign to that 
of ref.~[\use\DixonTASI].)
Define as well $J(1;P;k_y) = \pol_P\cdot \pol_1$.
A sum over intermediate polarizations will be understood implicitly in all products
of indexless currents or currents and amplitudes in this paper,
$$
J(\ldots;P;y) X(\ldots,-P,\ldots) 
= \sum_{{\rm polarizations\ }\sigma}J(\ldots;P^\sigma;y) 
X(\ldots,-P^{-\sigma},\ldots) 
\anoneqn$$

The $n$-point amplitude $A_n^\tree(1,\ldots,n)$ is given by removing
the propagator on the last, off-shell, leg of the contracted current, and then
taking the on-shell limit,
$$
A_n^\tree(1,\ldots,n) = \lim_{k_n^2\rightarrow 0} -i k_n^2 J(1,\ldots,n-1;n;0)
\eqn\Amplitude$$

\section{Splitting Amplitudes for Multiple Collinear Emission}
\tagsection\MultipleCollinearSection
\vskip 10pt

With the currents in hand, we are ready to begin the derivation of collinear
splitting amplitudes.
Let us begin by rederiving the collinear behavior of amplitudes as
two momenta become collinear; without loss of generality, we may take these
to be $k_1$ and $k_2$.  We are interested in the limit when 
$s_{12}$ becomes small compared to all other invariants in an $n$-point
amplitude (or equivalently, shrinks compared to the dot products of an arbitrary
reference momentum $q$ with $k_1$ and $k_2$~[\use\CataniSeymour]).
  The limit will be dominated by contributions which have
an explicit pole in this invariant.  The structure of the recurrence
relations tells us that the invariant always appears as a single pole,
inside $J(1,2;-(k_1\tp k_2);0)$.  We can also see that if we replace this
current by a polarization vector carrying the fused momentum 
$k_P = -(k_1+k_2)$, we will obtain an $(n-1)$-point amplitude from
eqn.~(\use\Amplitude).  The original amplitude thus factorizes in this limit,
$$
A_n^\tree(1,2,\ldots,n) \inlimit^{s_{12}\rightarrow 0}
J(1,2;k_P) A_{n-1}^\tree(1+2,3,\ldots,n)\,.
\anoneqn$$
In this equation, the notation `$1+2$' is a shorthand for $k_1+k_2$.
Gauge invariance ensures that only physical polarizations appear for the
fused leg, but we do need to sum over both,
$$
A_n^\tree(1,2,\ldots,n) \inlimit^{s_{12}\rightarrow 0}
 \sum_{\phpol\ \sigma_P}
 \Ctree_{\sigma_P}(1^{\sigma_1},2^{\sigma_2}) 
         A_{n-1}^\tree((1+2)^{-\sigma_P},3,\ldots,n)
   + {\rm finite}\,.
\anoneqn$$
The tree splitting amplitude is given by the singular limit of the current,
$$
 \Ctree_{\sigma_P}(1^{\sigma_1},2^{\sigma_2}) =
 J(1^{\sigma_1},2^{\sigma_2};k_P^{\sigma_P};0)\bigr|_{{\rm leading\ in\ } s_{12}}
\eqn\TreeSplitting$$
The helicity algebra will soften the full pole in $s_{12}$ to a square-root
singularity most conveniently expressed in terms of spinor products,
so that the splitting amplitude will scale as $1/\sqrt{\delta}$ when
the invariant shrinks by a factor of $\delta$.  The
leading, singular, part is gauge-invariant, in spite of the seemingly off-shell
nature of this object.

To derive a similar factorization and function when three color-adjacent
external momenta $k_1$, $k_2$, and $k_3$ 
become collinear simultaneously, we must extract all
singularities in $s_{12}$, $s_{23}$, and $t_{123} = s_{12}+s_{23}+s_{13}$,
with these invariants all of comparable order and much smaller than
other invariants of external momenta.  (The strongly-ordered
limit where some of the two-particle
invariants are much smaller again than $t_{123}$ is contained as a degenerate
case in the limit of interest; what we are {\it not\/} interested in is the limit where
$t_{123}$ becomes much smaller than any of the two-particle invariants
$s_{12}$, $s_{23}$, or $s_{13}$.)
Color-ordered amplitudes will not have singularities as the invariants
of non-adjacent legs, such as $s_{13}$, become collinear.  Furthermore,
as explained by Campbell and Glover~[\use\GloverCampbell],
we are interested only in the leading behavior, corresponding to 
singularities in {\it two\/} final-state integration variables, 
or equivalently that scale as 
 $1/\delta$ (the square of the scaling above)
when all the above invariants shrink by a factor of $\delta$.

We begin by extracting all singularities in $t_{123}$.  This invariant
appears only in the four-point current $J(1,2,3;P)$.  Next,
include the invariants $s_{12}$ and $s_{23}$, which
appear only in $J(1,2;P)$ and $J(2,3;P)$ respectively.  We might thus
be tempted to write down the following starting formula,
$$\eqalign{
A_n^\tree(1,2,\ldots,n) \inlimit^{1\parallel 2\parallel 3}
&J(1,2,3;k_P) A_{n-2}^\tree(1+2+3,\ldots,n)
\cr &\hskip 2mm
+J(1,2;k_P) A_{n-1}^\tree(1+2,3,\ldots,n)
\cr &\hskip 2mm
+J(2,3;k_P) A_{n-1}^\tree(1,2+3,\ldots,n)\,,
}\anoneqn$$
but this double-counts certain contributions, 
because $J(1,2,3;P)$ {\it also\/} contains
poles in $s_{12}$ and $s_{23}$, and $A_{n-1}^\tree$ contains a pole in
$t_{123}$.  Subtract off the double-counted contribution,
to obtain the corrected formula,
$$\eqalign{
A_n^\tree(1,2,\ldots,n) \inlimit^{1\parallel 2\parallel 3}
&J(1,2,3;k_P) A_{n-2}^\tree(1+2+3,\ldots,n)
\cr &\hskip 2mm 
+J(1,2;k_P) A_{n-1}^\tree(1+2,3,\ldots,n)
\cr &\hskip 2mm 
+J(2,3;k_P) A_{n-1}^\tree(1,2+3,\ldots,n)
\cr &\hskip 2mm 
 - \LP J(1,2,3;k_P)\RV_{s_{12} {\rm\ pole}}  A_{n-2}^\tree(1+2+3,\ldots,n)
\cr &\hskip 2mm 
	 - \LP J(1,2,3;k_P)\RV_{s_{23} {\rm\ pole}}  A_{n-2}^\tree(1+2+3,\ldots,n)\,,
}\eqn\CorrectedFactorization$$
We are not done, however, because the $(n-1)$-point amplitudes appearing
here contain additional singularities in the limit.  These may again
be expressed in terms of the three-point current,
$$\eqalign{
A_n^\tree(1,2,\ldots,n) \inlimit^{1\parallel 2\parallel 3}
&J(1,2,3;k_P) A_{n-2}^\tree(1+2+3,\ldots,n)
\cr &\hskip 2mm 
+J(1,2;k_R) J(1+2,3;k_P) A_{n-2}^\tree(1+2+3,\ldots,n)
\cr &\hskip 2mm 
+J(2,3;k_R) J(1,2+3;k_P) A_{n-2}^\tree(1+2+3,\ldots,n)
\cr &\hskip 2mm 
 - \LP J(1,2,3;k_P)\RV_{s_{12} {\rm\ pole}}  A_{n-2}^\tree(1+2+3,\ldots,n)
\cr &\hskip 2mm
 - \LP J(1,2,3;k_P)\RV_{s_{23} {\rm\ pole}}  A_{n-2}^\tree(1+2+3,\ldots,n)\,,
}\eqn\CorrectedFactorizationB$$
I emphasize that despite the appearance of sequential factorization in
the second and third terms, no assumption is
made about strong ordering of the limits; for example,
$s_{12}$ is {\it not\/} taken
to be much smaller than $s_{23}$ or $t_{123}$ in the second term.

The subtlety  which this section addresses arises from the fact
that the last two terms in eqn.~(\use\CorrectedFactorization) are
not gauge-invariant (because $k_P^2\neq 0$).  To see this explicitly,
it is convenient to decompose the propagator giving rise to (say) the
$s_{12}$ pole in a helicity basis~[\use\BernChalmers],
$$\eqalign{
\LP J(1,2,3;P)\RV_{s_{12} {\rm\ pole}}
= J^\mu&(1,2;-K_{1,2}) 
\cr &\times\biggl[
\pol_\mu^{(+)}(-K_{1,2};q)\pol_{\mu'}^{(-)}(K_{1,2};q)
%\cr&\hphantom{\times\biggl[}
+\pol_\mu^{(-)}(-K_{1,2};q)\pol_{\mu'}^{(+)}(K_{1,2};q)
\vphantom{ {q_\mu (K_{1,2})_{\mu'}\over q\cdot K_{1,2}} }
\cr&\hphantom{\times\biggl[}
-{(K_{1,2})_{\mu} q_{\mu'}\over q\cdot K_{1,2}}
-{q_\mu (K_{1,2})_{\mu'}\over q\cdot K_{1,2}}
\biggr] V_3^{\mu'\nu\rho}(-K_{1,2},3,P) 
  J_\nu(3;-k_3)\pol_{P\rho}
}\eqn\Decomposition$$
where $q$ is a null vector not collinear to any of the external momenta,
and $\pol_\mu^{(\sigma)}(K;q)$ is a polarization vector for a gluon of helicity
$\sigma$ carrying momentum $K$, with reference momentum $q$.
\iffalse
In the language of the spinor-helicity basis, it is the reference momentum
for the polarization vectors.  
\fi

The first two terms in brackets sum to the second 
term in eqn.~(\use\CorrectedFactorizationB), and cancel it.  A similar
cancellation occurs for the third term in eqn.~(\use\CorrectedFactorizationB).
The third term in eqn.~(\use\Decomposition) 
will vanish by gauge invariance, $K_{1,2}\cdot J(1,2;P) = 0$.
The last term is new, and survives; indeed, here the helicity algebra
in $q\cdot J(1,2;P)$
will not soften the $s_{12}$ pole, although $K_{1,2}\cdot V_3$
will give rise to a factor of $k_P^2 = t_{123}$ that will cancel off the 
$t_{123}$ pole, so that overall this term will scale like $1/\delta$.

The equation above is given in Feynman gauge.  
If we identify $q$ as the light-cone vector of light-cone gauge, then
the last two terms correspond precisely to the difference between the
spin-projectors of the gluon propagators in Feynman and light-cone
gauges.  That is, were we to work in light-cone gauge, the last two terms
in eqn.~(\use\Decomposition) would be absent, and the splitting
amplitude would be given (as we might naively have expected) by the
singular limit of the four-point current $J(1,2,3;P)$ alone.  
(A similar point about the convenience of a physical gauge for
derivations of collinear limits was made by Catani and 
Grazzini~[\use\CataniGrazzini] in the context
of a different formalism, 
and by Dixon~[\ref\LanceComment{L. Dixon, personal communication}].)

\def\lc{{\rm LC}}
In light-cone gauge, the three-vertex has the form,
$$
V_{3,\lc}^{\mu\nu\rho}(P_1,P_2,P_3) = {i\over\sqrt2}
\LB g^{\nu\rho} (P_1-P_2)^\mu + g^{\rho\mu} (P_2-P_3)^\nu 
    + g^{\mu\nu} (P_3-P_1)^\rho\RB\,,
\eqn\ThreePointVertexLC$$
the four-point vertex is unchanged, the gluon propagator's helicity projector is,
$$
d_{\mu\nu}(k) = -g_{\mu\nu} + {q^\mu k^\nu + k^\mu q^\nu\over q\cdot k}\,
\eqn\HelicityProjectorLC$$
(where $q$ is the light-cone momentum),
and the recurrence relations themselves have the same form as
in eqn.~(\use\RecurrenceRelations).  All currents appearing in the following
sections are understood to be in light-cone gauge.

Campbell and Glover 
extracted~[\use\GloverCampbell] the helicity-summed squared triple-collinear
splitting function from squared amplitudes. Catani and Grazzini
computed~[\use\CataniGrazzini]
a somewhat more general object, retaining the dependence on the polarization
of the parent parton, but still at the level of the squared matrix element
rather than at the amplitude level.  Neither set of authors
provided a factorizing form at the amplitude level.  
Del Duca, Frizzo, and Maltoni [\use\DDFM] derived splitting
amplitudes from six-point amplitudes.  
The splitting amplitudes which follow from the light-cone current above
are given in the appendix.  They
retain all relative phase and helicity-correlation information, and agree
with the expressions given in ref.~[\use\DDFM], up to a difference due
to normalization conventions, and a sign for the amplitudes with two positive
and two negative helicities.
(The overall phase of these splitting amplitudes depends on the phase
convention for the amplitudes; I follow the convention given by 
the Mangano and Parke review article~[\use\MP].)

Squaring the splitting amplitudes, and summing over all helicities, 
I find a squared splitting function in agreement with the $\e\rightarrow0$
limit of the expression given by Campbell and Glover [\use\GloverCampbell]
(up to a constant factor related to the different normalization of amplitudes).

% collinear-soft

\section{The Single-Emission Antenna Amplitude}
\tagsection\SingleEmissionSection
\vskip 10pt

As described in the introduction,
gauge-theory amplitudes have singularities not only when color-adjacent
momenta become collinear, of course, but also when external gluon momenta
become soft.  Factorization in both limits is given by universal 
functions~[\use\MP].
Catani and Seymour pointed out that for computational purposes,
it is desirable to unify the two different limits.  They wrote down
a function and a formalism for doing this for the color-summed
squared amplitude, for lone singular emission (that is, one collinear
pair or one soft gluon).  In a previous paper~[\use\SingleAntenna], I constructed a
function which unifies these two limits at the level of the color-ordered
amplitude.

The construction is based on the observation that one may treat
singular emission in gauge theories as occuring inside the `color antenna'
bounded by two hard colored particles, which I shall label $a$ and $b$.  
For the single-emission function, we need to consider the emission of
a single gluon carrying momentum $k_1$ inside the antenna.  There
are three sorts of singularities we should consider: \hfil\break
(a) $s_{a1} \rightarrow 0$, with $s_{1b}$ approaching a constant non-zero
limit, that is the collinear limit $k_1\parallel k_a$; \hfil\break
(b) $s_{1b} \rightarrow 0$, with $s_{a1}$ approaching a constant non-zero
limit, that is the collinear limit $k_1\parallel k_b$; \hfil\break
and (c) $s_{a1} {\rm\ and\ } s_{1b} \rightarrow 0$, that is the soft
limit $k_1\rightarrow 0$. \hfil\break
  In all cases, $s_{ab}$ defines a hard scale;
in the two collinear regions, the non-collinear hard momentum acts as
a reference momentum to define the collinear momentum fractions.

\def\ah{{\hat a}}
\def\bh{{\smash{{\hat b}}{}}}

To extract the antenna amplitude, we begin by extracting all contributions 
in the $n$-point amplitude
that have a pole in either $s_{a1}$ or $s_{1b}$,
$$
J(a,1;-(k_a+k_1);0) A_{n-1}(\ldots,k_a\tp k_1,b,\ldots)
+J(1,b;-(k_a+k_1);0) A_{n-1}(\ldots,a,k_b\tp k_1,\ldots)
\anoneqn$$
We introduce a complete set of polarization states for the unfused leg in each term,
 rewrite the product of polarization vectors as a two-point current, and introduce
new labels for the surviving hard momenta,
$$\eqalign{
&J(a,1;-(k_a\tp k_1);0) J(b;-k_b;0) 
   A_{n-1}(\ldots,-k_\ah = k_a\tp k_1,-k_\bh = k_b,\ldots)
\cr &
+J(a;-k_a;0) J(1,b;-(k_1\tp k_b);0)
   A_{n-1}(\ldots,-k_\ah = k_a,-k_\bh = k_b\tp k_1,\ldots).
}\anoneqn$$
At this point, the hatted momenta have different definitions in the two terms.
If we can find a suitable pair of {\it reconstruction\/} functions
$k_{\ah,\bh} = f_{\ah,\bh}(k_a,k_1,k_b)$, however, we can combine the two terms,
$$\eqalign{
&\LB J(a,1;-(k_a\tp k_1);0) J(b;-k_b;0) 
+J(a;-k_a;0) J(1,b;-(k_1\tp k_b);0)\RB\,
\cr&\hskip 15mm\times
A_{n-1}(\ldots,-k_\ah = f_\ah(k_a,k_1,k_b),
                -k_\bh = f_\bh(k_a,k_1,k_b),\ldots).
}\anoneqn$$
We also would like to have a manifestly gauge-invariant object inside
the brackets, or equivalently get rid of the implicit dependence on
the reference momentum $q$, so that we have a well-defined function in
between the different singular limits as well.  To do so, we would
like to put the off-shell arguments in the currents on-shell.  We can
do so, at the price of violating momentum conservation within each
current,
$$\eqalign{
&\LB J(a,1;\ah;k_b\tp k_\bh) J(b;\bh;k_\ah\tp k_a\tp k_1) 
+J(a;\ah;k_\bh\tp k_b\tp k_1) J(1,b;\bh;k_a\tp k_\ah)\RB\,
\cr&\hskip 15mm\times
   A_{n-1}(\ldots,-k_\ah = f_\ah(k_a,k_1,k_b),-k_\bh = f_\bh(k_a,k_1,k_b),\ldots)
\,.
}\eqn\AntennaPrep$$
(It is for this purpose that I added the additional argument to the current
in section~\use\RecurrenceSection.)
The factor in brackets is the antenna factorization amplitude,
$$
\Ant(\ah,\bh\leftarrow a,1,b) = 
J(a,1;\ah;k_b\tp k_\bh) J(b;\bh;k_a\tp k_1\tp k_\ah) 
+J(a;\ah;k_b\tp k_1\tp k_\bh) J(1,b;\bh;k_a\tp k_\ah)\,,
\eqn\AntennaFirstDef$$
with corresponding factorization,
$$
A_n(\ldots,a,1,b,\ldots) 
\inlimit^{k_1 {\rm\ singular}} \sum_{\phpol \lambda_{a,b}}
\Ant(\ah^{\lambda_a},\bh^{\lambda_b}\leftarrow a,1,b)
A_{n-1}(\ldots,-k_{\ah}^{-\lambda_a},-k_{\bh}^{-\lambda_b},\ldots).
\eqn\SingleEmissionFactorization$$

While there is a momentum excess inside each of the pair of currents
making up a term, the excesses cancel so that
overall, each term and hence the factor in brackets as a whole conserves
momentum,
$$
-(k_\ah+k_\bh) = k_a+k_1+k_b \equiv K\,.
\anoneqn$$
The reader might nonetheless wonder about the effect of this momentum 
excess on the uniqueness of the result; this question will be addressed in
the following section.

The antenna amplitude describes
the behavior of the amplitude when either $s_{1a}/s_{ab}\rightarrow 0$,
or $s_{1b}/s_{ab}\rightarrow 0$, or both; equivalently, when 
$\Delta(a,1,b)/s_{ab}^3 \rightarrow 0$, 
with $\Delta(p_1,\ldots,p_n)$ the Gram determinant,
$$
\Delta(p_1,\ldots,p_n) = \det( 2 p_i\cdot p_j )
\eqn\GramDet$$
(Note that the normalization here is non-standard.)  I will also make
use of the generalized Gram determinant,
$$
G\L {p_1,\ldots,p_n\atop q_1,\ldots,q_n}\R
= \det( 2 p_i\cdot q_j ),
\eqn\GeneralizedGramDet$$
which vanishes whenever two $p_i$ or two $q_i$ become collinear, or
when any momentum becomes soft.

We still need to specify the reconstruction functions $f_{\ah,\bh}$, however;
that is the subject of the next section.

\section{Single-Emission Reconstruction Function}
\tagsection\SingleReconstructionSection
\vskip 10pt

What makes a suitable reconstruction function? Sensible functions
will ensure that the reconstructed momenta are on shell, 
$k_\ah^2 = 0 = k_\bh^2$, as well as enforcing momentum conservation,
$k_\ah+k_\bh + k_a+k_1+k_b = 0$.  They must reduce to the appropriate forms
at the singular points,
$$\eqalign{
k_\ah &= -(k_a\tp k_1), k_\bh = -k_b,\qquad {\rm when\ } s_{a1} = 0, s_{1b}\neq 0\,;
\cr
k_\ah &= -k_a, k_\bh = -(k_1\tp k_b),\qquad {\rm when\ } s_{a1} \neq 0, s_{1b}= 0\,;
\cr
k_\ah &= -k_a, k_\bh = -k_b,\qquad {\rm when\ } s_{a1} = 0 =s_{1b}\,.
\cr}\anoneqn$$
Finally, they must ensure that the excess momentum within each of the
currents in eqn.~(\use\AntennaFirstDef) does not lead to additional terms which are
singular in any of the limits.  

To understand this last requirement better, examine the current
$J(a,1;\ah;k_b\tp k_\bh)$.  Because momentum is not conserved in this 
expression, different forms of the three-vertex -- for example, the form
given in eqn.~(\use\ThreePointVertexLC) and (say),
$$
{i\over\sqrt2}
\LB g^{\nu\rho} (P_1-P_2)^\mu + g^{\rho\mu} (2 P_2+P_1)^\nu 
    + g^{\mu\nu} (-P_2-2 P_1)^\rho\RB\,,
\anoneqn$$
will lead to different results.  However, the results will be equivalent
in the singular limits so long as they differ only by terms proportional
to $s_{a1}$.  When divided by the $s_{a1}$ in the current, these will
give rise only to finite terms.
Such finite terms are anyway not universal, and are
implicitly omitted from the antenna amplitude.  

The difference will be a sum of terms, each proportional to 
$(k_b\tp k_\bh)\cdot \pol_{a,1,\ah}$ or $(k_b\tp k_\bh)\cdot (k_a\tm k_1)$, 
each of which should be of order
$s_{a1}$ (or higher) as $s_{1a}\rightarrow 0$.

The first two requirements, for null momenta and momentum conservation,
can be satisfied by the following general form~[\use\SingleAntenna],
$$\eqalign{
k_\ah &= -{1\over 2 (K^2-s_{1b})} \LB {(1+\rho) K^2}
            - 2 s_{1b} r_1\RB\, k_a
          - r_1 k_1 
    -{1\over 2 (K^2 - s_{a1})}\LB (1-\rho) K^2 -2 s_{1a} r_1\RB\, k_b\,,\cr
k_\bh &= -{1\over 2 (K^2 - s_{1b})} \LB (1-\rho)K^2 -2 s_{1b} (1-r_1)\RB\, k_a
          - (1-r_1) k_1 
\cr &\hskip 30mm
    -{1\over 2 (K^2 - s_{a1})}\LB (1+\rho)K^2 - 2 s_{1a} (1-r_1)\RB\, k_b\,,\cr
}\eqn\ReconstructionA$$
for arbitrary $r_1$, where $K = k_a+k_1+k_b$ and
$$
\rho = \sqrt{1+ {4 r_1 (1-r_1) s_{1a} s_{1b}\over K^2 s_{ab}}}\,.
\anoneqn$$
Note that $\rho\rightarrow 1$ in all singular limits, so these functions
also satisfy the third requirement, of appropriate reduction in the different
singular limits, so long as $r_1$ is not singular.

What about the last requirement, of avoiding spurious terms due to a momentum
excess?  The momentum excess in $J(a,1,\ah;k_b+k_\bh)$ is,
$$
k_b+k_\bh =
(1-\rho)K^2 \Bigl[{k_b\over 2 (K^2 - s_{a1})}-{k_a\over 2 (K^2 - s_{1b})}
            \Bigr]
 +{s_{1b} (1-r_1)\over K^2 - s_{1b}} \, k_a
          - (1-r_1) k_1 
    -{s_{a1} r_1\over K^2 - s_{a1}}\, k_b\,,
\anoneqn$$
Now, as $s_{a1}\rightarrow 0$, $\rho \sim 1 + {\cal O}(s_{a1} r_1 (1-r_1))$, 
so the
terms proportional to $1-\rho$ satisfy the requirement even without taking
into account any suppression from $r_1 (1-r_1)$.  The same is true
of the very last term, explicitly proportional to $s_{a1}$.  This leaves us
with
$$
(1-r_1)\Bigl[ { s_{1b} \over K^2 - s_{1b}}\, k_a  -  k_1 \Bigr]
\eqn\LastTerms$$
for which the requirement
will clearly be satisfied if $r_1 \sim 1+{\cal O}(s_{a1})$ in this limit,
for example,
$$
r_1 = {s_{1b}\over s_{1a}+s_{1b}}
\eqn\rChoice$$
will do.  (Given this requirement on $r_1$, $\rho$ in fact goes like
$1+{\cal O}(s_{a1}^2)$ in the limit.  As I shall discuss in a later section,
$r_1\sim 1+{\cal O}(\sqrt{s_{a1}})$ would in fact be sufficient to ensure
that eqn.~(\use\LastTerms) leads only to subleading contributions, though
{\it not\/} for the seemingly obvious reason that the helicity algebra
knocks down the $1/s_{a1}$ pole to a square-root singularity, because
that doesn't happen in these additional terms.)
With this choice, eqns.~(\use\ReconstructionA) are invariant under
the simultaneous exchanges $\ah\leftrightarrow \bh, a\leftrightarrow b$,
so that the other term in eqn.~(\use\AntennaFirstDef), involving the
excess $k_a+k_\ah$ in the limit $s_{1b}\rightarrow 0$, also avoids introducing
additional or ambiguous terms into the antenna function.
This also implies that we can ignore the momentum excess in the definition
of the antenna amplitude, simplifying it to
$$
\Ant(\ah,\bh\leftarrow a,1,b) = 
J(a,1;\ah;0) J(b;\bh;0)+J(a;\ah;0) J(1,b;\bh;0)\,.
\eqn\AntennaDef$$

The choice~(\use\rChoice) for the
reconstruction functions is not unique, and other choices for
$r_1$ are possible.  Indeed, while
the choice $r_1 = \LB s_{1b}/(s_{a1}+s_{1b})\RB^{1/17}$, for example,
would fail to satisfy this scaling requirement as $s_{1b}\rightarrow 0$,
the choice $r_1 = \LB s_{1b}/(s_{a1}+s_{1b})\RB^{2}$ would be 
satisfactory (though asymmetric). 

The Catani--Seymour forms of the fused momenta for $k_a\parallel k_1$
can be obtained by setting $r_1=1$ in eqn.~(\use\ReconstructionA), 
while those for $k_b\parallel k_1$ can be
obtained by setting $r_1=0$.  Of course, these restricted forms are not appropriate
in the opposite limits, while eqn.~(\use\ReconstructionA) interpolates smoothly between
the two forms, approaching the Catani--Seymour forms in each of the two limits,
as well as in the soft limit, $k_1\rightarrow 0$.

\section{General Reconstruction Function}
\tagsection\GeneralReconstructionSection
\vskip 10pt

In order to go beyond the emission of a lone gluon, we need to 
generalize the constructs of the previous section to handle more
soft or collinear partons.  We seek a pair of reconstruction functions,
now a function of the $n+2$ momenta $k_a,k_1,\ldots,k_n,k_b$ with $k_{a,b}$
the surviving hard momenta.  As before, we want to keep the reconstructed
momenta $k_{\ah,\bh}$ massless, and to conserve momentum,
$$
K \equiv -(k_\ah+k_\bh) = k_a + k_1 + \ldots k_n + k_b\,.
\anoneqn$$
We will also want the reconstruction functions to have the appropriate
limits not only when all the numbered momenta become singular, but also
in strongly-ordered limits, when a subset of these momenta become 
collinear with each other, or when a subset becomes much softer than
other momenta.  As for the single-emission case, there will be 
scaling constraints arising from the need to leave the leading singularity
in the antenna amplitude unchanged; I defer a discussion of them until
the next section.

The constraints of masslessness and momentum conservation can be
satisfied by the following functional forms,
\def\tR{{\widetilde R}}
$$\hskip -15pt\eqalign{
	k_\ah &= -{1\over 2( K^2\tm t_{1\cdots nb})} 
            \Biggl[ {(1\tp\rho)} K^2
            +2 R\tc (k_a\tm k_b\tm K)
            +{1\over s_{ab}} G\L {k_a,k_b\atop R,K_{1,n}}\R
%            - \sum_{j=1}^n \biggl( s_{jb}-s_{aj}+2 k_j\tc K
%                       -{1\over s_{ab}} G\L {k_a,k_b\atop k_j,K_{1,n}}\R
%                     \biggr) r_j
            \Biggr]\, k_a
            -R
\cr &\hskip 5mm 
%   + \sum_{j=1}^n r_j k_j
%\cr &\hskip 5mm
-{1\over 2(K^2\tm t_{a1\cdots n})} 
            \Biggl[ {(1\tm\rho)} K^2
             +2 R\tc (k_b\tm k_a\tm K)
             +{1\over s_{ab}}G\L {k_b,k_a\atop R,K_{1,n}}\R
%            - \sum_{j=1}^n \biggl( s_{aj}-s_{jb}+2 k_j\tc K
%                       -{1\over s_{ab}} G\L {k_b,k_a\atop k_j,K_{1,n}}\R
%                    \biggr) r_j 
            \Biggr]\, k_b
\cr
k_\bh &= -{1\over 2(K^2\tm t_{1\cdots nb})} 
            \Biggl[ {(1\tm\rho)} K^2
            +2 \tR\tc (k_a\tm k_b\tm K)
            +{1\over s_{ab}} G\L {k_a,k_b\atop \tR,K_{1,n}}\R
%            - \sum_{j=1}^n \biggl( s_{jb}-s_{aj}+2 k_j\tc K
%                       -{1\over s_{ab}} G\L {k_a,k_b\atop k_j,K_{1,n}}\R
%                     \biggr) (1\tm r_j) 
            \Biggr]\, k_a
             -\tR
\cr &\hskip 5mm 
%+ \sum_{j=1}^n (1\tm r_j) k_j
%\cr &\hskip 5mm
-{1\over 2(K^2\tm t_{a1\cdots n})} 
            \Biggl[ {(1\tp\rho)} K^2
             +2 \tR\tc (k_b\tm k_a\tm K)
             +{1\over s_{ab}}G\L {k_b,k_a\atop \tR,K_{1,n}}\R
%            - \sum_{j=1}^n \biggl( s_{aj}-s_{jb}+2 k_j\tc K
%                       -{1\over s_{ab}} G\L {k_b,k_a\atop k_j,K_{1,n}}\R
%                   \biggr) (1\tm r_j) 
             \Biggr]\, k_b
}\eqn\GeneralReconstructionFunction$$
where 
$t_{1\cdots nb} = (k_1+\cdots+k_n+k_b)^2$ (so that
$K^2-t_{1\cdots nb} = 2 k_a\cdot(K_{1,n}+k_b)$);
$t_{a1\cdots n} = (k_a+k_1+\cdots+k_n)^2$;
$$
R = \sum_{j=1}^n k_j r_j\,,\qquad
\tR = K_{1,n}-R = \sum_{j=1}^n k_j (1-r_j)\,;\qquad
\anoneqn$$
and
$$
\rho = \LB 1 + {2 \,G\L{a,R,b\atop a,\tR,b}\R\over K^2 s_{ab}^2}
             + {\Delta(a,R,K,b)\over (K^2)^2 s_{ab}^2}\RB^{1/2}\,.
\anoneqn$$

Let us examine the limits of these functions when various combinations of momenta
become collinear or soft.  If two adjacent numbered legs, say $j$ and
\hbox{$j\tp 1$}, become collinear, then the reconstruction functions for 
$(k_a,k_1,\ldots,k_j,k_{j\tp 1},\ldots,k_n,k_b)$ change smoothly into those for
$(k_a,k_1,\ldots,k_j\tp k_{j\tp1},\ldots,k_n,k_b)$ so long as 
$r_{j\tp 1}\rightarrow r_j$ (and none of the $r_i$ are singular).  
Similarly, the reconstruction
functions reduce smoothly when $k_j$ becomes soft, so long as $r_j$ is not singular
in the limit.

In the limit when $k_1$ becomes collinear to $k_a$, define $k_A \equiv
k_a+k_1$, and take $z_a$ and $z_1$ to be the momentum fractions of $k_a$ 
and $k_1$ with respect to $k_A$.  Let primed variables represent sums 
omitting $k_1$,
e.g. $R'$; then
$$\eqalign{
\rho &\rightarrow \rho' = \LB 1 + {2 \,G\L{A,R',b\atop A,\tR',b}\R\over K^2 s_{Ab}^2}
             + {\Delta(A,R',K,b)\over (K^2)^2 s_{Ab}^2}\RB^{1/2}\,.
\cr
k_\ah &\rightarrow 
   -{1\over 2 z_a ( K^2\tm t_{2\cdots nb})} 
            \Biggl[ {(1\tp\rho')} K^2
            +2 R'\tc (k_A\tm k_b\tm K)
            +{1\over s_{Ab}} G\L {k_A,k_b\atop R',K_{2,n}}\R
\cr&\hphantom{    -{1\over 2 z_a ( K^2\tm t_{2\cdots nb})} \Biggl[}\hskip-7mm
            -2 z_1 r_1 k_A\tc (k_b+K)
            -2 z_1 R'\tc k_A
            +{z_1 r_1\over s_{Ab}} G\L {k_A,k_b\atop k_A,K_{2,n}}\R
            +{z_1\over s_{Ab}} G\L {k_A,k_b\atop R',k_A}\R
            \Biggr]\, (z_a k_A) 
\cr &\hskip 5mm 
            - z_1 r_1 k_A  -R'
-{1\over 2(K^2\tm t_{A2\cdots n})} 
            \Biggl[ {(1\tm\rho')} K^2
             +2 R'\tc (k_b\tm k_A\tm K)
             +{1\over s_{Ab}}G\L {k_b,k_A\atop R',K_{2,n}}\R
\cr&\hphantom{    -{1\over 2 z_a ( K^2\tm t_{2\cdots nb})} \Biggl[}\hskip2mm
             +2 z_1 r_1 k_A\tc (k_b-K)
             +2 z_1 R'\tc k_A
             +{z_1 r_1\over s_{Ab}}G\L {k_b,k_A\atop k_A,K_{2,n}}\R
             +{z_1\over s_{Ab}}G\L {k_b,k_A\atop R',k_A}\R
            \Biggr]\, k_b
\cr&=
   -{1\over 2 ( K^2\tm t_{2\cdots nb})} 
            \Biggl[ {(1\tp\rho')} K^2
            +2 R'\tc (k_A\tm k_b\tm K)
            +{1\over s_{Ab}} G\L {k_A,k_b\atop R',K_{2,n}}\R
            \Biggr]\, k_A -R'
\cr &\hskip 5mm 
   +{z_1 r_1 k_A\tc (k_b+K+K_{2,n})\over ( K^2\tm t_{2\cdots nb})} k_A
            - z_1 r_1 k_A 
     -{z_1 k_A\tc (k_b-K+K_{2,n})\over 2(K^2\tm t_{A2\cdots n})} r_1 k_b
\cr &\hskip 5mm 
-{1\over 2(K^2\tm t_{A2\cdots n})} 
            \Biggl[ {(1\tm\rho')} K^2
             +2 R'\tc (k_b\tm k_A\tm K)
             +{1\over s_{Ab}}G\L {k_b,k_A\atop R',K_{2,n}}\R
            \Biggr]\, k_b\,;
}\anoneqn$$
the terms proportional to $r_1$ on the penultimate line cancel, leaving
exactly the form required for the reconstruction function from
$(k_A,k_2,\ldots,k_n,k_b)$.  The derivation for $k_\bh$ is similar.
Because there are no singularities, this generalizes
to a collection of momenta $\{k_j,\ldots,k_{j\tp l}\}$ becoming collinear.  The
reconstruction functions thus have the correct form in any strongly-ordered limit.

\def\hr{{\hat r}}
The general limit in which we are interested involves more than one momentum
becoming singular: a subset of momenta in $\{k_1,\ldots,k_j\}$ will become collinear
with $k_a$; a subset of momenta in $\{k_{j\tp1},\ldots,k_n\}$ will become collinear
with $k_b$; and all remaining momenta will become 
soft\footnote{${}^{\dagger}$}{As explained by Campbell and Glover~[\use\GloverCampbell], 
it is sufficient to examine the limits of amplitudes for configurations 
where the collinear momenta within each of the two sets are color-connected.}.
The soft momenta disappear
quietly from the expressions for $k_{\ah,\bh}$ (so long as none of the $r_i$ are
singular), and $\rho\rightarrow 1$.  Define
$$\eqalign{
k_A&\equiv k_a\tp k_1+\cdots+k_j\,,\cr
k_B&\equiv k_{j\tp1}+\cdots+k_n+k_b\,,\cr
z_{l,m} &\equiv \sum_{j=l}^m z_j\,,\cr
\hr_{l,m} &\equiv \sum_{j=l}^m r_j z_j\,,\cr
}\anoneqn$$
where $z_1\cdots z_j$ and $z_{j\tp1}\cdots z_n$ refer to momentum fractions
with respect to $k_A$ and $k_B$ respectively.
We have the limit
$$\eqalign{
k_\ah &\rightarrow
-{1\over 2 s_{AB}} 
            \Biggl[ 2 K^2
            - \hr_{j\tp1,n} z_{1,j} s_{AB}
            - \hr_{1,j}(z_b+ 1) s_{AB}
\cr&\hphantom{ \rightarrow-{1\over 2 s_{AB}} \Biggl[  }\hskip 2mm
            +{\hr_{1,j} z_{j\tp1,n} \over s_{AB}} G\L {k_A,k_B\atop k_A,k_B}\R
            +{\hr_{j\tp1,n} z_{1,j} \over s_{AB}} G\L {k_A,k_B\atop k_B,k_A}\R
            \Biggr]\, k_A
            -\hr_{1,j} k_A -\hr_{j\tp1,n} k_B
\cr &\hskip 5mm 
-{1\over 2 s_{AB}}
            \Biggl[ 2 K^2
             -\hr_{1,j} z_{j\tp1,n} s_{AB}
             -\hr_{j\tp1,n} (z_a+1) s_{AB}
\cr&\hphantom{ \rightarrow -{1\over 2 s_{AB}} \Biggl[  }\hskip 2mm
             +{\hr_{1,j} z_{j\tp1,n} \over s_{AB}} G\L {k_B,k_A\atop k_A,k_B}\R
             +{\hr_{j\tp1,n} z_{1,j} \over s_{AB}} G\L {k_B,k_A\atop k_B,k_A}\R
            \Biggr]\, k_B
\cr
&= -[1- \hr_{1,j}]\, k_A -\hr_{1,j} k_A -\hr_{j\tp1,n} k_B
-[-\hr_{j\tp1,n}]\,k_B
\cr &= -k_A\,,
\cr
}\anoneqn$$
as desired.

For later purposes, we will also need the leading corrections
to the Gram determinants in the approach to the limit.
For this purpose, introduce two small parameters, to scale the invariants
involving $a$ and $b$ respectively,
$$\eqalign{
s_{a1}, t_{a12}, \ldots, t_{a1\cdots j} &\propto \delta_a\,,\cr
s_{nb}, \ldots, t_{(j\tp1)\cdots nb} &\propto \delta_b\,,\cr
}\eqn\ScalingParameters$$
though of course the two must be of the same order.

It is clear that expressions like 
$G\L{a,K_{1,j},b\atop a,K_{j\tp1,n},b}\R$ vanish in the limit; but how
quickly do they do so?  To understand this, first examine a simpler vanishing
object in the limit $k_a \parallel k_1$,
$$
G\L{a,1,b\atop a,q,b}\R = s_{ab} (s_{1b} s_{aq}-s_{1q} s_{ab}+s_{1a} s_{bq})
\eqn\SimpleG$$
with $q$ an arbitrary null vector.  The last term inside the parentheses
is of ${\cal O}(\delta_a)$, and the first two terms clearly cancel in the
limit.  To understand the size of the corrections, it is convenient to
rewrite them in terms of spinor products,
$$
s_{1b} s_{aq}-s_{1q} s_{ab} =
\spa1.b \spa{a}.q \spb1.b\spb{a}.q
-\spa1.q \spa{a}.b \spb1.q\spb{a}.b\,,
\anoneqn$$
and to apply the Schouten identity several times,
$$
s_{1b} s_{aq}-s_{1q} s_{ab} =
\spa1.a\spa{b}.q\spb1.b\spb{a}.q
+\spb1.a\spb{b}.q\spa1.b\spa{a}.q
-s_{1a} s_{bq}\,,
\anoneqn$$
so that overall the expression in eqn.~(\use\SimpleG) is manifestly of 
${\cal O}(\sqrt\delta_a)$ in the limit.  Each collinear pair will clearly
contribute a factor of $\sqrt\delta$, so that
$$\eqalign{
G\L{K_{1,j},a\atop q,b}\R &\sim 
 {\cal O}(\sqrt{\delta_a})\,,\cr
G\L{K_{j\tp1,n},b\atop q,a}\R &\sim 
 {\cal O}(\sqrt{\delta_b})\,,\cr
G\L{a,K_{1,j},b\atop a,K_{j\tp1,n},b}\R &\sim 
 {\cal O}(\sqrt{\delta_a\delta_b})\,,\cr
G\L {a,K_{1,j},K_{j\tp1,n},b\atop a,K_{1,j},K_{j\tp1,n},b}\R &\sim 
 {\cal O}(\delta_a \delta_b)\,.
}\anoneqn$$

\section{Antenna Amplitude for Double Emission}
\tagsection\DoubleEmissionSection
\vskip 10pt

I turn next to the construction of the antenna amplitude itself.  In
this section, I will consider the emission of two singular gluons,
deferring the derivation of a general form to the next section.  
We are interested in the following
singular configurations, \hfil\break
(a) $t_{a12}, s_{a1}, s_{12}\longrightarrow 0$, with $t_{12b}$ and $s_{2b}$
approaching different constant limits, that is the multiply-collinear limit 
$k_{1,2} \parallel k_a$;\hfil\break
(b) $s_{a1}, s_{2b}\longrightarrow 0$, with $t_{a12}$, $t_{12b}$, and $s_{12}$
approaching different constant limits, that is the double collinear limit 
$k_{1} \parallel k_a$ and $k_2\parallel k_b$;\hfil\break
(c) $t_{12b}, s_{2b}, s_{12}\longrightarrow 0$, with $t_{a12}$ and $s_{a1}$
approaching different constant limits, that is the multiply-collinear limit 
$k_{1,2} \parallel k_b$;\hfil\break
(d) $t_{a12}, s_{a1}, s_{12}, s_{2b}\longrightarrow 0$, with $t_{12b}$ 
approaching a constant limit, that is the collinear-soft limit 
$k_{1} \parallel k_a$ and $k_2$ soft;\hfil\break
(e) $t_{12b}, s_{2b}, s_{12}, s_{a1}\longrightarrow 0$, with $t_{a12}$
approaching a constant limit, that is the collinear-soft limit 
$k_{2} \parallel k_b$ and $k_1$ soft;\hfil\break
(f) All invariants 
$t_{a12}, s_{a1}, s_{12}, s_{2b}, t_{12b}\longrightarrow 0$, 
that is the double-soft limit where $k_{1,2}$ are both soft.  This is
the only case where one of the invariants ($s_{12}$) is much smaller
than the others.\hfil\break
In all cases, $s_{ab}$ again defines a hard scale, and the non-collinear 
momentum in each region acts as a reference momentum to define the collinear 
momentum fractions.  (As pointed out by Campbell and Glover~[\use\GloverCampbell],
we need not concern ourselves with the limit 
$k_{2} \parallel k_a$ where $k_1$ is soft, nor with the limit
$k_{1} \parallel k_b$ where $k_2$ is soft, because the amplitude, while singular
in these limits, will not be sufficiently so to yield any 
poles in $(D-4)$
from integrating over the phase space of singular emission.)

In general, none of the small invariants are significantly smaller
than the others, though strongly-ordered limits are included as
degenerate cases.  In all cases, however, the two-particle invariants
will vanish at least as quickly as the three-particle invariants:
$s_{a1}/t_{a12}$, for example, will be bounded above by a constant.
(If all particles are in the final state, that constant is 1; if some
are in the initial state, it can be larger than 1, but in any event we
are {\it not\/} dealing with configurations where $t_{a12}\rightarrow
0$ without the two-particle invariants $s_{a1}$, $s_{12}$ getting
small.)  We want to extract all terms that scale as $1/\delta$ or more
singular when the invariant shrinks with $\delta$, dropping less
singular terms.  Equivalently, we are interested in terms singular
when all of the ratios $\Delta(a,1,2,b)/(\Delta(a,1,b) s_{ab}),
\Delta(a,1,2,b)/(\Delta(a,2,b) s_{ab}),
\Delta(a,1,b)/s_{ab}^3$, and $\Delta(a,2,b)/s_{ab}^3$ tend to zero.

Begin by extracting all terms in the $n$-point amplitude which have 
singularities
in either $t_{a12}$; $s_{a1}$ and $s_{2b}$; or $t_{12b}$.  As explained in 
section~\use\MultipleCollinearSection, if we work in light-cone gauge, then all
poles in $s_{a1}$ or $s_{2b}$ in the regions where one of the three-particle
invariants vanishes will be contained in the first or last of these
contributions.  This
gives us the following form,
$$\eqalign{
&J(a,1,2;-(k_a\tp k_1\tp k_2);0) A_{n-2}(\ldots,k_a\tp k_1\tp k_2,b,\ldots)
\cr &
+J(a,1;-(k_a+k_1);0) J(2,b;-(k_2+k_b);0)  
  A_{n-2}(\ldots,k_a\tp k_1,k_b\tp k_2,\ldots)
\cr &
+J(1,2,b;-(k_1+k_2+k_b);0) A_{n-2}(\ldots,a,k_b\tp k_1\tp k_2,\ldots)
}\anoneqn$$
I do {\it not\/} extract a contribution proportional to $J(1,2;\cdot;\cdot)$; 
while
such a term indeed gives rise to an $s_{12}$ pole, it cannot give rise to
singularities in other invariants, and hence would yield only a subleading
 contribution.  That is, 
if we examine those diagrams containing only three-point
vertices, and ignore the helicity algebra which will soften the poles,
we need to extract all terms that have poles in two invariants; one 
invariant will not suffice.

As in the single-emission case, introduce a complete set of polarization
states for the unfused leg in the first and last terms, rewriting the
product of polarization vectors as a two-point current; and introduce
new labels for the surviving hard momenta, yielding
$$\eqalign{
&J(a,1,2;-(k_a\tp k_1\tp k_2);0) J(b;-k_b;0) 
   A_{n-2}(\ldots,-k_\ah = k_a\tp k_1\tp k_2,-k_\bh=k_b,\ldots)
\cr &
+J(a,1;-(k_a\tp k_1);0) J(2,b;-(k_2\tp k_b);0)  
   A_{n-2}(\ldots,-k_\ah=k_a\tp k_1,-k_\bh=k_b\tp k_2,\ldots)
\cr &
+J(a;-k_a;0) J(1,2,b;-(k_1\tp k_2\tp k_b);0)
   A_{n-2}(\ldots,-k_\ah=k_a,-k_\bh=k_b\tp k_1\tp k_2,\ldots)
}\anoneqn$$
Using the reconstruction functions defined in the previous section,
and again putting the hatted momenta on-shell (at the price of allowing
momentum to flow between the two currents in each term),
we can combine terms to obtain an antenna factorization amplitude,
$$\eqalign{
\Ant(\ah,\bh\leftarrow a,1,2,b) =
&J(a,1,2;\ah;k_b\tp k_\bh) J(b;\bh;k_a\tp k_1\tp k_2\tp k_\ah) 
\cr&
+J(a,1;\ah;k_b\tp k_2\tp k_\bh) J(2,b;\bh;k_a\tp k_1\tp k_\ah)  
\cr&
+J(a;\ah;k_b\tp k_1\tp k_2\tp k_\bh) J(1,2,b;\bh;k_a\tp k_\ah)\,,
}\eqn\DoubleEmissionAntennaFirstDef$$
and corresponding factorization in any singular limit,
$$
\Ant(\ah,\bh\leftarrow a,1,2,b) 
   A_{n-2}(\ldots,-k_\ah=f_\ah(k_a,k_1,k_2,k_b)
                 ,-k_\bh=f_\bh(k_a,k_1,k_2,k_b),\ldots)\,.
\eqn\DoubleEmissionAntennaFactorization$$
In this equation, the summation over physical polarizations of $\ah$ and $\bh$
is implicit.

We must still impose the requirement that the momentum excess in each
current leads only to subleading terms.  This will lead to constraints
on the coefficients $r_i$.  
In the last term of eqn.~(\use\DoubleEmissionAntennaFirstDef),
we must examine the behavior of $k_a\tp k_\ah$ as $k_{1,2}$ become
collinear to $k_b$ (the argument will also hold when either or both
become soft),
$$\eqalign{
% Expansions checked 07/22/02
k_a\tp k_\ah &\sim 
\Biggl[{1-\rho\over2}
       + {t_{12b}\over 4 k_a\tc (k_b+K_{1,2})\,s_{ab}} \biggl\{
              {(1-\rho) s_{ab}}
              -2 s_{ab}
              +s_{a2} {s_{1b}\over t_{12b}} (r_1-r_2)
              -s_{a1} {s_{2b}\over t_{12b}} (r_1-r_2)
\cr&\hphantom{ \sim \Biggl[{1-\rho\over2}
       + {t_{12b}\over 2 k_a\tc (k_b+K_{1,2})\,s_{ab}} \biggl\{ }
              +s_{ab} {s_{12}\over t_{12b}} (r_1+r_2)
              +2 s_{ab} {s_{1b}\over t_{12b}} r_1
              +2 s_{ab} {s_{2b}\over t_{12b}} r_2\biggr\}
             \Biggr] k_a 
\cr&\hphantom{\sim}
- r_1 k_1 - r_2 k_2 
\cr&\hphantom{\sim}
       +{1\over 2 s_{ab}}\Biggl[{2 (\rho-1)\,k_a\tc (k_b+K_{1,2})}
               +2{s_{a1}} r_1
               +2{s_{a2}} r_2
\cr&\hphantom{\sim +\Biggl[}
               +{t_{12b}\over s_{ab}} \biggl\{
                (1-\rho) (s_{a1}+s_{a2})
                +2 (\rho-1) {s_{12}\over t_{12b}} k_a\tc (k_b+K_{1,2})
              -s_{a1} {s_{2b}\over t_{12b}} (r_1+r_2)
\cr&\hphantom{ \sim +\Biggl[ +{t_{12b}\over 2 s_{ab}^2} \biggl\{ }
              -s_{a2} {s_{1b}\over t_{12b}} (r_1+r_2)
              +s_{ab} {s_{12}\over t_{12b}} (r_1+r_2)
              -2 s_{a1} {s_{1b}\over t_{12b}} r_1
              -2 s_{a2} {s_{2b}\over t_{12b}} r_1\biggr\}\Biggr] k_b
\cr&\hphantom{\sim}
+{\cal O}(t_{12b}^2).
}\anoneqn$$
Also,
$$\eqalign{
\rho &\sim 1+{t_{12b}\over 2 k_a\tc (k_b+K_{1,2})\,s_{ab}}
\biggl\{ 2 s_{a1} r_1 (1-r_1) {s_{1b}\over t_{12b}}
+ 2 s_{a2} r_2 (1-r_2) {s_{2b}\over t_{12b}}
\cr&\hphantom{ \sim 1+{t_{12b}\over k_a\tc (k_b+K_{1,2})\,s_{ab}}\biggl\{ }
+ (r_1+r_2- 2 r_1 r_2) \biggl( s_{a2} {s_{1b}\over t_{12b}}
+s_{a1} {s_{2b}\over t_{12b}} - s_{ab} {s_{12}\over t_{12b}}
\biggl)
\biggr\} + {\cal O}(t_{12b}^2)\,.
}\anoneqn$$
As noted above, for the limits in which we are interested,
the ratios $s_{1b}/t_{12b}$, $s_{2b}/t_{12b}$, and 
$s_{12}/t_{12b}$ are all bounded in regions where $t_{12b}\rightarrow 0$.
Most of the terms in $k_a+k_\ah$ therefore kill off the leading
$t_{12b}$ pole in $ J(1,2,b;\bh;\cdot)$, and leave only subleading 
contributions.  The surviving terms are,
$$
-r_1 k_1 - r_2 k_2 + {s_{a1}\over s_{ab}} r_1 k_b
+ {s_{a2}\over s_{ab}} r_2 k_b\,.
\eqn\Residual$$
The choice 
$$
r_j = {k_j\cdot (K_{j\tp1,n}+k_b)\over k_j\cdot K}
    = {t_{j\cdots nb}-t_{(j\tp1)\cdots nb}\over 2 k_j\cdot K}
\eqn\DoubleRChoice$$
ensures that these vanish like a full power of a small invariant
as as $t_{12b}\rightarrow 0$,
$$\eqalign{
r_1 &= {t_{12b}\over 2 k_1\cdot K} \L 1- {s_{2b}\over t_{12b}}\R\,,\cr
r_2 &= {t_{12b}\over 2 k_2\cdot K} {s_{2b}\over t_{12b}}\,,\cr
}\anoneqn$$
so that the remaining terms also kill off the leading pole, leaving
only subleading contributions to the antenna amplitude.   In fact,
a closer examination of the terms in eqn.~(\use\Residual) shows that
a square-root vanishing would have been sufficient: we recognize
the dot product
of that vector with any other vector $w$ as
$$
-{r_1\over s_{ab}} G\L{k_b, k_1\atop k_a, w}\R
-{r_2\over s_{ab}} G\L{k_b, k_2\atop k_a, w}\R\,,
\anoneqn$$
which, as discussed in the previous section, already vanishes like the square root of
a small invariant in the limit.  Similarly, in the single-emission
case, taking $r_1\sim \sqrt{s_{a1}}$ in the limit would have been
sufficient to avoid changing the leading singularity.

A similar argument holds for the first term in 
eqn.~(\use\DoubleEmissionAntennaFirstDef), and the choice
of $r_i$ in eqn.~(\use\DoubleRChoice) 
will lead to contributions of ${\cal O}(t_{a12})$ from the
excess momentum, that is, subleading contributions.

In the middle term of eqn.~(\use\DoubleEmissionAntennaFirstDef), we need
to examine the behavior of $k_a+k_\ah+k_1$ and $k_b+k_\bh+k_2$
as $k_1$ becomes collinear to $k_a$, and $k_2$ to $k_b$,
$$\eqalign{
%% checked 07/22/02
k_\ah&+k_a+k_1 \sim
\cr &{1\over 2 s_{ab} (s_{a2}+s_{ab})}
 \biggl[ (s_{1b} s_{a2}-s_{12} s_{ab})
                \biggl(1-{s_{a1}\over s_{a2}+s_{ab}}\biggr)
        +(1-\rho) s_{ab}(s_{12}+s_{1b}) 
                \biggl(1-{s_{a1}\over s_{a2}+s_{ab}}\biggr)
\cr&\hphantom{ {1\over 2 s_{ab} (s_{a2}+s_{ab})} \biggl[ }
        +(1-\rho) s_{ab} (s_{a2}+s_{ab})
                \biggl(1+{s_{2b}\over s_{a2}+s_{ab}}\biggr)
\cr&\hphantom{ {1\over 2 s_{ab} (s_{a2}+s_{ab})} \biggl[ }
        +(r_1-1) \biggl(1-{s_{a1}\over s_{a2}+s_{ab}}\biggr)
                    (s_{1b} s_{a2}+s_{12} s_{ab}+2 s_{1b} s_{ab})
\cr&\hphantom{ {1\over 2 s_{ab} (s_{a2}+s_{ab})} \biggl[ }
        +r_2 (s_{12} s_{ab}-s_{1b} s_{a2})
                \biggl(1-{s_{a1}\over s_{a2}+s_{ab}}\biggr)
        +2 (r_2-1) s_{2b} s_{ab}\biggr] k_a
        +(1-r_1) k_1
\cr&
%% checked 07/22/02
 +{1\over 2 s_{ab} (s_{1b}+s_{ab})}\biggl[ 
        (s_{12} s_{ab}-s_{1b} s_{a2})
                \biggl(1-{s_{2b}\over s_{1b}+s_{ab}}\biggr)
        +(\rho-1) s_{ab} (s_{12}+s_{a2}) 
                \biggl(1-{s_{2b}\over s_{1b}+s_{ab}}\biggr)
\cr&\hphantom{ +{1\over 2 s_{ab} (s_{1b}+s_{ab})} \biggl[ }
        +(\rho-1) s_{ab} (s_{1b}+s_{ab}) 
                \biggl(1+{s_{a1}\over s_{1b}+s_{ab}}\biggr)
\cr&\hphantom{ +{1\over 2 s_{ab} (s_{1b}+s_{ab})} \biggl[ }
        +(r_1-1) (s_{12} s_{ab}-s_{1b} s_{a2})
                \biggl(1-{s_{2b}\over s_{1b}+s_{ab}}\biggr)
        +2 r_1 s_{a1} s_{ab}
\cr&\hphantom{ +{1\over 2 s_{ab} (s_{1b}+s_{ab})} \biggl[ }
        +r_2 (s_{1b} s_{a2} + s_{12} s_{ab} + 2 s_{a2} s_{ab})
                \biggl(1-{s_{2b}\over s_{1b}+s_{ab}}\biggr)
        \biggr] k_b-r_2 k_2\,,
}\anoneqn$$
$$\eqalign{
k_\bh&+k_b+k_2 \sim
%% checked 07/22/02
\cr &{1\over 2 s_{ab} (s_{2a}+s_{ab})}
 \biggl[ (s_{12} s_{ab}-s_{1b} s_{a2})
                \biggl(1-{s_{a1}\over s_{a2}+s_{ab}}\biggr)
        +(\rho-1) (s_{12}+s_{1b})
                \biggl(1-{s_{a1}\over s_{a2}+s_{ab}}\biggr)
\cr&\hphantom{ {1\over 2 s_{ab} (s_{1b}+s_{ab})} \biggl[ }
        +(\rho-1) (s_{a2}+s_{ab})
                \biggl(1+{s_{2b}\over s_{a2}+s_{ab}}\biggr)
\cr&\hphantom{ {1\over 2 s_{ab} (s_{1b}+s_{ab})} \biggl[ }
        +(1-r_1) (s_{1b} s_{a2}+s_{12} s_{ab} +2 s_{1b} s_{ab})
                \biggl(1-{s_{a1}\over s_{a2}+s_{ab}}\biggr)
\cr&\hphantom{ {1\over 2 s_{ab} (s_{1b}+s_{ab})} \biggl[ }
        +2 (1-r_2)s_{2b} s_{ab}
        -r_2 (s_{12} s_{ab}-s_{1b} s_{a2})
                \biggl(1-{s_{a1}\over s_{a2}+s_{ab}}\biggr)
 \biggr] k_a - (1-r_1) k_1
%% checked 07/22/02
\cr &+{1\over 2 s_{ab} (s_{1b}+s_{ab})}
 \biggl[ (s_{1b} s_{a2}-s_{12} s_{ab})
                \biggl(1-{s_{2b}\over s_{1b}+s_{ab}}\biggr)
        +(1-\rho) s_{ab}(s_{12}+s_{a2}) 
                \biggl(1-{s_{2b}\over s_{1b}+s_{ab}}\biggr)
\cr&\hphantom{ {1\over 2 s_{ab} (s_{a2}+s_{ab})} \biggl[ }
        +(1-\rho) s_{ab} (s_{1b}+s_{ab})
                \biggl(1+{s_{a1}\over s_{1b}+s_{ab}}\biggr)
\cr&\hphantom{ {1\over 2 s_{ab} (s_{a2}+s_{ab})} \biggl[ }
        +(r_1-1) (s_{12} s_{ab}-s_{1b} s_{a2})
                \biggl(1-{s_{2b}\over s_{1b}+s_{ab}}\biggr)
        -2 r_1 s_{a1} s_{ab}
\cr&\hphantom{ {1\over 2 s_{ab} (s_{a2}+s_{ab})} \biggl[ }
        -r_2 \biggl(1-{s_{2b}\over s_{1b}+s_{ab}}\biggr)
                    (s_{1b} s_{a2}+s_{12} s_{ab}+2 s_{a2} s_{ab})
        \biggr] k_b + r_2 k_2.
}\anoneqn$$

Any term proportional to {\it either\/} $s_{a1}$ or $s_{2b}$ will kill
off the pole in 
$J(a,1;\ah;k_b\tp k_2\tp k_\bh)$ or $J(2,b;\bh;k_a\tp k_1\tp k_\ah)$
respectively,
giving rise only to subleading contributions.  Also, using 
eqn.~(\use\DoubleRChoice), we find
$$\eqalign{
r_1-1 &= -{k_1\cdot(k_a+k_1)\over k_1\cdot K} = {s_{a1}\over 2 k_1\cdot K}\,,
\cr
r_2 &= {s_{2b}\over 2 k_2\cdot K}\,,
}\anoneqn$$
so terms proportional to $(r_1-1)$ or $r_2$ can be dropped as well.
This leaves us with,
$$\eqalign{
k_\ah&+k_a+k_1 \sim
\cr &{1\over 2 s_{ab} (s_{a2}+s_{ab})}
 \biggl[ (s_{1b} s_{a2}-s_{12} s_{ab})
        +(1-\rho) s_{ab}(s_{12}+s_{1b}+s_{a2}+s_{ab})\biggr] k_a
\cr&
 +{1\over 2 s_{ab} (s_{1b}+s_{ab})}\biggl[ 
        (s_{12} s_{ab}-s_{1b} s_{a2})
        +(\rho-1) s_{ab} (s_{12}+s_{a2}+s_{1b}+s_{ab}) \biggr] k_b\,,
\cr
k_\bh&+k_b+k_2 \sim
\cr &{1\over 2 s_{ab} (s_{2a}+s_{ab})}
 \biggl[ (s_{12} s_{ab}-s_{1b} s_{a2})
        +(\rho-1) (s_{12}+s_{1b}+s_{a2}+s_{ab}) \biggr] k_a 
\cr &+{1\over 2 s_{ab} (s_{1b}+s_{ab})}
 \biggl[ (s_{1b} s_{a2}-s_{12} s_{ab})
        +(1-\rho) s_{ab}(s_{12}+s_{a2}+s_{1b}+s_{ab}) \biggr] k_b\,.
}\anoneqn$$
We recognize $(s_{1b} s_{a2}-s_{12} s_{ab})$
as $G({a,1,b\atop a,2,b})/s_{ab}$, which as discussed in the previous section,
is~\hbox{$\sim \sqrt{s_{a1} s_{2b}}$}
in the limit; likewise,
$$
\rho-1 \sim {(s_{12} s_{ab}-s_{a2} s_{1b})
               (s_{1b} s_{2a} + s_{12} s_{ab} + 2 (s_{2a}+s_{1b}) s_{ab} 
                +2 s_{ab}^2)
               \over 2 s_{ab}^2 (s_{a2}+s_{12}+s_{1b}+s_{ab})^2}
\sim \sqrt{s_{a1} s_{2b}}\,,
\eqn\ExcessA$$
so that the surviving terms in both 
$k_\ah+k_a+k_1$ and $k_\bh+k_b+k_2$ are all of ${\cal O}(\sqrt{s_{a1} s_{2b}})$.

Next, split $J(a,1;\ah;k_b\tp k_2\tp k_\bh)$ and $J(2,b;\bh;k_a\tp k_1\tp k_\ah)$
each into a `canonical'
term and an `excess' term.  The former is
 defined as the current with the excess momenta set to zero,
while the latter contains the excess momenta,
$$\eqalign{
J^{\rm canon}(a,1;\ah;k_b\tp k_2\tp k_\bh) &= J(a,1;\ah;0)\,,\cr
J^{\rm excess}(a,1;\ah;k_b\tp k_2\tp k_\bh) &= 
  J(a,1;\ah;k_b\tp k_2\tp k_\bh)-J(a,1;\ah;0)\,.\cr
}\anoneqn$$
Because the off-shell momentum does not appear on the right-hand side
of the recurrence relation~(\use\RecurrenceRelations), the canonical
term is in fact the same as the current with the hatted argument
replaced by the off-shell momentum, respectively $-(k_a+k_1)$ or
$-(k_2+k_b)$ in the two currents.  In the canonical terms, while an
$s_{a1}$ or $s_{2b}$ pole appears in the formal expression for the
currents, evaluation of the polarization vectors will soften the
singularity to a square-root one.  In the excess terms, no such
softening will necessarily occur.  The product of the two canonical
terms gives us a term in the antenna amplitude; we want to show that
either the product of a canonical and an excess term, or the product
of the two excess terms, yield only subleading terms in this singular
region.  Since each excess term is $\sim \sqrt{s_{a1} s_{2b}}$ in the
limit, the product indeed kills off both poles, as desired.  In the
product of an excess term and a canonical term, the strength of the
canonical term's pole will be reduced only by a square root of the
pole invariant; but here, this suffices because the canonical term
only has a square-root singularity rather than a full pole in the
invariant.

Thus, in each of the regions where a given term in the antenna
amplitude contributions, the excess momentum transferred between the
currents gives rise to no corrections to the leading singularity.  We
can thus use a simplified definition, ignoring this excess momentum,
$$\eqalign{
\Ant(\ah,\bh\leftarrow a,1,2,b) =
&J(a,1,2;\ah;0) J(b;\bh;0) 
\cr&
+J(a,1;\ah;0) J(2,b;\bh;0)  
\cr&
+J(a;\ah;0) J(1,2,b;\bh;0)\,.
}\eqn\DoubleEmissionAntennaDef$$

\section{Antenna Amplitude for Multiple Emission}
\tagsection\MultipleEmissionSection
\vskip 10pt

The construction and arguments of the previous section generalize straightforwardly to 
the case of multiple singular emission.  In the general case, with emission of
$m$ singular gluons, we are interested in all terms with singularities in at least
$m$ 
invariants.  To extract these, we isolate all terms with poles in 
$t_{a1\cdots m}$; $t_{a1\cdots (m\tm1)}$ and $s_{mb}$;
$t_{a1\cdots (m\tm2)}$ and $t_{(m\tm1)mb}$; and so on through
terms with poles in $t_{1\cdots mb}$.  As in the double-emission case, these terms will 
necessarily have singularities in additional variables.  For example the term isolated
via simultaneous poles in $t_{a1\cdots j}$ and $t_{(j\tp1)\cdots mb}$ will also contain
poles in $s_{a1}, t_{a12}, \ldots, t_{a1\cdots (j\tm1)}$.  This extraction yields
the following form,
$$\eqalign{
J&(a,1,\ldots,m;-(k_a\tp K_{1,m});0)
   A_{n-m}(\ldots,k_a\tp K_{1,m},b,\ldots)
\cr &+\sum_{j=1}^{m\tm1}
J(a,1,\ldots,j;-(k_a\tp K_{1,j});0)
J(j\tp1,\ldots,m,b;-(K_{j\tp1,m} \tp k_b);0)
\cr &\hskip 20mm\times
   A_{n-m}(\ldots,k_a\tp K_{1,j},k_b\tp K_{j\tp1,m},\ldots)
\cr &+
J(1,\ldots,m,b;-(k_1\tp\cdots\tp k_m\tp k_b);0)
   A_{n-m}(\ldots,a,k_b\tp K_{1,m},\ldots)
}\anoneqn$$

As in the single- and double-emission cases, introduce a complete set of polarization
states for the unfused leg in the first and last terms, rewriting the
product of polarization vectors as a two-point current.  Again introducing
new labels for the surviving hard momenta, we obtain for the factorization,
$$\eqalign{
\sum_{j=0}^m J&(a,1,\ldots,j;-(k_a\tp K_{1,j});0) 
             J(j\tp1,\ldots,m,b;-(k_b\tp K_{j\tp1,m});0) 
\cr&\hskip 10mm\times
   A_{n-m}(\ldots,-k_\ah = k_a\tp K_{1,j},
                  -k_\bh=k_b\tp K_{j\tp1,m},\ldots)
\cr
}\anoneqn$$
Using the reconstruction functions defined in section~\use\GeneralReconstructionSection,
and again putting the hatted momenta on-shell (at the price of allowing
momentum to flow between the two currents in each term),
we can combine terms to obtain a general antenna factorization amplitude,
$$\eqalign{
\Ant&(\ah,\bh\leftarrow a,1,\ldots,m,b) =
\cr &\hskip 10mm
\sum_{j=0}^m J(a,1,\ldots,j;\ah;k_b\tp k_\bh\tp K_{j\tp1,m})
             J(j\tp1,\ldots,m,b;\bh;k_a\tp k_\ah\tp K_{1,j}) 
\cr
}\eqn\MultipleEmissionAntennaFirstDef$$
and corresponding factorization in singular limits,
$$
\Ant(\ah,\bh\leftarrow a,1,\ldots,m,b) 
   A_{n-m}(\ldots,-k_\ah=f_\ah(k_a,k_1,\ldots,k_m,k_b),
                  -k_\bh=f_\bh(k_a,k_1,\ldots,k_m,k_b),\ldots)\,,
\eqn\MultipleEmissionAntennaFactorization$$
where again the summation over physical polarizations of $\ah$ and $\bh$
is implicit.

We can now verify that the choice~(\use\DoubleRChoice) for the coefficients $r_i$
ensures that the momentum excesses in this equation lead only to subleading
terms.  In a generic term, $0<j< m$, we must examine the behavior of
$k_a\tp k_\ah\tp k_1\tp\cdots\tp k_j$ as $k_{j\tp1,\ldots,m}$ become collinear
to $k_b$.  In terms of the small parameters $\delta_{a,b}$ introduced in 
eqn.~(\use\ScalingParameters), the coefficients
$r_i\sim 1+{\cal O}(\delta_a)$ for $1\leq i\leq j$ and
$r_i\sim {\cal O}(\delta_b)$ for $j\tp1\leq i\leq n$.  Define $\delta r_i$ via
$$
r_i =\left\{\eqalign{1+\delta r_i\,,\qquad &1\leq i\leq j;\cr
 \delta r_i\,,\qquad &j\tp1\leq i\leq n;\cr}\right.
\anoneqn$$
and $\delta R_{l,m} = \Sigma_{j=l}^m \delta r_j k_j$.

We then see that
$$\eqalign{
\rho^2 &\sim  1 
             + {2\over K^2 s_{ab}^2} \LB 
                  G\L{a,K_{1,j},b\atop a,K_{j\tp1,n},b}\R
                  -G\L{a,K_{1,j},b\atop a,\delta R_{1,j},b}\R
                  -G\L{a,K_{1,j},b\atop a,\delta R_{j\tp1,n},b}\R
\RP\cr &\hphantom{ \sim  1 + {2\over K^2 s_{ab}} \LB
                   \vphantom{G\L{a,K_{1,j},b\atop a,K_{j\tp1,n},b}\R}\RP  }\LP
                  +G\L{a,\delta R_{1,j},b\atop a,K_{j\tp1,n},b}\R
                  +G\L{a,\delta R_{1,j},b\atop a,\delta R_{1,j},b}\R
                  +G\L{a,\delta R_{1,j},b\atop a,\delta R_{j\tp1,n},b}\R
                  \RB
\cr &\hphantom{\sim 1}
             + {1\over (K^2)^2 s_{ab}^2}\LB
    G\L {a,K_{1,j},K_{j\tp1,n},b\atop a,K_{1,j},K_{j\tp1,n},b}\R
    +2 G\L {a,K_{1,j},K_{1,n},b\atop a,\delta R_{1,j},K_{1,n},b}\R
    +2 G\L {a,K_{1,j},K_{1,n},b\atop a,\delta R_{j\tp1,n},K_{1,n},b}\R
\RP\cr &\hphantom{ \sim  1 + {2\over K^2 s_{ab}} \LB
                   \vphantom{G\L{a,K_{1,j},b\atop a,K_{j\tp1,n},b}\R}\RP  }\LP
    +G\L {a,\delta R_{1,j},K_{1,n},b\atop a,\delta R_{1,j},K_{1,n},b}\R
    +2 G\L {a,\delta R_{1,j},K_{1,n},b\atop a,\delta R_{j\tp1,n},K_{1,n},b}\R
    +G\L {a,\delta R_{j\tp1,n},K_{1,n},b\atop a,\delta R_{j\tp1,n},K_{1,n},b}\R
%    G\L {a,R,K_{1,n},b\atop a,R,K_{1,n},b}\R
             \RB\,\cr
&\sim 1 + {\cal O}(\sqrt{\delta_a\delta_b})\,,
}\eqn\rhoExpandA$$

so that
$$\eqalign{
k_\ah &+ k_a + K_{1,j} \sim \cr
&  -{1\over 2( K^2\tm t_{1\cdots nb})} 
            \Biggl[ (\rho-1) K^2 +2 t_{1\cdots nb}
            -2 K_{1,j}\tc (2 k_b\tp K_{1,n})
            -2 \delta R_{1,j}\tc (2 k_b\tp K_{1,n})
\cr &\hphantom{ \sim -{1\over 2( K^2\tm t_{1\cdots nb})} \Biggl[}
            -2 \delta R_{j\tp1,n}\tc (2 k_b\tp K_{1,n})
            +{1\over s_{ab}} G\L {k_a,k_b\atop K_{1,j},K_{j\tp1,n}}\R
            +{1\over s_{ab}} G\L {k_a,k_b\atop \delta R_{1,j},K_{1,n}}\R
\cr &\hphantom{ \sim -{1\over 2( K^2\tm t_{1\cdots nb})} \Biggl[}
            +{1\over s_{ab}} G\L {k_a,k_b\atop \delta R_{j\tp1,n},K_{1,n}}\R
            \Biggr]\, k_a
            +\delta R_{1,j} + \delta R_{j\tp1,n}
\cr &\hskip 5mm 
-{1\over 2(K^2\tm t_{a1\cdots n})} 
            \Biggl[ {(1\tm\rho)} K^2
             -2 K_{1,j}\tc (2 k_a\tp K_{1,n})
             -2 \delta R_{1,j}\tc (2 k_a\tp K_{1,n})
\cr &\hphantom{ \sim -{1\over 2( K^2\tm t_{1\cdots nb})} \Biggl[}
             -2 \delta R_{j\tp1,n}\tc (2 k_a\tp K_{1,n})
             +{1\over s_{ab}}G\L {k_b,k_a\atop K_{1,j},K_{1,n}}\R
             +{1\over s_{ab}}G\L {k_b,k_a\atop \delta R_{1,j},K_{1,n}}\R
\cr &\hphantom{ \sim -{1\over 2( K^2\tm t_{1\cdots nb})} \Biggl[}
             +{1\over s_{ab}}G\L {k_b,k_a\atop \delta R_{j\tp1,n},K_{1,n}}\R
            \Biggr]\, k_b
\cr
}$$

$$\eqalign{&=  -{1\over 2( K^2\tm t_{1\cdots nb})} 
            \Biggl[ (\rho-1) K^2 
            +2 (k_b+K_{j\tp1,n})^2
            +2 K_{1,j}\tc K_{j\tp1,n}
            -{4 k_a\tc K_{j\tp1,n}\, k_b\tc K_{1,j}\over s_{ab}}
\cr &\hphantom{ \sim -{1\over 2( K^2\tm t_{1\cdots nb})} \Biggl[}
            +{4 k_a\tc K_{1,j}\, k_b\tc K_{j\tp1,n}\over s_{ab}} 
            -2 \delta R_{1,j}\tc (2 k_b\tp K_{1,n})
            -2 \delta R_{j\tp1,n}\tc (2 k_b\tp K_{1,n})
\cr &\hphantom{ \sim -{1\over 2( K^2\tm t_{1\cdots nb})} \Biggl[}
            +{4 k_a\tc K_{1,j}\,k_b\tc\delta R_{1,j}\over s_{ab}} 
            +{4 k_a\tc K_{j\tp1,n}\,k_b\tc\delta R_{1,j}\over s_{ab}} 
            -{4 k_a\tc \delta R_{1,j}\,k_b\tc K_{1,j}\over s_{ab}}
\cr &\hphantom{ \sim -{1\over 2( K^2\tm t_{1\cdots nb})} \Biggl[}
            -{4 k_a\tc \delta R_{1,j}\,k_b\tc K_{j\tp1,n}\over s_{ab}}
            +{4 k_a\tc K_{1,j}\,k_b\tc\delta R_{j\tp1,n}\over s_{ab}} 
            +{4 k_a\tc K_{j\tp1,n}\,k_b\tc\delta R_{j\tp1,n}\over s_{ab}} 
\cr &\hphantom{ \sim -{1\over 2( K^2\tm t_{1\cdots nb})} \Biggl[}
            -{4 k_a\tc \delta R_{j\tp1,n}\,k_b\tc K_{1,j}\over s_{ab}}
            -{4 k_a\tc \delta R_{j\tp1,n}\,k_b\tc K_{j\tp1,n}\over s_{ab}}
            \Biggr]\, k_a
            +\delta R_{1,j} + \delta R_{j\tp1,n}
\cr &\hskip 5mm 
-{1\over 2(K^2\tm t_{a1\cdots n})} 
            \Biggl[ {(1\tm\rho)} K^2
             -2 K_{1,j}\tc (2 k_a\tp K_{1,j})
             -2 K_{1,j}\tc K_{j\tp1,n}
             -2 \delta R_{1,j}\tc (2 k_a\tp K_{1,n})
\cr &\hphantom{ \sim -{1\over 2( K^2\tm t_{1\cdots nb})} \Biggl[}
             -2 \delta R_{j\tp1,n}\tc (2 k_a\tp K_{1,n})
             -{4 k_b\tc K_{j\tp1,n}\,k_a\tc K_{1,j}\over s_{ab}}
             +{4 k_b\tc K_{1,j}\,k_a\tc K_{j\tp1,n}\over s_{ab}}
\cr &\hphantom{ \sim -{1\over 2( K^2\tm t_{1\cdots nb})} \Biggl[}
             -{4 k_b\tc K_{1,j}\,k_a\tc \delta R_{1,j}\over s_{ab}}
             -{4 k_b\tc K_{j\tp1,n}\,k_a\tc \delta R_{1,j}\over s_{ab}}
             +{4 k_b\tc \delta R_{1,j}\,k_a\tc K_{1,j}\over s_{ab}}
\cr &\hphantom{ \sim -{1\over 2( K^2\tm t_{1\cdots nb})} \Biggl[}
             +{4 k_b\tc \delta R_{1,j}\,k_a\tc K_{j\tp1,n}\over s_{ab}}
             -{4 k_b\tc K_{1,j}\,k_a\tc \delta R_{j\tp1,n}\over s_{ab}}
             -{4 k_b\tc K_{j\tp1,n}\,k_a\tc \delta R_{j\tp1,n}\over s_{ab}}
\cr &\hphantom{ \sim -{1\over 2( K^2\tm t_{1\cdots nb})} \Biggl[}
             +{4 k_b\tc \delta R_{j\tp1,n}\,k_a\tc K_{1,j}\over s_{ab}}
             +{4 k_b\tc \delta R_{j\tp1,n}\,k_a\tc K_{j\tp1,n}\over s_{ab}}
            \Biggr]\, k_b
\cr
&=  {\cal O}(\sqrt{\delta_a\delta_b})
            +{\cal O}(\delta_a)
            +{\cal O}(\delta_b)\,,
\cr
}$$
with similar result for $k_b+k_\bh+K_{j\tp1,m}$.  A term that is of order $\delta_a$
or $\delta_b$ will kill off the leading pole in either 
$J(a,1,\ldots,j;\ah;k_b\tp k_\bh\tp K_{j\tp1,m})$ or 
$J(j\tp1,\ldots,m,b;\bh;k_a\tp k_\ah\tp K_{1,j})$, respectively.
As in the double-emission case, the terms of order $\sqrt{\delta_a \delta_b}$ will
either combine to kill off the leading poles in both currents, or else will kill
off the leading, helicity-softened, square-root singularity in the `canonical' part
of the other current.  

In the case $j=0$, we must consider the behavior of 
$$\eqalign{
k_a+k_\ah &\sim
 -{(\rho\tm1)\over 2} \, k_a
 +{\rho\tm1\over 2} \, k_b + {\cal O}(\delta_b)\,;\cr
}\anoneqn$$
in this special case, $\rho-1\sim {\cal O}(\delta_b^2)$, so that terms containing
$k_a+k_\ah$ will kill off the leading singularity in the current
$J(1,\ldots,m,b;\bh;k_a\tp k_\ah)$.  A similar argument holds for the case $j=m$.

In all cases, the additional terms due to the excess momentum transferred between
the two currents in each term will give no corrections
to the leading singularity.  We can thus set the excess momentum to zero in order
to arrive at our final, simplified formula for
the antenna amplitude for emission of $m$ singular gluons,
$$\eqalign{
\Ant&(\ah,\bh\leftarrow a,1,\ldots,m,b) =
%\cr &\hskip 10mm
\sum_{j=0}^m J(a,1,\ldots,j;\ah;0)
             J(j\tp1,\ldots,m,b;\bh;0)\,.
\cr
}\eqn\MultipleEmissionAntennaDef$$

\section{Antenna Amplitudes for Specific Helicities}
\tagsection\DoubleEmissionHelicitySection
\vskip 10pt

Using eqn.~(\use\AntennaDef) and 
the spinor-helicity basis~[\use\Spinor], I obtain the following
explicit forms for the antenna helicity amplitudes,
$$\eqalign{
%%%%% begin Ant1 : Ant[PPPPP]
\Ant(\ah^+,\bh^+\leftarrow a^+,1^+,b^+) &= 0,\cr
%%%%% end Ant1 : Ant[PPPPP]
%%%%% begin Ant1 : Ant[PPMPP]
\Ant(\ah^+,\bh^+\leftarrow a^-,1^+,b^+) &= 0,\cr
%%%%% end Ant1 : Ant[PPMPP]
%%%%% begin Ant1 : Ant[PPPMP]
\Ant(\ah^+,\bh^+\leftarrow a^+,1^-,b^+) &= 0,\cr
%%%%% end Ant1 : Ant[PPPMP]
%%%%% begin Ant1 : Ant[PPPPM]
\Ant(\ah^+,\bh^+\leftarrow a^+,1^+,b^-) &= 0,\cr
%%%%% end Ant1 : Ant[PPPPM]
%%%%% begin Ant1 : Ant[PPMMP]
\Ant(\ah^+,\bh^+\leftarrow a^-,1^-,b^+) &= 
\frac{{\spa{a}.{1}}^3}{\spa{a}.{b}\,{\spa{\ah}.{\bh}}^2\,\spa{1}.{b}}
,\cr
%%%%% end Ant1 : Ant[PPMMP]
%%%%% begin Ant1 : Ant[PPMPM]
\Ant(\ah^+,\bh^+\leftarrow a^-,1^+,b^-) &= 
\frac{{\spa{a}.{b}}^3}{\spa{a}.{1}\,{\spa{\ah}.{\bh}}^2\,\spa{1}.{b}}
,\cr
%%%%% end Ant1 : Ant[PPMPM]
%%%%% begin Ant1 : Ant[PPPMM]
\Ant(\ah^+,\bh^+\leftarrow a^+,1^-,b^-) &= 
  -\Ant(\bh^+,\ah^+\leftarrow b^-,1^-,a^+)\cr
&= 
\frac{{\spa{1}.{b}}^3}{\spa{a}.{b}\,\spa{a}.{1}\,{\spa{\ah}.{\bh}}^2}
,\cr
%%%%% end Ant1 : Ant[PPPMM]
%%%%% begin Ant1 : Ant[PPMMM]
\Ant(\ah^+,\bh^+\leftarrow a^-,1^-,b^-) &= 
-\frac{{\spb{\ah}.{\bh}}^2}{\spb{a}.{b}\,\spb{a}.{1}\,\spb{1}.{b}}
,\cr
%
%%%%% end Ant1 : Ant[PPMMM]
%%%%% begin Ant1 : Ant[PMPPP]
\Ant(\ah^+,\bh^-\leftarrow a^+,1^+,b^+) &= 0,\cr
%%%%% end Ant1 : Ant[PMPPP]
%%%%% begin Ant1 : Ant[PMMPP]
\Ant(\ah^+,\bh^-\leftarrow a^-,1^+,b^+) &= 
\frac{{\spa{a}.{\bh}}^4}{\spa{a}.{b}\,\spa{a}.{1}\,{\spa{\ah}.{\bh}}^2\,\spa{1}.{b}}
,\cr
%%%%% end Ant1 : Ant[PMMPP]
%%%%% begin Ant1 : Ant[PMPMP]
\Ant(\ah^+,\bh^-\leftarrow a^+,1^-,b^+) &= 
-\frac{{\spa1.{\bh}}^3\,\spb{\ah}.{b}}
       {\spa{a}.{b}\,\spa{a}.{1}\,{\spa{\ah}.{\bh}}^2\,\spb{\ah}.{\bh}}
,\cr
%%%%% end Ant1 : Ant[PMPMP]
%%%%% begin Ant1 : Ant[PMPPM]
\Ant(\ah^+,\bh^-\leftarrow a^+,1^+,b^-) &= 0,\cr
%%%%% end Ant1 : Ant[PMPPM]
%%%%% begin Ant1 : Ant[PMMMP]
\Ant(\ah^+,\bh^-\leftarrow a^-,1^-,b^+) &= 
-\frac{{\spb{\ah}.{b}}^4}{\spb{a}.{b}\,\spb{a}.{1}\,{\spb{\ah}.{\bh}}^2\,\spb{1}.{b}}
,\cr
%%%%% end Ant1 : Ant[PMMMP]
%%%%% begin Ant1 : Ant[PMMPM]
\Ant(\ah^+,\bh^-\leftarrow a^-,1^+,b^-) &= 
-\frac{{\spb{a}.{1}}^3}{\spb{a}.{b}\,{\spb{\ah}.{\bh}}^2\,\spb{1}.{b}}
,\cr
%%%%% end Ant1 : Ant[PMMPM]
%%%%% begin Ant1 : Ant[PMPMM]
\Ant(\ah^+,\bh^-\leftarrow a^+,1^-,b^-) &= 0,\cr
%%%%% end Ant1 : Ant[PMPMM]
%%%%% begin Ant1 : Ant[PMMMM]
\Ant(\ah^+,\bh^-\leftarrow a^-,1^-,b^-) &= 0.\cr
%%%%% end Ant1 : Ant[PMMMM]
}\anoneqn$$
The remaining helicity amplitudes can be obtained via parity or
reflection antisymmetry.  In deriving these forms, I have used
identities such as
$$
{\spa{q}.{\ah}\over\spa{q}.a} = {\spa{b}.{\ah}\over\spa{b}.a}
 + {\spa{q}.b\spa{\ah}.{a}\over \spa{q}.a\spa{b}.a},
\anoneqn$$
and have dropped non-universal terms (terms insufficiently singular in the various
limits).

From eqn.~(\use\DoubleEmissionAntennaDef), 
I obtain the $2\leftarrow 4$ antenna helicity amplitudes,
$$\eqalign{
%%%%% begin Ant2 : Ant[PPPPPP]
\Ant(\ah^+,\bh^+\leftarrow a^+,1^+,2^+,b^+) &= 0\,,\cr
%%%%% end Ant2 : Ant[PPPPPP]
%%%%% begin Ant2 : Ant[PPMPPP]
\Ant(\ah^+,\bh^+\leftarrow a^-,1^+,2^+,b^+) &= 0\,,\cr
%%%%% end Ant2 : Ant[PPMPPP]
%%%%% begin Ant2 : Ant[PPPMPP]
\Ant(\ah^+,\bh^+\leftarrow a^+,1^-,2^+,b^+) &= 0\,,\cr
%%%%% end Ant2 : Ant[PPPMPP]
%%%%% begin Ant2 : Ant[PPPPMP]
\Ant(\ah^+,\bh^+\leftarrow a^+,1^+,2^-,b^+) &= 0\,,\cr
%%%%% end Ant2 : Ant[PPPPMP]
%%%%% begin Ant2 : Ant[PPPPPM]
\Ant(\ah^+,\bh^+\leftarrow a^+,1^+,2^+,b^-) &= 0 \,,\cr
%
%%%%% end Ant2 : Ant[PPPPPM]
%%%%% begin Ant2 : Ant[PPMMPP]
\Ant(\ah^+,\bh^+\leftarrow a^-,1^-,2^+,b^+) &= 
-\frac{{\spa{a}.{1}}^2\spa{\ah}.{1}\spb{\ah}.{b}}{\spa{a}.{b}\,{\spa{\ah}.{\bh}}^2\spa{1}.{2}\spa{2}.{b}\spb{a}.{b}}
\,,\cr
%%%%% end Ant2 : Ant[PPMMPP]
%%%%% begin Ant2 : Ant[PPMPMP]
\Ant(\ah^+,\bh^+\leftarrow a^-,1^+,2^-,b^+) &= 
\frac{{\spa{a}.{2}}^4}{\spa{a}.{b}\spa{a}.{1}\,{\spa{\ah}.{\bh}}^2\spa{1}.{2}\spa{2}.{b}}
\,,\cr
%%%%% end Ant2 : Ant[PPMPMP]
%%%%% begin Ant2 : Ant[PPMPPM]
\Ant(\ah^+,\bh^+\leftarrow a^-,1^+,2^+,b^-) &= 
\frac{{\spa{a}.{b}}^3}{\spa{a}.{1}\,{\spa{\ah}.{\bh}}^2\spa{1}.{2}\spa{2}.{b}}
\,,\cr
%%%%% end Ant2 : Ant[PPMPPM]
%%%%% begin Ant2 : Ant[PPPMMP]
\Ant(\ah^+,\bh^+\leftarrow a^+,1^-,2^-,b^+) &= 
\frac{{\spa{\ah}.{2}}^3\,{\spa{1}.{\bh}}^3}{\spa{a}.{b}\spa{a}.{1}\,{\spa{\ah}.{\bh}}^5\spa{2}.{b}}
\,,\cr
%%%%% end Ant2 : Ant[PPPMMP]
%%%%% begin Ant2 : Ant[PPPMPM]
\Ant(\ah^+,\bh^+\leftarrow a^+,1^-,2^+,b^-) &=
   \Ant(\bh^+,\ah^+\leftarrow b^-,2^+,1^-,a^+)\cr
  &= 
\frac{{\spa{1}.{b}}^4}{\spa{a}.{b}\spa{a}.{1}\,{\spa{\ah}.{\bh}}^2\spa{1}.{2}\spa{2}.{b}}
\,,\cr
%%%%% end Ant2 : Ant[PPPMPM]
\Ant(\ah^+,\bh^+\leftarrow a^+,1^+,2^-,b^-) &=
   \Ant(\bh^+,\ah^+\leftarrow b^-,2^-,1^+,a^+)\cr 
  &= 
%%%%% begin Ant2 : Ant[PPPPMM]
\frac{{\spa{2}.{b}}^3}{\spa{a}.{b}\,\spa{a}.{1}\,{\spa{\ah}.{\bh}}^2\,\spa{1}.{2}}
\,,\cr
%%%%% end Ant2 : Ant[PPPPMM]
%%%%% begin Ant2 : Ant[PPMMMP]
\Ant(\ah^+,\bh^+\leftarrow a^-,1^-,2^-,b^+) &= 
\cr &\hskip -40mm
\frac{{\spa{a}.{1}}^2\,{\spa{a}.{2}}^2\spb{a}.{b}}{\left( s_{ab} + s_{a2} \right) \,s_{2b}\spa{a}.{b}\,{\spa{\ah}.{\bh}}^2} 
+ \frac{\spa{a}.{2}\spa{\ah}.{2}\spb{\ah}.{b}\left( s_{\ah1}+s_{\ah2}\right) }{\spa{a}.{b}\,{\spa{\ah}.{\bh}}^2\spa{2}.{b}\spb{a}.{b}\spb{a}.{1}\spb{1}.{2}} 
\cr &\hskip -40mm
+ \frac{\spa{a}.{\bh}\spa{a}.{2}\spa{\ah}.{1}\spb{\ah}.{b}\spb{b}.{\bh}}{\spa{a}.{b}\,{\spa{\ah}.{\bh}}^2\spa{2}.{b}\spb{a}.{b}\spb{1}.{2}\spb{2}.{b}} 
+ \frac{\spa{a}.{\bh}\spa{1}.{2}\spb{\ah}.{b}\,{\spb{b}.{\bh}}^2}{\spa{\ah}.{\bh}\spa{2}.{b}\spb{a}.{b}\spb{1}.{2}\spb{2}.{b}\,t_{12b}}
\,,\cr
%%%%% end Ant2 : Ant[PPMMMP]
%%%%% begin Ant2 : Ant[PPMMPM]
\Ant(\ah^+,\bh^+\leftarrow a^-,1^-,2^+,b^-) &= 
\cr &\hskip -40mm
-\frac{{\spa{a}.{b}}^2\spa{1}.{b}\spb{a}.{2}\,{\spb{\ah}.{\bh}}^2}{s_{ab}\,{s_{\ah\bh}}^2\spa{2}.{b}\spb{a}.{1}} 
- \frac{{\spa{a}.{b}}^2\spa{1}.{b}\spb{a}.{\bh}\spb{\ah}.{\bh}\spb{\ah}.{2}}{s_{ab}\,s_{\ah\bh}\spa{1}.{2}\spa{2}.{b}\spb{a}.{1}\spb{1}.{2}} 
- \frac{{\spa{a}.{b}}^2\spa{\ah}.{1}\spb{a}.{\bh}\,{\spb{\ah}.{2}}^2}{\spa{a}.{1}\,{\spa{\ah}.{\bh}}^2\spa{2}.{b}\spb{a}.{b}\spb{a}.{1}\spb{\ah}.{\bh}\spb{1}.{2}} 
\cr &\hskip -40mm
+ \frac{{\spa{a}.{1}}^2\spa{\ah}.{b}\spb{a}.{\bh}\,{\spb{\ah}.{2}}^2}{\left( s_{ab} + s_{a2} \right) \,{\spa{\ah}.{\bh}}^2\spa{2}.{b}\spb{a}.{b}\spb{\ah}.{\bh}\spb{2}.{b}} 
+ \frac{\left( s_{1b}-s_{a1} \right) \spa{a}.{b}\spa{1}.{b}\spb{\ah}.{\bh}\spb{\ah}.{2}\spb{2}.{\bh}}{s_{ab}\,s_{\ah\bh}\spa{1}.{2}\spa{2}.{b}\spb{a}.{1}\spb{1}.{2}\spb{2}.{b}} 
\cr &\hskip -40mm
+ \frac{\spa{a}.{1}\spa{\ah}.{b}\spb{a}.{\bh}\,{\spb{\ah}.{2}}^2}{\spa{\ah}.{\bh}\spa{1}.{2}\spb{a}.{b}\spb{a}.{1}\spb{1}.{2}\,t_{a12}} 
+ \frac{\spa{a}.{\bh}\,{\spa{1}.{b}}^2\spb{\ah}.{b}\,{\spb{2}.{\bh}}^2}{\spa{\ah}.{\bh}\spa{1}.{2}\spa{2}.{b}\spb{a}.{b}\spb{1}.{2}\spb{2}.{b}\,t_{12b}}
\,,\cr
%%%%% end Ant2 : Ant[PPMMPM]
}\anoneqn$$

$$\eqalign{
%%%%% begin Ant2 : Ant[PPMPMM]
\Ant(\ah^+,\bh^+\leftarrow a^-,1^+,2^-,b^-) &=
   \Ant(\bh^+,\ah^+\leftarrow b^-,2^-,1^+,a^-) = 
\cr &\hskip -40mm
\frac{\spa{a}.{\bh}\,{\spa{2}.{b}}^2\,\spb{\ah}.{b}\,{\spb{1}.{\bh}}^2}{\left( s_{ab} + s_{1b} \right) \,\spa{a}.{1}\,{\spa{\ah}.{\bh}}^2\,\spb{a}.{b}\,\spb{a}.{1}\,\spb{\ah}.{\bh}} 
- \frac{{\spa{a}.{b}}^2\,\spa{a}.{2}\,{\spb{\ah}.{\bh}}^2\,\spb{1}.{b}}{s_{ab}\,{s_{\ah\bh}}^2\,\spa{a}.{1}\,\spb{2}.{b}} 
- \frac{{\spa{a}.{b}}^2\,\spa{a}.{2}\,\spb{\ah}.{b}\,\spb{\ah}.{\bh}\,\spb{1}.{\bh}}{s_{ab}\,s_{\ah\bh}\,\spa{a}.{1}\,\spa{1}.{2}\,\spb{1}.{2}\,\spb{2}.{b}} 
\cr &\hskip -40mm
+ \frac{\left( s_{a2} \tm s_{2b} \right) \,\spa{a}.{b}\,\spa{a}.{2}\,\spb{\ah}.{\bh}\,\spb{\ah}.{1}\,\spb{1}.{\bh}}{s_{ab}\,s_{\ah\bh}\,\spa{a}.{1}\,\spa{1}.{2}\,\spb{a}.{1}\,\spb{1}.{2}\,\spb{2}.{b}} 
- \frac{{\spa{a}.{b}}^2\,\spa{2}.{\bh}\,\spb{\ah}.{b}\,{\spb{1}.{\bh}}^2}{\spa{a}.{1}\,{\spa{\ah}.{\bh}}^2\,\spa{2}.{b}\,\spb{a}.{b}\,\spb{\ah}.{\bh}\,\spb{1}.{2}\,\spb{2}.{b}} 
\cr &\hskip -40mm
+ \frac{{\spa{a}.{2}}^2\,\spa{\ah}.{b}\,\spb{a}.{\bh}\,{\spb{\ah}.{1}}^2}{\spa{a}.{1}\,\spa{\ah}.{\bh}\,\spa{1}.{2}\,\spb{a}.{b}\,\spb{a}.{1}\,\spb{1}.{2}\,t_{a12}} 
+ \frac{\spa{a}.{\bh}\,\spa{2}.{b}\,\spb{\ah}.{b}\,{\spb{1}.{\bh}}^2}{\spa{\ah}.{\bh}\,\spa{1}.{2}\,\spb{a}.{b}\,\spb{1}.{2}\,\spb{2}.{b}\,t_{12b}}
\,,\cr
%%%%% end Ant2 : Ant[PPMPMM]
\Ant(\ah^+,\bh^+\leftarrow a^+,1^-,2^-,b^-) &=
   \Ant(\bh^+,\ah^+\leftarrow b^-,2^-,1^-,a^+)  = 
\cr &\hskip -40mm
%%%%% begin Ant2 : Ant[PPPMMM]
\frac{{\spa{1}.{b}}^2\,{\spa{2}.{b}}^2\,\spb{a}.{b}}{s_{a1}\,\left( s_{ab} + s_{1b} \right) \,\spa{a}.{b}\,{\spa{\ah}.{\bh}}^2} 
- \frac{\spa{\ah}.{b}\,\spa{1}.{b}\,\spa{2}.{\bh}\,\spb{a}.{\ah}\,\spb{a}.{\bh}}{\spa{a}.{b}\,\spa{a}.{1}\,{\spa{\ah}.{\bh}}^2\,\spb{a}.{b}\,\spb{a}.{1}\,\spb{1}.{2}} 
\cr &\hskip -40mm
- \frac{\spa{1}.{b}\,\spa{1}.{\bh}\,\spb{a}.{\bh}\,\left( \spa{1}.{\bh}\,\spb{1}.{\bh} + \spa{2}.{\bh}\,\spb{2}.{\bh} \right) }{\spa{a}.{b}\,\spa{a}.{1}\,{\spa{\ah}.{\bh}}^2\,\spb{a}.{b}\,\spb{1}.{2}\,\spb{2}.{b}} 
+ \frac{\spa{\ah}.{b}\,\spa{1}.{2}\,{\spb{a}.{\ah}}^2\,\spb{a}.{\bh}}{\spa{a}.{1}\,\spa{\ah}.{\bh}\,\spb{a}.{b}\,\spb{a}.{1}\,\spb{1}.{2}\,t_{a12}}
\,,\cr
%%%%% end Ant2 : Ant[PPPMMM]
%%%%% begin Ant2 : Ant[PPMMMM]
\Ant(\ah^+,\bh^+\leftarrow a^-,1^-,2^-,b^-) &= 
\frac{{\spb{\ah}.{\bh}}^2}{\spb{a}.{b}\,\spb{a}.{1}\,\spb{1}.{2}\,\spb{2}.{b}}
\,,\cr
%%%%% end Ant2 : Ant[PPMMMM]
%
}\anoneqn$$

and
$$\eqalign{
%%%%% begin Ant2 : Ant[PMPPPP]
\Ant(\ah^+,\bh^-\leftarrow a^+,1^+,2^+,b^+) &= 0\,,\cr
%%%%% end Ant2 : Ant[PMPPPP]
%%%%% begin Ant2 : Ant[PMMPPP]
\Ant(\ah^+,\bh^-\leftarrow a^-,1^+,2^+,b^+) &= 
\frac{{\spa{a}.{\bh}}^4}{\spa{a}.{b}\spa{a}.{1}\,{\spa{\ah}.{\bh}}^2\spa{1}.{2}\spa{2}.{b}}
\,,\cr
%%%%% end Ant2 : Ant[PMMPPP]
%%%%% begin Ant2 : Ant[PMPMPP]
\Ant(\ah^+,\bh^-\leftarrow a^+,1^-,2^+,b^+) &= 
\frac{{\spa{1}.{\bh}}^4}{\spa{a}.{b}\spa{a}.{1}\,{\spa{\ah}.{\bh}}^2\spa{1}.{2}\spa{2}.{b}}
\,,\cr
%%%%% end Ant2 : Ant[PMPMPP]
%%%%% begin Ant2 : Ant[PMPPMP]
\Ant(\ah^+,\bh^-\leftarrow a^+,1^+,2^-,b^+) &= 
- \frac{{\spa{2}.{\bh}}^3\spb{\ah}.{b}}{\spa{a}.{b}\spa{a}.{1}\,{\spa{\ah}.{\bh}}^2\spa{1}.{2}\spb{\ah}.{\bh}}
\,,\cr
%%%%% end Ant2 : Ant[PMPPMP]
%%%%% begin Ant2 : Ant[PMPPPM]
\Ant(\ah^+,\bh^-\leftarrow a^+,1^+,2^+,b^-) &= 0\,,\cr
%%%%% end Ant2 : Ant[PMPPPM]
%%%%% begin Ant2 : Ant[PMMMPP]
\Ant(\ah^+,\bh^-\leftarrow a^-,1^-,2^+,b^+) &= 
\cr &\hskip -40mm
\frac{\left( s_{\ah\bh} \tm s_{a1} \tm s_{2b} \right) \spa{a}.{\bh}\spa{1}.{\bh}\spb{\ah}.{b}\spb{\ah}.{2}}{s_{ab} s_{\ah\bh}\spa{1}.{2}\spa{2}.{b}\spb{a}.{1}\spb{1}.{2}} 
+ \frac{\spa{a}.{\bh}\spa{a}.{1}\spb{\ah}.{b} {\spb{\ah}.{2}}^2}{\spa{a}.{b}\spa{1}.{2}\spb{a}.{1}\spb{\ah}.{\bh}\spb{1}.{2}\,t_{a12}} 
%\cr &\hskip -40mm
+ \frac{\spa{a}.{\bh}\,{\spa{1}.{\bh}}^2\spb{\ah}.{b}\spb{2}.{b}}{\spa{\ah}.{\bh}\spa{1}.{2}\spa{2}.{b}\spb{a}.{b}\spb{1}.{2}\,t_{12b}}
\,,\cr
%%%%% end Ant2 : Ant[PMMMPP]
}\anoneqn$$

$$\eqalign{
%%%%% begin Ant2 : Ant[PMMPMP]
\Ant(\ah^+,\bh^-\leftarrow a^-,1^+,2^-,b^+) &= 
\cr &\hskip -40mm
\frac{\spa{a}.{\bh}\,{\spa{2}.{\bh}}^2\spb{\ah}.{b}\,{\spb{1}.{b}}^2}{\left( s_{ab} + s_{1b} \right) \spa{a}.{1}\,{\spa{\ah}.{\bh}}^2\spb{a}.{b}\spb{a}.{1}\spb{\ah}.{\bh}} 
- \frac{\spa{a}.{\bh}\,{\spa{a}.{2}}^2\,{\spb{\ah}.{b}}^3}{\spa{a}.{b}\spa{a}.{1}\spa{\ah}.{\bh}\spa{1}.{2}\spb{a}.{b}\,{\spb{\ah}.{\bh}}^2\spb{2}.{b}} 
\cr &\hskip -40mm
+ \frac{\spa{a}.{\bh}\,{\spa{a}.{2}}^2\spb{\ah}.{b}\,{\spb{\ah}.{1}}^2}{\left( s_{ab} + s_{a2} \right) \spa{a}.{b}\spa{\ah}.{\bh}\spa{2}.{b}\,{\spb{\ah}.{\bh}}^2\spb{2}.{b}} 
- \frac{{\spa{a}.{\bh}}^2\spa{a}.{2}\,{\spb{\ah}.{b}}^2\spb{1}.{b}}{s_{ab}\,s_{\ah\bh}\spa{a}.{1}\spa{1}.{2}\spb{1}.{2}\spb{2}.{b}} 
\cr &\hskip -40mm
+ \frac{\left( s_{12} -s_{a1}- s_{2b} \right) \spa{a}.{\bh}\spa{a}.{2}\spa{2}.{\bh}\spb{\ah}.{b}\spb{\ah}.{1}\spb{1}.{b}}{s_{ab}\,s_{\ah\bh}\spa{a}.{1}\spa{1}.{2}\spa{2}.{b}\spb{a}.{1}\spb{1}.{2}\spb{2}.{b}} 
- \frac{{\spa{a}.{\bh}}^3\spb{\ah}.{b}\,{\spb{1}.{b}}^2}{\spa{a}.{b}\spa{a}.{1}\,{\spa{\ah}.{\bh}}^2\spb{a}.{b}\spb{\ah}.{\bh}\spb{1}.{2}\spb{2}.{b}} 
\cr &\hskip -40mm
+ \frac{\spa{a}.{\bh}\,{\spa{a}.{2}}^2\spb{\ah}.{b}\,{\spb{\ah}.{1}}^2}{\spa{a}.{b}\spa{a}.{1}\spa{1}.{2}\spb{a}.{1}\spb{\ah}.{\bh}\spb{1}.{2}\,t_{a12}} 
+ \frac{\spa{a}.{\bh}\,{\spa{2}.{\bh}}^2\spb{\ah}.{b}\,{\spb{1}.{b}}^2}{\spa{\ah}.{\bh}\spa{1}.{2}\spa{2}.{b}\spb{a}.{b}\spb{1}.{2}\spb{2}.{b}\,t_{12b}}
\,,\cr
%
%%%%% end Ant2 : Ant[PMMPMP]
%%%%% begin Ant2 : Ant[PMMPPM]
\Ant(\ah^+,\bh^-\leftarrow a^-,1^+,2^+,b^-) &= 
\cr &\hskip -40mm
\frac{\spa{a}.{b}\,{\spa{a}.{\bh}}^2\,{\spb{a}.{2}}^2\,{\spb{\ah}.{1}}^2}{\left( s_{ab} + s_{a2} \right) \,s_{2b}\,{\spa{\ah}.{\bh}}^2\spb{a}.{b}\,{\spb{\ah}.{\bh}}^2} 
- \frac{{\spa{a}.{\bh}}^2\spa{b}.{\bh}\spb{a}.{2}\spb{\ah}.{1}}{{\spa{\ah}.{\bh}}^2\spa{1}.{2}\spa{2}.{b}\spb{a}.{b}\spb{\ah}.{\bh}\spb{2}.{b}} 
- \frac{{\spa{a}.{\bh}}^2\spb{a}.{2}
%(-1)\left( \spa{a}.{b}\spb{\ah}.{b} + \spa{a}.{\bh}\spb{\ah}.{\bh} \right) 
\sand{a}.{b\tp \bh}.{\ah}
\spb{\ah}.{2}}{\spa{a}.{1}\,{\spa{\ah}.{\bh}}^2\spa{1}.{2}\spb{a}.{b}\,{\spb{\ah}.{\bh}}^2\spb{2}.{b}} 
\cr &\hskip -40mm
+ \frac{\spa{a}.{\bh}\,{\spa{b}.{\bh}}^2\spb{\ah}.{b}\spb{1}.{2}}{\spa{\ah}.{\bh}\spa{1}.{2}\spa{2}.{b}\spb{a}.{b}\spb{2}.{b}\,t_{12b}}
\,,\cr
%%%%% end Ant2 : Ant[PMMPPM]
%%%%% begin Ant2 : Ant[PMPMMP]
\Ant(\ah^+,\bh^-\leftarrow a^+,1^-,2^-,b^+) &= 
\cr &\hskip -40mm
\frac{{\spa{1}.{b}}^2\,{\spa{2}.{\bh}}^2\spb{a}.{b}\,{\spb{\ah}.{b}}^2}{s_{a1}\left( s_{ab} + s_{1b} \right) \spa{a}.{b}\,{\spa{\ah}.{\bh}}^2\,{\spb{\ah}.{\bh}}^2} 
+ \frac{\spa{1}.{b}\spa{2}.{\bh}\spb{a}.{\ah}\,{\spb{\ah}.{b}}^2}{\spa{a}.{b}\spa{a}.{1}\spa{\ah}.{\bh}\spb{a}.{1}\,{\spb{\ah}.{\bh}}^2\spb{1}.{2}} 
- \frac{\spa{1}.{b}\spa{1}.{\bh}\,{\spb{\ah}.{b}}^2
% (-1)\left( \spa{a}.{\bh}\spb{a}.{b} + \spa{\ah}.{\bh}\spb{\ah}.{b} \right) 
\sand{\bh}.{a+\ah}.b
}
{\spa{a}.{b}\spa{a}.{1}\,{\spa{\ah}.{\bh}}^2\,{\spb{\ah}.{\bh}}^2\spb{1}.{2}\spb{2}.{b}} 
\cr &\hskip -40mm
+ \frac{\spa{a}.{\bh}\spa{1}.{2}\,{\spb{a}.{\ah}}^2\spb{\ah}.{b}}{\spa{a}.{b}\spa{a}.{1}\spb{a}.{1}\spb{\ah}.{\bh}\spb{1}.{2}\,t_{a12}}
\,,\cr
%%%%% end Ant2 : Ant[PMPMMP]
}\anoneqn$$

$$\eqalign{
\Ant(\ah^+,\bh^-\leftarrow a^+,1^-,2^+,b^-) &= 
%%%%% begin Ant2 : Ant[PMPMPM]
- \frac{\spa{1}.{b}\,{\spa{1}.{\bh}}^2\spb{a}.{2}\,{\spb{\ah}.{2}}^2}{\spa{a}.{b}\spa{a}.{1}\,{\spa{\ah}.{\bh}}^2\spb{a}.{b}\,{\spb{\ah}.{\bh}}^2\spb{2}.{b}}
\,,\cr
%%%%% end Ant2 : Ant[PMPMPM]
%%%%% begin Ant2 : Ant[PMPPMM]
\Ant(\ah^+,\bh^-\leftarrow a^+,1^+,2^-,b^-) &= 0\,,\cr
%%%%% end Ant2 : Ant[PMPPMM]
%%%%% begin Ant2 : Ant[PMMMMP]
\Ant(\ah^+,\bh^-\leftarrow a^-,1^-,2^-,b^+) &= 
\frac{{\spb{\ah}.{b}}^4}{\spb{a}.{b}\spb{a}.{1}\,{\spb{\ah}.{\bh}}^2\spb{1}.{2}\spb{2}.{b}}
\,,\cr
%%%%% end Ant2 : Ant[PMMMMP]
%%%%% begin Ant2 : Ant[PMMMPM]
\Ant(\ah^+,\bh^-\leftarrow a^-,1^-,2^+,b^-) &= 
\frac{{\spb{\ah}.{2}}^4}{\spb{a}.{b}\spb{a}.{1}\,{\spb{\ah}.{\bh}}^2\spb{1}.{2}\spb{2}.{b}}
\,,\cr
%%%%% end Ant2 : Ant[PMMMPM]
%%%%% begin Ant2 : Ant[PMMPMM]
\Ant(\ah^+,\bh^-\leftarrow a^-,1^+,2^-,b^-) &= 
- \frac{\spa{a}.{\bh}\,{\spb{\ah}.{1}}^3}{\spa{\ah}.{\bh}\spb{a}.{b}\,{\spb{\ah}.{\bh}}^2\spb{1}.{2}\spb{2}.{b}} 
\,,\cr
%%%%% end Ant2 : Ant[PMMPMM]
%%%%% begin Ant2 : Ant[PMPMMM]
\Ant(\ah^+,\bh^-\leftarrow a^+,1^-,2^-,b^-) &= 0\,,\cr
%%%%% end Ant2 : Ant[PMPMMM]
%%%%% begin Ant2 : Ant[PMMMMM]
\Ant(\ah^+,\bh^-\leftarrow a^-,1^-,2^-,b^-) &= 0\,,\cr
%%%%% end Ant2 : Ant[PMMMMM]
}\anoneqn$$

The reader may verify that these functions reproduce the appropriate triple-collinear,
products of collinear splitting amplitudes, mixed soft-collinear, or double-soft
amplitudes in the various limits listed in section~\use\DoubleEmissionSection.

\section{The Antenna Amplitudes Squared in Dimensional Regularization}
\tagsection\DimensionalRegularizationSection
\vskip 10pt

If we square the expressions for the single-emission antenna functions
given in the previous section, and sum over the helicities of legs $a,1,b$ while
averaging over the helicities of $\ah$ and $\bh$, 
we obtain the following expression in four dimensions,
$$
%%%%% begin Ant1sq : Ant1sq
|\Ant_1|^2 = 2 {\left( K^2 (s_{a1}+s_{1b}) + s_{ab}^2\right)^2
           \over s_{a1} s_{1b} s_{ab} (K^2)^2}
%%%%% end Ant1sq : Ant1sq
\eqn\SingleEmissionSquared$$
The same expression turns out to hold away from $D=4$, 
in the conventional dimensional regularization (CDR)
scheme~[\use\CollinsBook].
(In the CDR scheme, there are $D-2$ gluon helicities in contrast to the
FDH scheme~[\use\FDH,\use\TwoloopRegulator] in which the number of
gluon helicities is kept fixed at 2.)

For the double-emission antenna function, I find for the helicity-summed and -averaged
square in four dimensions,
$$
%%%%% begin Ant2sq : Ant2sq
|\Ant_2|^2 = {1\over4} \LB A_1(\ah,a,1,2,b,\bh)
             +A_2(\ah,a,1,2,b,\bh)+A_2(\bh,b,2,1,a,\ah)\RB\,,
%%%%% end Ant2sq : Ant2sq
\eqn\DoubleEmissionSquared
$$
where
$$\eqalign{
%%%%% begin A1 : A1[ah_,a_,s1_,s2_,b_,bh_]
A_1&(\ah,a,1,2,b,\bh) = \cr
&\frac{16\,s_{a1}\,s_{2b}}{s_{12}^2\,t_{a12}\,t_{12b}}
-\frac{16 s_{\ah\bh}}{s_{12}\,t_{a12}\,t_{12b}}
\biggl(2 - \frac{\left( s_{a1} + s_{2b} \right)}{{s_{\ah\bh}}}
+\frac{2\,\left( s_{a1} + s_{2b} \right)^2 }{{s_{\ah\bh}}^2}\biggr)
-\frac{32}{t_{a12}\,t_{12b}}
\cr &+\frac{1}{s_{a1}\,s_{2b}}
\biggl(42 
+ \frac{8\,s_{ab}^2}{{s_{\ah\bh}}^2} 
- \frac{36\,s_{ab}}{s_{\ah\bh}} 
- \frac{70\,s_{\ah\bh}}{s_{ab}} 
+ \frac{46\,t_{a12}}{s_{ab}} 
+ \frac{32\,s_{ab}\,t_{a12}}{{s_{\ah\bh}}^2} 
- \frac{16\,t_{a12}}{s_{\ah\bh}} 
+ \frac{16\,t_{a12}^2}{{s_{\ah\bh}}^2} 
- \frac{32\,t_{a12}^2}{s_{ab}\,s_{\ah\bh}} 
\cr &\hskip 15mm
+ \frac{16\,t_{a12}^3}{s_{ab}\,{s_{\ah\bh}}^2} 
+ \frac{46\,t_{12b}}{s_{ab}} 
+ \frac{32\,s_{ab}\,t_{12b}}{{s_{\ah\bh}}^2} 
- \frac{16\,t_{12b}}{s_{\ah\bh}} 
+ \frac{16\,t_{12b}^2}{{s_{\ah\bh}}^2} 
- \frac{32\,t_{12b}^2}{s_{ab}\,s_{\ah\bh}} 
+ \frac{16\,t_{12b}^3}{s_{ab}\,{s_{\ah\bh}}^2}\biggr)
\cr &
+\frac{s_{\ah\bh}}{s_{a1}\,s_{12}\,s_{2b}}
\biggl(\frac{28\,s_{ab}}{s_{\ah\bh}}
+ \frac{8\,s_{ab}^3}{{s_{\ah\bh}}^3} 
- \frac{20\,s_{ab}^2}{{s_{\ah\bh}}^2} - 16 + \frac{8\,{s_{\ah\bh}}}{s_{ab}}
\biggr)
+\frac{2\,{s_{\ah\bh}}^2}{s_{a1}\,s_{2b}\,t_{a12}\,t_{12b}}
\biggl(4\, + \frac{2\,s_{12}}{s_{\ah\bh}} + \frac{s_{12}^2}{s_{ab}s_{\ah\bh}}\biggr)
\cr &
+\frac{1}{s_{12}^2}
\biggl(
-34 - \frac{10\,s_{ab}^2}{{s_{\ah\bh}}^2} - \frac{14\,s_{ab}\,s_{a1}}{{s_{\ah\bh}}^2} - \frac{8\,{s_{a1}}^2}{{s_{\ah\bh}}^2} + \frac{36\,s_{ab}}{s_{\ah\bh}} + \frac{38\,s_{a1}}{s_{\ah\bh}} - \frac{14\,s_{ab}\,s_{2b}}{{s_{\ah\bh}}^2} + \frac{38\,s_{2b}}{s_{\ah\bh}} - \frac{8\,s_{2b}^2}{{s_{\ah\bh}}^2}\biggr)\,,
%%%%% end A1 : A1[ah_,a_,s1_,s2_,b_,bh_]
}\anoneqn$$
and
$$\hskip -5mm\eqalign{
%%%%% begin A2 : A2[ah_,a_,s1_,s2_,b_,bh_]
A_2&(\ah,a,1,2,b,\bh) = \cr
&\frac{4\,{s_{\ah\bh}}^4}{s_{a1}^2\,s_{12}^2\,t_{a12}^2}
\biggl( \frac{s_{a2}^2\,{s_{\ah1}}^2}{{s_{\ah\bh}}^4} 
       + \frac{s_{a1}^2\,{s_{\ah2}}^2}{{s_{\ah\bh}}^4} 
       + \frac{s_{a\ah}^2\,s_{12}^2}{{s_{\ah\bh}}^4} \biggr)
-\frac{8\,s_{\ah\bh}}{s_{a1}\,t_{a12}\,t_{12b}}
\biggl(
\frac{\left( s_{\ah\bh} - s_{2b} \right)^3}{{s_{\ah\bh}}^3} 
   + \frac{{s_{\ah\bh}}^2 - s_{2b}^2}{{s_{\ah\bh}}^2}
\biggr)
\cr &
+\frac{4}{s_{a1}^2}
\biggl( -9 - \frac{2\,s_{a\bh}^2}{{s_{\ah\bh}}^2} - \frac{10\,s_{a\bh}}{s_{\ah\bh}} - \frac{4\,s_{1b}\,s_{1\bh}}{s_{\ah\bh}\,\left( s_{ab} + s_{1b} \right) } - \frac{4\,s_{ab}\,s_{2b}}{{s_{\ah\bh}}^2} + \frac{10\,s_{2b}}{s_{\ah\bh}} - \frac{3\,s_{2b}^2}{{s_{\ah\bh}}^2} + \frac{2\,s_{1b}^2\,s_{2b}^2}{{s_{\ah\bh}}^2\,{\left( s_{ab} + s_{1b} \right) }^2} 
\cr & \hskip 12mm
\vphantom{  \frac{2\,s_{1b}^2\,s_{2b}^2}{{s_{\ah\bh}}^2\,{\left( s_{ab} + s_{1b} \right) }^2} }
+ \frac{2\,s_{a\bh}\,s_{1\bh}}{{s_{\ah\bh}}^2}
+ \frac{2\,s_{a\bh}\,s_{1\bh}\,t_{\ah a1}}{s_{2b}\,t_{a12}\,s_{\ah\bh}}
\biggr)
-\frac{8\,\left( s_{\ah\bh} - s_{1\bh} \right) \,s_{1\bh}\,t_{a12}}{s_{a1}^2\,{s_{\ah\bh}}^2\,s_{12}}
+\frac{s_{a1}\,\left( 11\,t_{12b} -16\,s_{2b} \right) }{s_{\ah\bh}\,s_{12}^2\,t_{a12}}
-\frac{4\,s_{ab}\,s_{\ah1}}{s_{a1}\,s_{12}\,s_{2b}\,t_{a12}}
\cr &
+\frac{2}{s_{a1}\,s_{12}}
\biggl( -3 + \frac{s_{ab}^2}{{s_{\ah\bh}}^2} + \frac{8\,s_{ab}}{s_{\ah\bh}} - \frac{8\,s_{\ah\bh}}{s_{ab}} + \frac{12\,s_{2b}}{s_{ab}} + \frac{9\,s_{ab}\,s_{2b}}{{s_{\ah\bh}}^2} - \frac{5\,s_{2b}}{s_{\ah\bh}} + \frac{12\,s_{2b}^2}{{s_{\ah\bh}}^2} - \frac{8\,s_{2b}^2}{s_{ab}\,s_{\ah\bh}} + \frac{4\,s_{2b}^3}{s_{ab}\,{s_{\ah\bh}}^2} 
\cr & \hskip 17mm
+ \frac{4\,\left( {s_{\ah\bh}}^2 + s_{1b}^2 \right) \,t_{12b}}{{s_{\ah\bh}}^2\,\left( s_{ab} + s_{1b} \right) } \biggr)
-\frac{4\,s_{1\bh}^2\,t_{a12}^2}{s_{a1}^2\,{s_{\ah\bh}}^2\,s_{12}^2}
+\frac{4\,s_{\ah\bh}}{s_{a1}\,s_{2b}\,t_{a12}}
\biggl(\frac{s_{ab}}{s_{\ah\bh}} - \frac{2\,s_{ab}^2}{{s_{\ah\bh}}^2} 
      - 4 + \frac{2\,s_{\ah\bh}}{s_{ab}}\biggr)
\cr &
+\frac{s_{\ah\bh}}{s_{a1}\,s_{12}^2}
\biggl( \frac{-8\,s_{2b}\,t_{a12}}{{s_{\ah\bh}}^2} - \frac{8\,s_{1b}\,s_{2b}\,t_{a12}}{{s_{\ah\bh}}^3} - \frac{8\,s_{ab}\,t_{a12}\,t_{12b}}{{s_{\ah\bh}}^3} + \frac{16\,t_{a12}\,t_{12b}}{{s_{\ah\bh}}^2} - \frac{8\,s_{2b}\,t_{a12}\,t_{12b}}{{s_{\ah\bh}}^3} + \frac{3\,s_{2b}^2\,t_{a12}\,t_{12b}}{{s_{\ah\bh}}^4} 
\cr & \hskip 17mm
- \frac{6\,s_{2b}\,t_{a12}\,t_{12b}^2}{{s_{\ah\bh}}^4} + \frac{3\,t_{a12}\,t_{12b}^3}{{s_{\ah\bh}}^4} \biggr)
-\frac{8\,s_{ab}\,t_{\ah a1}\,s_{1b}}{s_{a1}^2\,s_{\ah\bh}\,s_{2b}\,t_{a12}}
-\frac{3\,{s_{\ah\bh}}^2}{s_{a1}\,s_{12}^2\,t_{a12}}
\biggl(\frac{s_{\ah1}^2}{{s_{\ah\bh}}^2} + \frac{s_{a\bh}\,s_{\ah1}^2}{{s_{\ah\bh}}^3}
\biggr)
\cr &
+\frac{2}{s_{12}\,t_{a12}}
\biggl( 16 + \frac{8\,s_{2b}}{s_{ab}} + \frac{4\,s_{2b}}{s_{\ah\bh}} - \frac{4\,s_{2b}^2}{s_{ab}\,s_{\ah\bh}} + \frac{4\,s_{2b}^2}{s_{ab}\,\left( s_{ab} + s_{1b} \right) } - \frac{8\,\left( s_{1b} + s_{2b} \right) }{s_{ab} + s_{1b}} - \frac{4\,s_{1b}\,{\left( s_{1b} + s_{2b} \right) }^2}{{s_{\ah\bh}}^2\,\left( s_{ab} + s_{1b} \right) } 
\cr & \hskip 17mm
- \frac{4\,t_{12b}}{s_{ab}} + \frac{3\,t_{12b}}{s_{\ah\bh}} + \frac{4\,t_{12b}^2}{{s_{\ah\bh}}^2} \biggr)
+\frac{8\,s_{\ah\bh}}{s_{a1}^2\,t_{a12}}
\biggl(\frac{-2\,s_{12}\,s_{1b}}{(s_{ab} + s_{1b})\,s_{\ah\bh}} 
+ \frac{s_{12}\,\left( s_{\ah\bh} - 2 s_{1\bh} + s_{2b} \right) }{{s_{\ah\bh}}^2}
 \biggr)
\cr &
+\frac{4\,s_{\ah\bh}}{s_{12}\,s_{2b}\,t_{a12}}
\biggl( \frac{-2\,s_{a1}^3}{{s_{\ah\bh}}^3} + \frac{4\,s_{a1}^2}{{s_{\ah\bh}}^2} + \frac{2\,s_{a1}\,{s_{a2}}^2}{\left( s_{ab} + s_{a2} \right) \,{s_{\ah\bh}}^2} + \frac{s_{ab}}{s_{\ah\bh}} - \frac{2\,s_{a1}}{s_{\ah\bh}} + \frac{2\,s_{ab}\,t_{12b}^2}{{s_{\ah\bh}}^3} + \frac{2\,t_{12b}^2}{s_{ab}\,s_{\ah\bh}} \biggr)
\cr &
-\frac{1}{s_{a1}\,t_{a12}}
\biggl(
\frac{16\,s_{\ah\bh}}{s_{ab}} 
+ \frac{16 s_{1b}\left( s_{\ah\bh} \tp 2s_{1b} \right)}{s_{\ah\bh}\,\left(s_{ab}\tp s_{1b} \right) } 
- \frac{24\,s_{2b}}{s_{ab}} + \frac{32\,s_{2b}}{s_{\ah\bh}} 
+ \frac{32\,s_{2b}}{s_{ab} + s_{1b}} + \frac{16\,s_{2b}^2}{s_{ab}\,s_{\ah\bh}} 
- \frac{8\,s_{2b}^3}{s_{ab}\,{s_{\ah\bh}}^2} - \frac{59\,t_{12b}}{s_{\ah\bh}}
\biggr)\,.
%%%%% end A2 : A2[ah_,a_,s1_,s2_,b_,bh_]
}\anoneqn$$

Outside of four dimensions, in the CDR scheme there are 
the following additional contributions,
$$
%%%%% begin Ant2sqE : Ant2sqE
\delta|\Ant_2|^2 = -2\e \LB E_1(\ah,a,1,2,b,\bh)
             +E_2(\ah,a,1,2,b,\bh)+E_2(\bh,b,2,1,a,\ah)\RB\,,
%%%%% end Ant2sqE : Ant2sqE
\eqn\DoubleEmissionSquaredEps
$$
where
$$\eqalign{
%%%%% begin E1 : E1[ah_,a_,s1_,s2_,b_,bh_]
E_1&(\ah,a,1,2,b,\bh) = 
\frac{2\,\left( s_{a1} \tp s_{12} \right) \,\left( s_{12} \tp s_{2b} \right) }{s_{12}^2\,t_{a12}\,t_{12b}}
+\frac{1}{s_{12}^2}\,,\cr
%%%%% end E1 : E1[ah_,a_,s1_,s2_,b_,bh_]
%% 
%%%%% begin E2 : E2[ah_,a_,s1_,s2_,b_,bh_]
E_2&(\ah,a,1,2,b,\bh) = \cr
&\frac{{\left( s_{a1}^2 \tp s_{a1}\,s_{12} \tp s_{12}^2 \right) }^2}{s_{a1}^2\,s_{12}^2\,t_{a12}^2}
-\frac{2\,\left( s_{a1} \tp s_{12} \right) \,\left( s_{a1}^2 \tp s_{12}^2 \right) }{s_{a1}^2\,s_{12}^2\,t_{a12}}
+\frac{2\,s_{ab}\,\left( s_{a1} \tp s_{12} \right) }{s_{a1}^2\,\left( s_{ab} \tp s_{1b} \right) \,t_{a12}}
%\cr &
+\frac{s_{1b}^2}{s_{a1}^2 {\left( s_{ab} \tp s_{1b} \right) }^2} \,.
%%%%% end E2 : E2[ah_,a_,s1_,s2_,b_,bh_]
}\anoneqn$$

\section{Summary}
\vskip 10pt

This paper provides a general formula~(\use\MultipleEmissionAntennaDef)
 for a function, an {\it antenna amplitude\/}, summarizing all multiply-collinear,
mixed collinear-soft, and multiple-soft singularities in the emission of color-connected
gluons in an arbitrary tree-level amplitude in gauge theory.  Products of such functions
over separate color-connected sets
will then describe all singular limits of tree-level amplitudes.  I have also evaluated
all the antenna helicity amplitudes for single- and double-gluon emission, and given
expressions for the helicity-summed and -averaged forms, both in and away from four 
dimensions.  The latter functions, integrated appropriately over the phase space
for singular emission, would provide functions that will cancel off 
virtual singularities in loop amplitudes.
Integrals over singular phase space of 
the single-emission functions~(\use\SingleEmissionSquared)
would combine the soft and collinear integrals of Giele and Glover~[\use\GieleGlover],
and be equivalent to the integrated dipole functions of Catani and 
Seymour~[\use\CataniSeymour].  Integrals of the double-emission 
functions~(\use\DoubleEmissionSquared,\use\DoubleEmissionSquaredEps) would provide
the corresponding ingredient in NNLO calculations.  The final-state integration
would be sufficient for the simplest NNLO process, $e^+ e^- \rightarrow 3{\rm\ jets}$.
Although this paper presented results only for pure-gluon processes, the master
formul\ae~(\use\AntennaDef,\use\DoubleEmissionAntennaDef,%
\use\MultipleEmissionAntennaDef) carry over to mixed quark-gluon
antenna amplitudes simply with the replacement of gluon currents
by the quark equivalents as appropriate, and the reconstruction 
functions~(\use\ReconstructionA,\use\rChoice,\use\GeneralReconstructionFunction,%
\use\DoubleRChoice) carry over unchanged.  Supersymmetry 
identities~[\use\SusyIdentities,\use\MP] can of course
also be used to relate the quark and gluon antenna amplitudes.

The construction of the antenna amplitude relies on the reconstruction functions,
which combine sets of momenta into a pair of massless momenta, with smooth limits
as additional gluons' momenta become soft or collinear.  These reconstruction functions
(or more precisely their inverses) can be used to generate near-singular phase-space
configurations numerically in an efficient manner, and thus should prove useful in the 
writing of numerical programs for higher-order calculations as well.  
Ref.~[\use\NumericalPhaseSpace] gives an example
of a similar remapping.

I thank Z.~Bern and L.~Dixon for helpful discussions and comments, and the
Department of Physics at UCLA, where the some of the calculations herein were 
done, for its hospitality.

\appendix{Triply-Collinear Splitting Amplitudes}
\tagappendix\TripleCollinearAppendix

In this appendix, I evaluate the light-cone current $J^\lc(1,2,3;P)$ in order
to obtain the triple-collinear splitting amplitudes.  This requires the evaluation
of $\pol_\mu^{(+)}(P;q)$ where $P$ is not null.  We can evaluate this
expression by multiplying
by $\pol_\mu^{(-)}(P;q)\cdot q'$ and dividing by 
$\pol_\mu^{(-)}(k_i;q)\cdot q'$, where $q'$ is another null reference
four-vector ($q\cdot q' \neq 0$, $q'\cdot k_i\neq 0$), and $k_i$ is one
of the momenta becoming collinear to $P$ in the limit.  The collinear
form of the Schouten identity,
$$
 \sqrt{z_3} \spa1.2  + \sqrt{z_1}\spa2.3 - \sqrt{z_2}\spa1.3 = 0\,,
\eqn\Schouten$$
where $z_i$ are the momentum fractions of the $k_i$ ($z_1+z_2+z_3=1$),
and its partner with the bracket product are useful in simplifying the expressions.

Recall that the ordinary collinear splitting amplitudes~[\use\MP] are,
$$\eqalign{
%%%%% begin Ctree : split[PPP]
 \Ctree_{+}(1^+,2^+;z) &= 0\,,
%%%%% end Ctree : split[PPP]
\cr
%%%%% begin Ctree : split[PMP]
 \Ctree_{+}(1^-,2^+;z) &= {z^2\over\sqrt{z (1-z)} \spa1.2}\,,
%%%%% end Ctree : split[PMP]
\cr
%%%%% begin Ctree : split[PPM]
 \Ctree_{+}(1^+,2^-;z) &= -\Ctree_{+}(2^-,1^+;1-z)
\cr &= {(1-z)^2\over\sqrt{z (1-z)} \spa1.2}\,,
%%%%% end Ctree : split[PPM]
\cr
%%%%% begin Ctree : split[PMM]
 \Ctree_{+}(1^-,2^-;z) &= -{1\over\sqrt{z (1-z)} \spb1.2}\,,
%%%%% end Ctree : split[PMM]
}\anoneqn$$

The triple-collinear splitting amplitudes are,
$$\eqalign{
%%%%% begin Ctree : split[PPPP]
 \Ctree_{+}(1^+,2^+,3^+;z_1,z_2) &= 0\,,
%%%%% end Ctree : split[PPPP]
\cr
%%%%% begin Ctree : split[PMPP]
 \Ctree_{+}(1^-,2^+,3^+;z_1,z_2) &=  {z_1^2\over \sqrt{z_1 z_3} \spa1.2 \spa2.3}
%%%%% end Ctree : split[PMPP]
\cr
%%%%% begin Ctree : split[PPMP]
 \Ctree_{+}(1^+,2^-,3^+;z_1,z_2) &= {z_2^2\over \sqrt{z_1 z_3} \spa1.2 \spa2.3}
%%%%% end Ctree : split[PPMP]
\cr
%%%%% begin Ctree : split[PPPM]
 \Ctree_{+}(1^+,2^+,3^-;z_1,z_2) &=  \Ctree_{+}(3^-,2^+,1^+;z_3,z_2)
\cr &= {z_3^2\over \sqrt{z_1 z_3} \spa1.2 \spa2.3}
%%%%% end Ctree : split[PPPM]
\cr
%%%%% begin Ctree : split[PMMP]
 \Ctree_{+}(1^-,2^-,3^+;z_1,z_2) &= 
- {\spa1.2 (\sqrt{z_1}\spb1.3 + \sqrt{z_2}\spb2.3)^2
   \over\spa2.3\spb1.2\spb2.3 t_{123}}
%\cr &\hskip 8mm
- {z_1 z_3\over (1\tm z_1)\spa2.3\spb2.3} 
-{(1\tm z_3)^2\over \sqrt{z_1 z_3}\spa2.3\spb1.2}
%%%%% end Ctree : split[PMMP]
\cr
%%%%% begin Ctree : split1[PMPM]
 \Ctree_{+}(1^-,2^+,3^-;z_1,z_2) &= 
- {\spa1.3^2 (\sqrt{z_1} \spb1.2 - \sqrt{z_3}\spb2.3)^2
   \over \spa1.2 \spa2.3 \spb1.2 \spb2.3 t_{123}}
 - {\sqrt{z_2}\spa1.3 (z_1^{3/2}\spb1.2 + z_3^{3/2}\spb2.3)
    \over \sqrt{z_1 z_3}\spa1.2\spa2.3\spb1.2\spb2.3}
\cr &\hskip 8mm
+ {z_2 z_3\over (1\tm z_3)\spa1.2\spb1.2} 
+ {z_1 z_2\over (1\tm z_1) \spa2.3 \spb2.3}
%%%%% end Ctree : split1[PMPM]
\cr
&= 
%%%%% begin Ctree : split[PMPM]
- {\spa1.3^2 (\sqrt{z_1} \spb1.2 - \sqrt{z_3}\spb2.3)^2
   \over \spa1.2 \spa2.3 \spb1.2 \spb2.3 t_{123}}
 - {z_3^2 \over \sqrt{z_1 z_3}\spa2.3\spb1.2}
 - {z_1^2 \over \sqrt{z_1 z_3}\spa1.2\spb2.3}
\cr &\hskip 8mm
- {z_1 z_3\over (1\tm z_3)\spa1.2\spb1.2} 
- {z_1 z_3\over (1\tm z_1) \spa2.3 \spb2.3}
%%%%% end Ctree : split[PMPM]
\cr
%
%%%%% begin Ctree : split[PPMM]
 \Ctree_{+}(1^+,2^-,3^-;z_1,z_2) &= \Ctree_{+}(3^-,2^-,1^+;z_3,z_2)
\cr &= -{\spa2.3 (\sqrt{z_2} \spb1.2 + \sqrt{z_3} \spb1.3)^2
         \over \spa1.2\spb1.2\spb2.3 t_{123}}
%\cr &\hskip 8mm
- {z_1 z_3\over (1\tm z_3) \spa1.2 \spb1.2}
- {(1\tm z_1)^2\over \sqrt{z_1 z_3} \spa1.2 \spb2.3}\,,
%%%%% end Ctree : split[PPMM]
\cr
%%%%% begin Ctree : split[PMMM]
 \Ctree_{+}(1^-,2^-,3^-;z_1,z_2) &= {1\over\sqrt{z_1 z_3}\spb1.2\spb2.3}\cr
%%%%% end Ctree : split[PMMM]
}\eqn\TripleCollinearSplittingAmplitude$$
The relations between the splitting amplitudes listed above follow from
the reflection properties of the Berends--Giele current.
The remaining amplitudes can be obtained by parity.  The photon decoupling
identity leads to the following relations,
$$\eqalign{
 \Ctree_{+}(1^{\sigma_1},2^{\sigma_2},3^{\sigma_3};z_1,z_2) 
+ \Ctree_{+}(2^{\sigma_2},1^{\sigma_1},3^{\sigma_3};z_2,z_1) 
+ \Ctree_{+}(2^{\sigma_2},3^{\sigma_3},1^{\sigma_1};z_2,z_3) &= 0\,,\cr
}\anoneqn$$
whose validity for the expressions in 
eqn.~(\use\TripleCollinearSplittingAmplitude) is left to the reader.

In the strongly-ordered limit $s_{12}\ll s_{13}, s_{23}, t_{123}$, 
we expect the factorization
$$
\Ctree_{\sigma_P}(1^{\sigma_1},2^{\sigma_2},3^{\sigma_3};z_1,z_2) 
\longrightarrow \sum_{\rho=\pm}
\Ctree_{\sigma_P}((1\tp2)^{\rho},3^{\sigma_3};z_1+z_2) 
\Ctree_{-\rho}(1^{\sigma_1},2^{\sigma_2};\textstyle{z_1\over z_1+z_2})
+\cdots\,,
\anoneqn$$
where the omitted terms are less singular.

For some of the splitting amplitudes, this limit is straightforward 
($R=k_1+k_2$, $z_R=z_1+z_2$):
$$\eqalign{
 \Ctree_{+}(1^+,2^-,3^+;z_1,z_2) &\rightarrow
 {(1-\textstyle{z_1\over z_1+z_2})^2\over 
    \sqrt{\textstyle{z_1\over z_1+z_2}(1-{z_1\over z_1+z_2})}\spa1.2} 
    {z_R^2\over\sqrt{z_R (1-z_R) \vphantom{ \textstyle{z_1\over z_1+z_2} }} \spa{R}.3}
\cr& = \Ctree_{+}(R^-,3^+;z_R)\Ctree_{+}(1^+,2^-;\textstyle{z_1\over z_1+z_2})\,.
}\anoneqn$$
For others, it is more delicate, because we must use the Schouten 
identities~(\use\Schouten)
to reduce the strength of the leading $1/(\spa1.2\spb1.2)$ pole before applying
limiting relations for $k_1\parallel k_2$,
$$\eqalign{
 \Ctree_{+}&(1^-,2^+,3^-;z_1,z_2) \longrightarrow
- { z_3\spa1.3^2\spb2.3
   \over \spa1.2 \spa2.3 \spb1.2 s_{R3}}
 - {z_3\sqrt{z_2}\spa1.3
    \over \sqrt{z_1}\spa1.2\spa2.3\spb1.2}
+ {z_2 z_3\over (1\tm z_3)\spa1.2\spb1.2} 
\cr &\hphantom{ (1^-,2^+,3^-;z_1,z_2) \longrightarrow }\hskip 5mm
+ {2\sqrt{z_1 z_3} \spa1.3^2 
   \over \spa1.2 \spa2.3 s_{R3}}
 - {z_1\sqrt{z_2}\spa1.3  
    \over \sqrt{z_3}\spa1.2\spa2.3\spb2.3}
+ \cdots
\cr&=
- { z_1 z_3\spa2.3\spb2.3
   \over z_2\spa1.2 \spb1.2 s_{R3}}
- {z_3 \over \spa1.2\spb1.2}
+ {z_2 z_3\over (1\tm z_3)\spa1.2\spb1.2} 
\cr &\hphantom{\longrightarrow}\hskip 5mm
- { 2\sqrt{z_1} z_3^{3/2}\spb2.3
   \over z_2 \spb1.2 s_{R3}}
 - {z_3^{3/2} \over \sqrt{z_1}\spa2.3\spb1.2}
\cr &\hphantom{\longrightarrow}\hskip 5mm
+ {2\sqrt{z_1 z_3} \spa1.3^2 
   \over \spa1.2 \spa2.3 s_{R3}}
 - {z_1\sqrt{z_2}\spa1.3  
    \over \sqrt{z_3}\spa1.2\spa2.3\spb2.3}
+ \cdots
\cr&=
- { z_1 z_3\spa2.3\spb2.3
   \over z_2\spa1.2 \spb1.2 s_{R3}}
+ {z_1 z_3 \spa2.3\spb2.3\over (1\tm z_3)\spa1.2\spb1.2 s_{R3}} 
+ {z_1 z_3 \spa1.3\spb1.3\over (1\tm z_3)\spa1.2\spb1.2 s_{R3}} 
\cr &\hphantom{\longrightarrow}\hskip 5mm
- { 2\sqrt{z_1} z_3^{3/2}\spb2.3
   \over z_2 \spb1.2 s_{R3}}
 - {z_3^{3/2} \over \sqrt{z_1}\spa2.3\spb1.2}
\cr &\hphantom{\longrightarrow}\hskip 5mm
+ {2\sqrt{z_1 z_3} \spa1.3^2 
   \over \spa1.2 \spa2.3 s_{R3}}
 - {z_1\sqrt{z_2}\spa1.3  
    \over \sqrt{z_3}\spa1.2\spa2.3\spb2.3}
+ \cdots
\cr&=
- { z_1 z_3\spa2.3\spb2.3 \over z_2\spa1.2 \spb1.2 s_{R3}}
+ {z_1 z_3 \spa2.3\spb2.3\over (z_1+z_2)\spa1.2\spb1.2 s_{R3}} 
+ {z_1^{2} z_3 \spa2.3\spb2.3\over z_2 (z_1+z_2)\spa1.2\spb1.2 s_{R3}} 
\cr &\hphantom{\longrightarrow}\hskip 5mm
- { 2\sqrt{z_1} z_3^{3/2}\spb2.3
   \over z_2 \spb1.2 s_{R3}}
 - {z_3^{3/2} \over \sqrt{z_1}\spa2.3\spb1.2}
+ {z_1^{3/2} z_3^{3/2}\spb2.3\over z_2 (z_1+z_2)\spb1.2 s_{R3}} 
\cr &\hphantom{\longrightarrow}\hskip 5mm
+ {2\sqrt{z_1 z_3} \spa1.3^2 
   \over \spa1.2 \spa2.3 s_{R3}}
 - {z_1\sqrt{z_2}\spa1.3  
    \over \sqrt{z_3}\spa1.2\spa2.3\spb2.3}
+ {z_1^{3/2} z_3^{3/2} \spa2.3\over z_2 (z_1+z_2)\spa1.2 s_{R3}} 
+ \cdots
}\anoneqn$$
The coefficient of the leading pole cancels,
so that we can now use the limiting relations 
$\spa2.3\rightarrow \sqrt{z_2\over z_1+z_2}\spa{R}.3$, etc.,
since the corrections will give rise to terms non-singular as $s_{12}\rightarrow 0$:
$$\eqalign{
 \Ctree_{+}&(1^-,2^+,3^-;z_1,z_2) \longrightarrow
\cr &
 { 2 \sqrt{z_1} z_3^{3/2}
   \over \sqrt{z_2 (z_1+z_2)}\spb1.2 \spa{R}.3}
- {z_3^{3/2}\sqrt{z_1+z_2} \over \sqrt{z_1 z_2}\spb1.2\spa{R}.3}
- {z_1^{3/2} z_3^{3/2}\over \sqrt{z_2} (z_1+z_2)^{3/2}\spb1.2\spa{R}.3} 
\cr &
- {2\sqrt{z_3} z_1^{3/2} \over \sqrt{z_2(z_1+z_2)} \spa1.2 \spb{R}.3}
 - {z_1^{3/2}\sqrt{z_1+z_2}
    \over \sqrt{z_2 z_3}\spa1.2\spb{R}.3}
- {z_1^{3/2} z_3^{3/2}\over \sqrt{z_2} (z_1+z_2)^{3/2}\spa1.2 \spb{R}.3} 
+ \cdots
\cr &=
- {z_2^2 z_3^{2} \over \sqrt{z_1 z_2 z_3} (z_1+z_2)^{3/2}\spb1.2\spa{R}.3}
- {z_1^{2} \over \sqrt{z_1 z_2 z_3} {(z_1+z_2)}^{3/2} \spa1.2 \spb{R}.3}
+ \cdots
\cr& = 
\Ctree_{-}(1^-,2^+;\textstyle{z_1\over z_1+z_2})\Ctree_{+}(R^+,3^-;z_R)
+\Ctree_{+}(1^-,2^+;\textstyle{z_1\over z_1+z_2})\Ctree_{+}(R^-,3^-;z_R)
+ \cdots\cr
}\anoneqn$$

\listrefs
\bye